\documentclass[aps,twocolumn,floatfix]{revtex4}
\usepackage{graphicx} 
\usepackage{amsmath}
\usepackage{color}
\newcommand{\bea}{\begin{eqnarray}}
\newcommand{\eea}{\end{eqnarray}}
\newcommand{\bse}{\begin{subequations}}
\newcommand{\ese}{\end{subequations}}
\newcommand{\scna}{Sr(Co$_{1-x}$Ni$_x$)$_2$As$_2$}

\begin{document}

\title{Attractive Kronig-Penney Band Structures and Wave Functions}

\author{David C. Johnston}
\affiliation{Ames Laboratory and Department of Physics and Astronomy, Iowa State University, Ames, Iowa 50011, USA}

\begin{abstract}

The repulsive-potential Kronig-Penney (KP) model for a one-dimensional band structure is well known.  However, real metals contain positively-charged ions resulting in {\it attractive} potential wells seen by the metallic electrons.  Here we consider the latter case in detail.  The square-well version of the KP model is considered first, for which the band structure and wave functions for different potential-well depths are derived.  Then an extended treatment of the attractive Dirac-comb version of the KP model is presented.  For the nearly-free-electron case, the band structure exhibits a negative-energy band in addition to positive-energy bands.  The wave functions, electron densities of states, effective masses, and group velocities are derived for the positive-energy band states.  The wave functions of the negative-energy band states are also calculated and found to be quite different from the sinusoidal wave functions for the positive-energy band and band-gap states.  High-degeneracy bound states are found at negative energies and their wave functions are derived.

\end{abstract}

\date{\today}

\maketitle

\section{\label{Introduction} Introduction}

The repulsive Kronig-Penney (KP) model~\cite{Kronig1931} is a model in which an electron is in a one-dimensional (1D) lattice of negative ions with periodic repulsive potentials.  This simple model has had an enduring pedagogical influence in presenting the electronic band structure and associated properties  of solids as described in textbooks and other books~\cite{Grosso2014,MullerKirsten2012,Kittel2005, Ashcroft1976, Harrison1970, Blatt1968, Ziman1964}. Many papers have been published treating this model in different ways~\cite{Wolfe1978, Bahurmuz1981, Singh1983, Goni1986, Nussbaum1986, DominguezAdame1987, Cannas1991, Oseguera1992, Leung1993, Szmulowicz1997, Mishra2001, Mischok2014, Rowe2014, Pavelich2015, LeVot2016, Pavelich2016}.  For example, in addition to the usual method of solution for the wave function and electron energy using the time-independent Schr\"odinger equation by enforcing conditions on the wave function at the positions of the change in potential~\cite{Kronig1931}, the plane-wave diffraction method consists of Fourier-transforming both the electrostatic potential and the wave function in terms of reciprocal lattice vectors and solving for the coefficients~\cite{Bahurmuz1981, Singh1983}.  Following the advent of the computer, accurate numerical calculations of the band structure became widely accessible~\cite{Nussbaum1986}.  The relativistic version of the KP Dirac $\delta$-function model was also studied~\cite{DominguezAdame1987}.  An externally-excited semi-infinite KP model was examined in Ref.~\cite{Cannas1991} and a classical version of the KP model in Ref.~\cite{Oseguera1992}.  The Korringa-Kohn-Rostoker method and the tail-cancellation method were used to solve the KP problem in Refs.~\cite{Leung1993} and~\cite{Mishra2001}, respectively, and numerical matrix mechanics methods were used to solve the KP model and related models in Refs.~\cite{Pavelich2015, LeVot2016, Pavelich2016}.  The KP model was recently utilized to understand the spatially-resolved optical emission spectra of an organic microcavity modulated with a periodic 1D grating of metal stripes~\cite{Mischok2014}.  This model was also used to illustrate how the electrical resistivity of a semiconductor can be affected by an applied pressure (piezoresistance), in a review of this subject~\cite{Rowe2014}.  We are not aware of any detailed studies of the wave functions and probability densities associated with the electronic band structure.

\begin{figure}
\includegraphics[width=2.in]{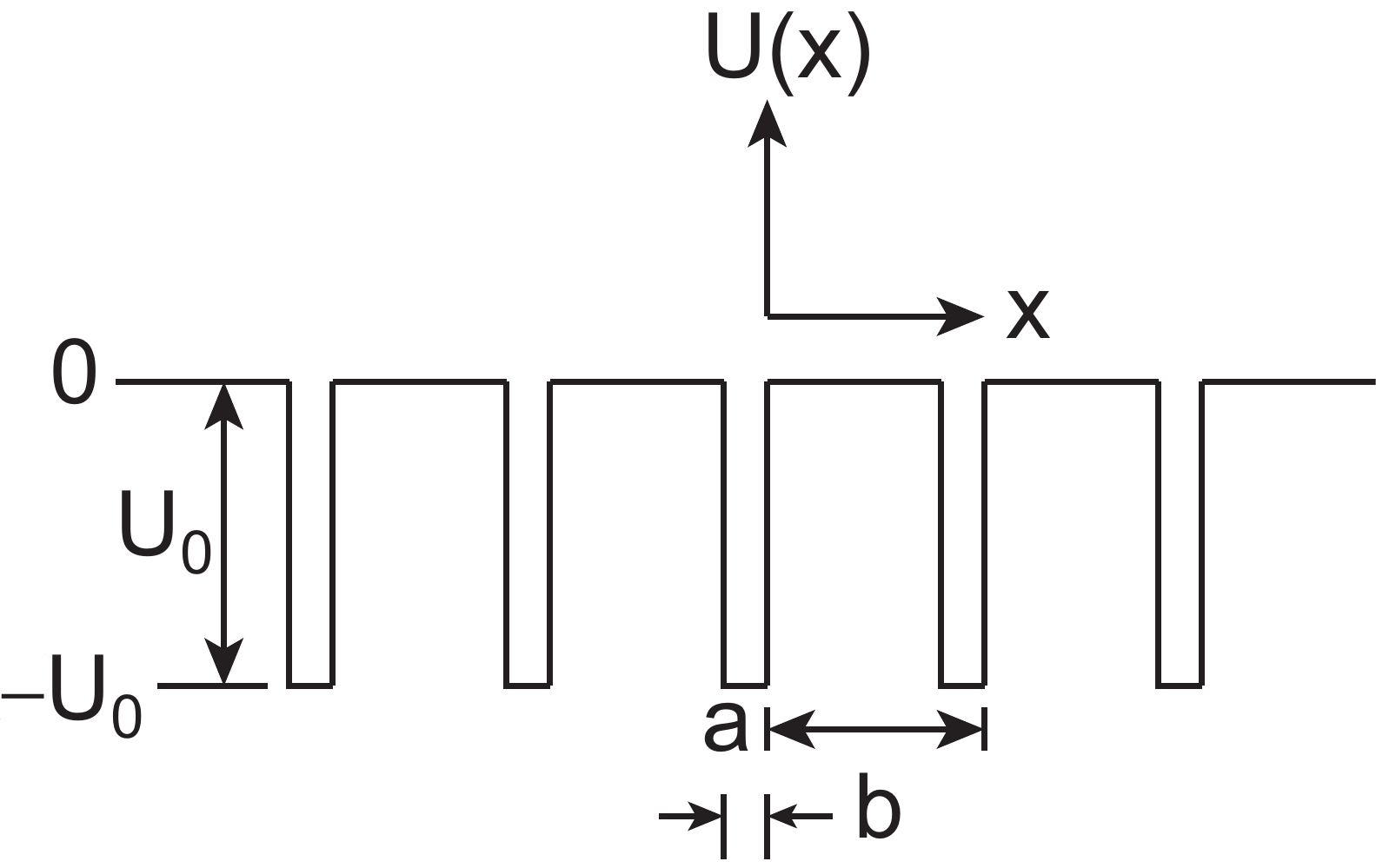}
\caption{One-dimensional KP model for the interaction of negatively-charged conduction electrons with positively-charged ions.  The distance between the centers of the ions (the lattice parameter) is $a$ and the width of the potential well with value $U(x) = -U_0$ around each ion is~$b$.  The potential energy of an electron in the region around each ion is negative (attractive) as shown.}
\label{Fig:Kronig_Penney_Model} 
\end{figure}

It seems that the original repulsive KP model with an electron in a periodic repulsive potential is unphysical, because the electron in that model would be repelled from the negatively-charged ions and the electron-ion system would be unstable.  Real materials contain positively charged ions along with the negatively-charged conduction electrons so that the solid is electrically neutral.  It is therefore of interest to study whether any new features appear in the band structure or other properties of this model.  For electrical neutrality, the number of negatively-charged electrons in the lattice is the same as the total charge on the positive ions.  However, since non-interacting electrons are considered here, the behavior of each electron is the same, and in the following we therefore only consider a single electron in the one-dimensinal positive-ion lattice.

Bloch's theorem~\cite{Bloch1929} applied to this one-dimensional problem reads
\bse
\label{Eqs:BlochTnm}
\bea
\psi(x) = e^{ikx}u(x),
\label{Eq:psi(x)u(x)}
\eea
where $\hbar k$ is the crystal momentum of the electron and the function $u(x)$ has the periodicity of the lattice.  Then
\bea
\psi(x + na) &=& e^{ik(x+na)} u(x+na) \nonumber\\
&=& e^{inka}e^{ikx} u(x)= e^{inka}\psi(x),\label{Eq:Psixplusa2}\hspace{0.3in}\\
\frac{d\psi(x)}{dx}(x+na) &=&  e^{inka}\frac{d\psi(x)}{dx}(x), \label{Eq:dpsixpma2}
\eea
\ese
where $n$ is a positive or negative integer.  Thus if $\psi(x)$ and $d\psi(x)/dx$ can be found in the interval $0\leq x \leq a$, these quantities for any $x$ outside that interval can be found from the former ones using Eqs.~(\ref{Eq:Psixplusa2}) and~(\ref{Eq:dpsixpma2}), respectively.

For real $k$ values which correspond to electron-band states, the probability density~${\cal P}$ is obtained from Eq.~(\ref{Eq:Psixplusa2}) as
\bea
{\cal P}(x+na) &=& \psi(x+na)\psi^*(x+na) \label{Eq:ProbDens}\\
&=& \psi(x)\psi^*(x) = {\cal P}(x).\nonumber
\eea
Thus ${\cal P}(x)$ is periodic with the lattice in this case.

Bloch's theorem only deals with propagating  (itinerant) electron states.  It does not describe bound-electron states that are localized to a finite number of adjacent unit cells.

Here we describe our studies of the attractive-potential KP model~\cite{Wiki2019}.  In Sec.~\ref{Sec:DispRelSquWell} the electronic band structure of the square-potential KP model is discussed.  Contrary to the repulsive potential case, we find one band of states at negative energy in the nearly-free-electron regime. With increasing depth of the potential well, an additional band that was at positive energy is pulled down into the negative-energy region.  In Sec.~\ref{Sec:Wave Functions} both the wave functions $\psi(x)$ and the periodic part $u(x)$ are derived and plotted. A special case of bound states in the square-ptential KP model is discussed in Appendix~\ref{Sec:KTBoundStates}.

Kronig and Penney also considered what happens when the periodic square repulsive potential is replaced by a repulsive Dirac comb.  This is the case of most interest in solid-state physics.  The band structure for the corresponding {\it attractive} Dirac comb potential is discussed in Sec.~\ref{Sec:deltaFcnExtended}, where the lowest-energy band is found to be at negative energy.  The wave functions for the positive-energy bands and energy-gap states are presented in Sec.~\ref{Sec:WveFcnsDiracPlus}. The electronic density of states versus energy for the positive-energy band structure is derived in Sec.~\ref{Sec:DOS} and the band effective mass versus energy in Sec.~\ref{Sec:Effective mass}.  The group velocity and band effective mass of an electron wave packet versus energy for the positive-energy bnds are derived and the influence of an electric field on the electron motion is discussed in Sec.~\ref{Sec:group velocity}. The wave functions for the negative-energy band states are calculated in Sec.~\ref{Sec:NegativeEnergyStates}, and are found to be qualitatively different from the sinusoidal wave functions for positive-energy band and energy-gap states.  Highly-degenerate wave functions for negative-energy bound states are derived and discussed in Sec.~\ref{Sec:KTBoundDiracCombStates}.  Concluding remarks are given in Sec.~\ref{Summary}.

\section{\label{Sec:DispRelSquWell} Attractive Square-Potential Kronig-Penney Model}

\subsection{Band Structure}

The attractive KP square potential versus position was shown above in Fig.~\ref{Fig:Kronig_Penney_Model}.  To solve for the energies and wave functions of the electron quantum states, one utilizes the time-independent Schr\"odinger equation, which can be written
\begin{equation}
-\frac{\partial^2\psi(x)}{\partial x^2} =\frac{2m}{\hbar^2}[E-U(x)]\psi(x).
\label{Eq:SchEqn2}
\end{equation}

\subsubsection{Region 1: $0\leq x\leq a-b$, $U(x)=0$}

Here $U(x)=0$ and Eq.~(\ref{Eq:SchEqn2}) is the same as for a free electron.  The crystal momentum (wave vector) of the electron in this region is designated as $k_1$, so Eq.~(\ref{Eq:SchEqn2}) gives
\begin{equation}
k_1^2 \equiv \frac{2mE}{\hbar^2} \geq 0
\label{Eq:k1Def}
\end{equation}
with general wave function 
\begin{equation}
\psi_1(x) = A e^{ik_1x} + A^\prime e^{-ik_1x}.
\label{Eq:psi1(x)}
\end{equation}
Using Eq.~(\ref{Eq:psi(x)u(x)}) one obtains the periodic function in Region~1 as 
\begin{equation}
u_1(x)=\frac{\psi_1(x)}{e^{ikx}} = Ae^{i(k_1-k)x} + A^\prime e^{-i(k_1+k)x}.
\label{Eq:u1}
\end{equation}

\vspace{0.1in}
\subsubsection{Region 2: $-b\leq x\leq 0$, $U(x)= -U_0$}

Here the electron wave vector has the value $k_2$ given by
\begin{subequations}
\label{Eqs:Reg2}
\begin{equation}
k_2^2 \equiv \frac{2m}{\hbar^2}(E+U_0)
\label{Eq:k2Def}
\end{equation}
with general wave function 
\begin{equation}
\psi_2(x) = B e^{ik_2 x} + B^\prime e^{-ik_2x}.
\label{Eq:psi2(x)}
\end{equation}
Again using Eq.~(\ref{Eq:psi(x)u(x)}) one obtains
\begin{equation}
u_2(x)=\frac{\psi_2(x)}{e^{ikx}} = Be^{i(k_2-k)x} + B^\prime e^{-i(k_2+k)x}.
\label{Eq:u2}
\end{equation}
\end{subequations}

\subsubsection{Applying Boundary Conditions on the Wave Function between Regions~1 and~2}

As usual for steplike potential energy changes, the wave functions and their derivatives must be continuous across a step.  In our problem, this is expressed as 
\begin{subequations}
\label{Eqs:BCs}
\bea
\psi_1(x=0) &=& \psi_2(x=0),\\
\frac{d\psi_1}{dx}(x=0) &=& \frac{d\psi_2}{dx}(x=0).
\eea
Also, since the potential energy is periodic, Bloch's theorem requires that 
\begin{eqnarray}
u_1(a-b) &=& u_2(-b),\\
\frac{d u_1}{dx}(a-b) &=& \frac{d u_2}{dx}(-b).
\end{eqnarray}
\end{subequations}
Using Eqs.~(\ref{Eq:psi1(x)}), (\ref{Eq:u1}), (\ref{Eq:psi2(x)}), and~(\ref{Eq:u2}), the four Eqs.~(\ref{Eqs:BCs}) yield the linear homogeneous matrix equation~\cite{Wiki2019}
\begin{widetext}
\begin{gather}
\begin{pmatrix}
1 & 1 & -1 & -1\\
k_1  & -k_1 & - k_2 & k_2\\
e^{i(a-b)(k_1-k)} & e^{-i(a-b)(k_1+k)} & -e^{-ib(k_2-k)} & -e^{ib(k_2+k)}\\
e^{i(a-b)(k_1-k)}(k_1-k) & -e^{-i(a-b)(k_1+k)}(k_1+k) & -e^{ib(k_2-k)}(k_2 - k) & e^{ib(k_2+k)}(k_2+k) 
\end{pmatrix}
\begin{pmatrix}
A \\
A^\prime \\
B \\
B^\prime 
\end{pmatrix}
=
\begin{pmatrix}
0 \\
0 \\
0 \\
0 
\end{pmatrix}
.
\label{Eq:pmatrix}
\end{gather}
To obtain nonzero solutions for the coefficients $A,\ A^\prime,\ B$ and~$B^\prime$, the determinant of the matrix must vanish.  This condition yields
\begin{equation}
\cos(ka) = \cos[k_1(a-b)]\cos(k_2b) - \left(\frac{k_1^2 + k_2^2}{2k_1k_2}\right)\sin[k_1(a-b)]\sin(k_2b).
\label{Eq:coska}
\end{equation}
This expression relates the crystal momentum $k$ of an electron to the kinetic and potential energies of the electron contained in $k_1$ and~$k_2$ according to Eqs.~(\ref{Eq:k1Def}) and~(\ref{Eq:k2Def}).

It is convenient to introduce the dimensionless reduced variables 
\begin{subequations}
\begin{equation}
b_a = b/a,\qquad \varepsilon = \frac{2ma^2E}{\hbar^2}, \qquad Q = \frac{2ma^2U_0}{\hbar^2}, \qquad k_1a = \sqrt{\varepsilon}, \qquad k_2a = \sqrt{\varepsilon+Q},
\label{Eq:SymbolDefs}
\end{equation}
where  Eqs.~(\ref{Eq:k1Def}) and~(\ref{Eq:k2Def}) were used to obtain the last two expressions.  Then Eq.~(\ref{Eq:coska}) becomes
\begin{equation}
\cos(ka) = \cos[\sqrt{\varepsilon}(1-b_a)]\cos\left(\sqrt{\varepsilon + Q}\,b_a\right) - \left[\frac{2\varepsilon + Q}{2{\sqrt{\varepsilon(\varepsilon + Q)}}}\right]\sin[\sqrt{\varepsilon}(1-b_a)]\sin\left(\sqrt{\varepsilon + Q}\,b_a\right).
\label{Eq:coska2}
\end{equation}
\end{subequations}
\end{widetext}

In the limit $U_0\to 0$ ($Q\to0$), Eq.~(\ref{Eq:coska2}) reduces to
\bea
\cos(ka) &=& \cos[\sqrt{\varepsilon}(1-b_a)]\cos\left(\sqrt{\varepsilon}\,b_a\right)\label{Eq:coska55}\\
&&  -\ \sin[\sqrt{\varepsilon}(1-b_a)]\sin\left(\sqrt{\varepsilon}\,b_a\right)\nonumber\\
&=& \cos(\sqrt{\varepsilon}).\nonumber
\eea
Thus one has
\bse
\label{Eqs:FreeElectronBand}
\bea
\varepsilon &=& (ka)^2. 
\label{Eq:coska2a}
\eea
Then using the second of Eqs.~(\ref{Eq:SymbolDefs}), Eq.~(\ref{Eq:coska2a}) gives the expected free-electron dispersion relation
\bea
E=\frac{\hbar^2k^2}{2m}.
\label{Eq:FEDR}
\eea
\ese

As will be shown, negative-energy propagating solutions exist for the attractive KP model. Furthermore, if one wishes to calculate the dispersion relation for the traditional repulsive KP model the sign of $Q$ is negative.  Depending on whether one or both of $\varepsilon$ and $Q$ are negative and whether $\varepsilon + Q$ is positive or negative, one or both of each of the sine and cosine functions in Eq.~(\ref{Eq:coska2}) become hyperbolic sine and hyperbolic cosine functions and their prefactors are modified accordingly.

\begin{figure}
\includegraphics[width=3in]{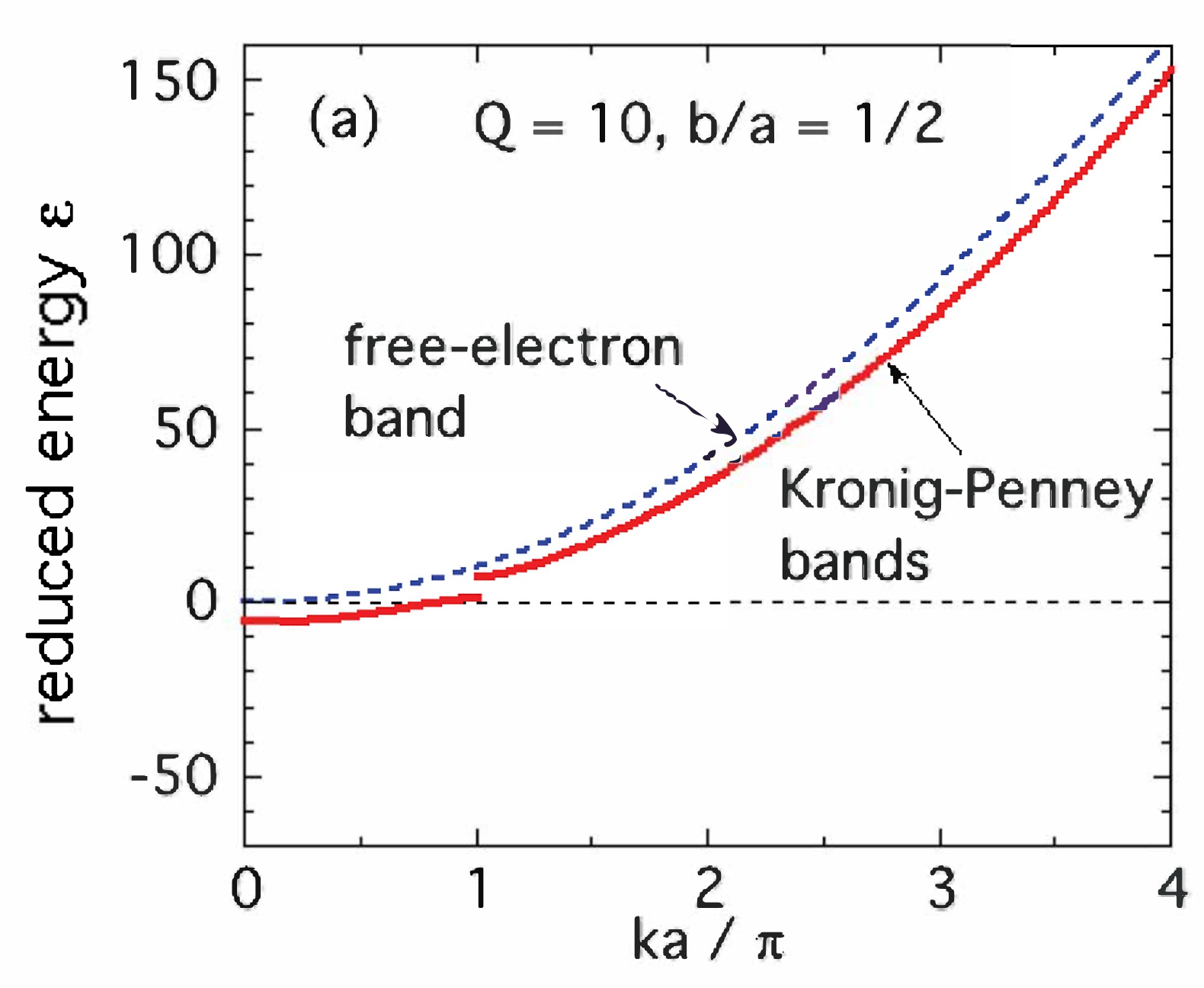}
\includegraphics[width=3in]{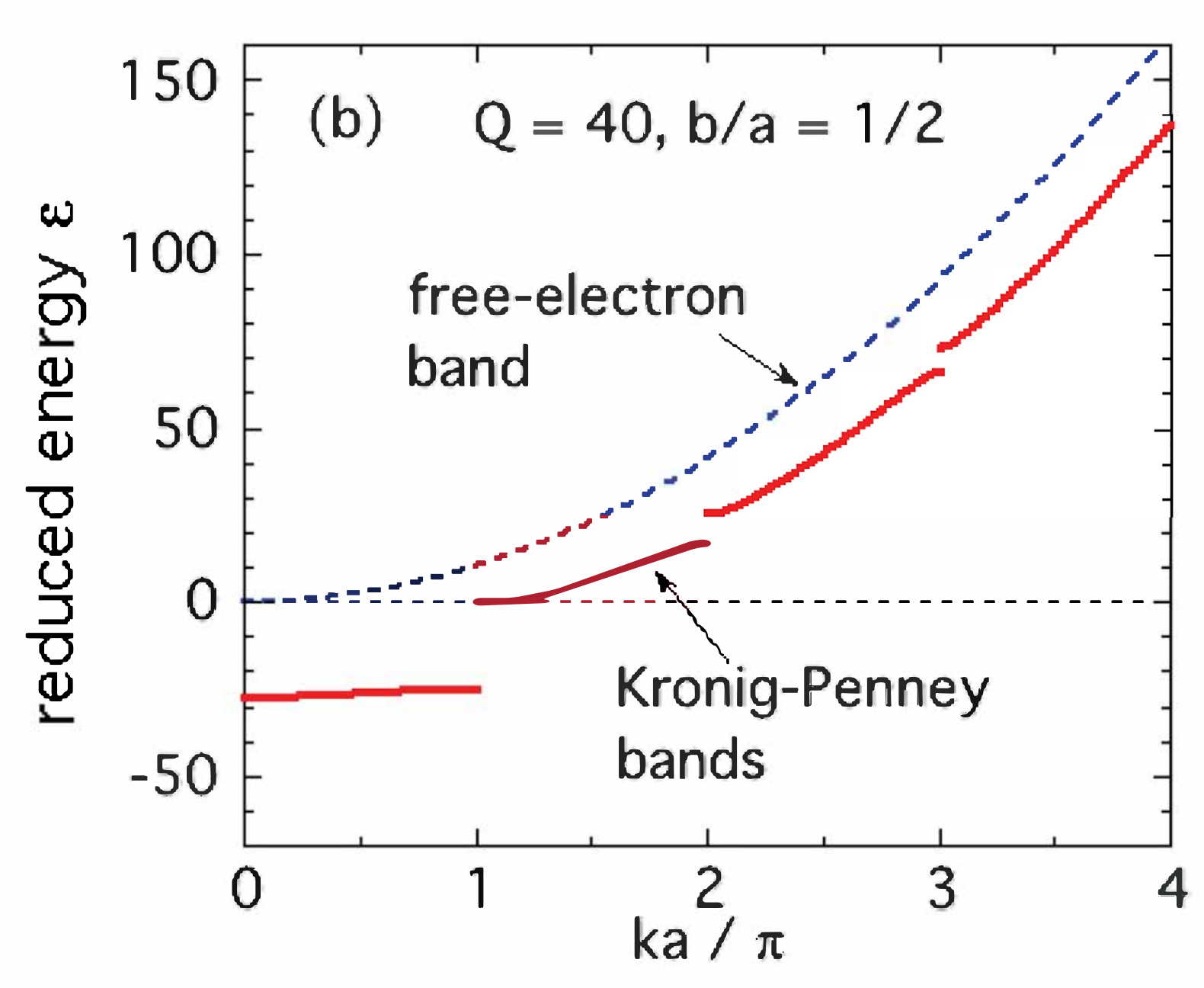}
\includegraphics[width=3in]{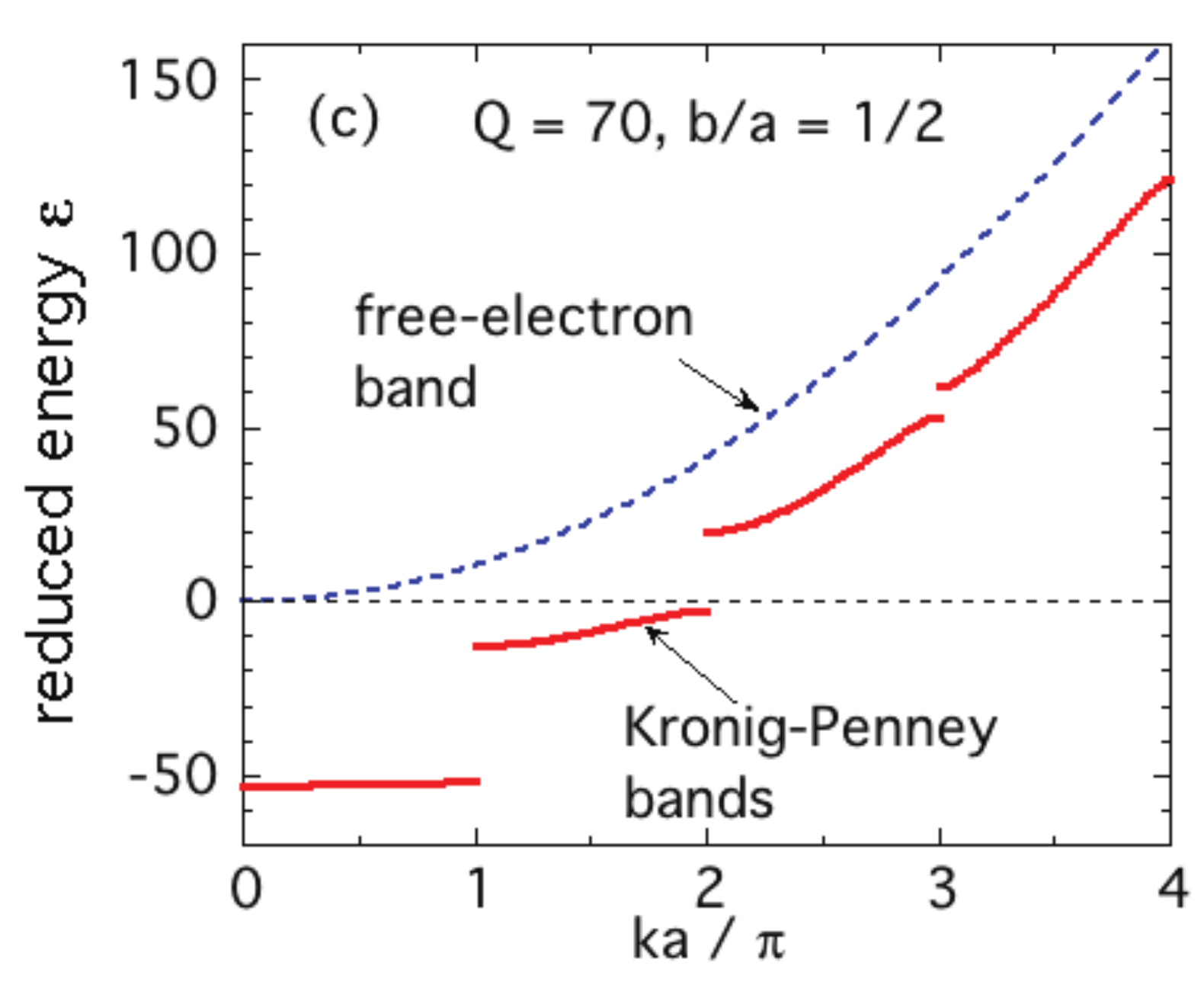}
\caption{Band Structure $\varepsilon$ versus $ka/\pi$ in the extended-zone scheme for the first four bands of the attractive KP model obtained using Eq.~(\ref{Eq:coska2}) with $b/a=1/2$ and the listed strengths~$Q$ of the potential well. Vertical band gaps occur for the bands at Brillouin-zone boundaries $ka = n\pi$ with integer~$n$.  Band gaps at $ka/\pi\geq 2$ for $Q=10$ are present but are too small to see clearly in panel~(a).  Also shown in each panel for comparison is the dispersion relation $\varepsilon = (ka)^2$ for free electrons from Eq.~(\ref{Eq:coska2a}).}
\label{Fig:KP_Evska_Q10_40_70}
\end{figure}

\begin{figure}
\includegraphics[width=3in]{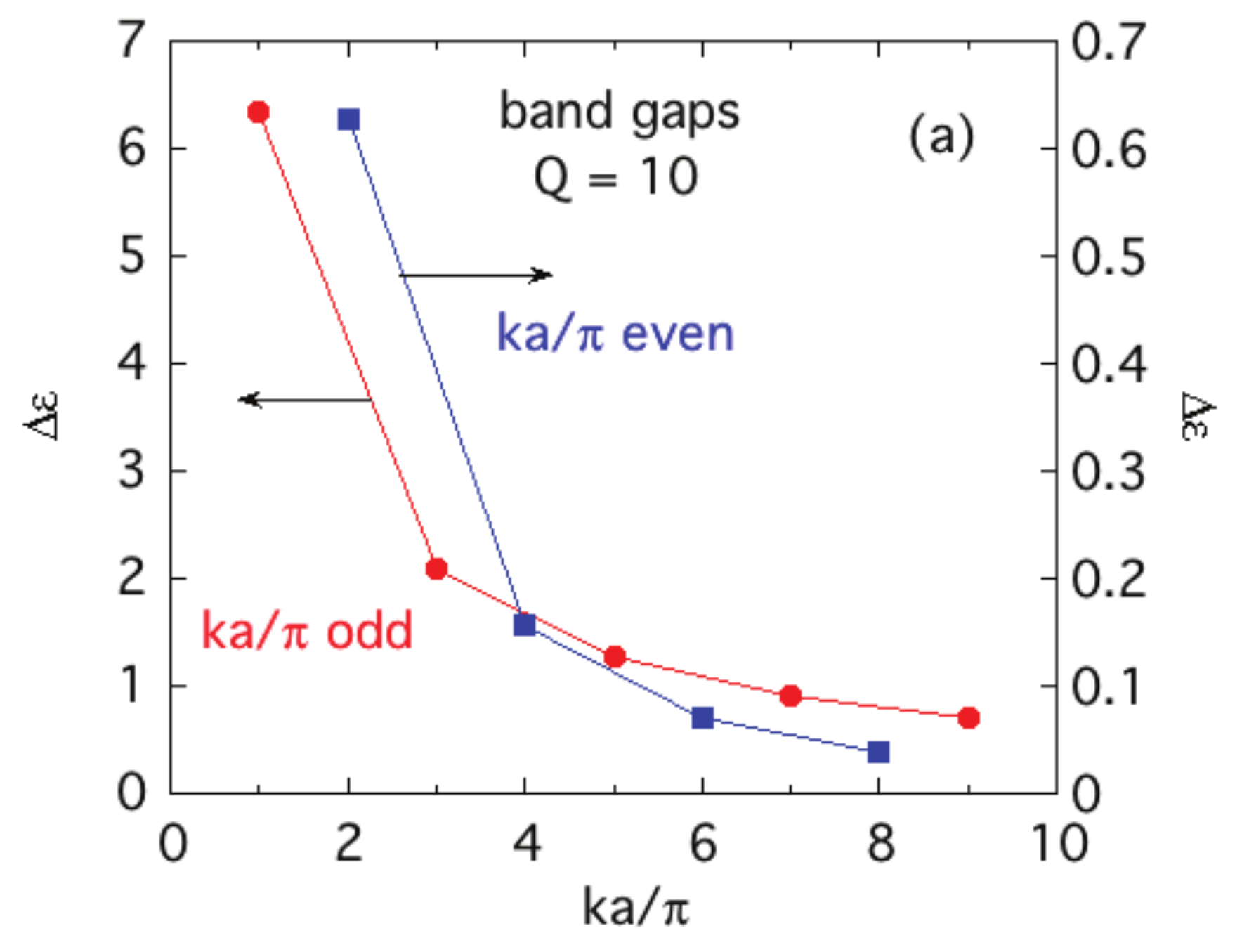}
\includegraphics[width=3in]{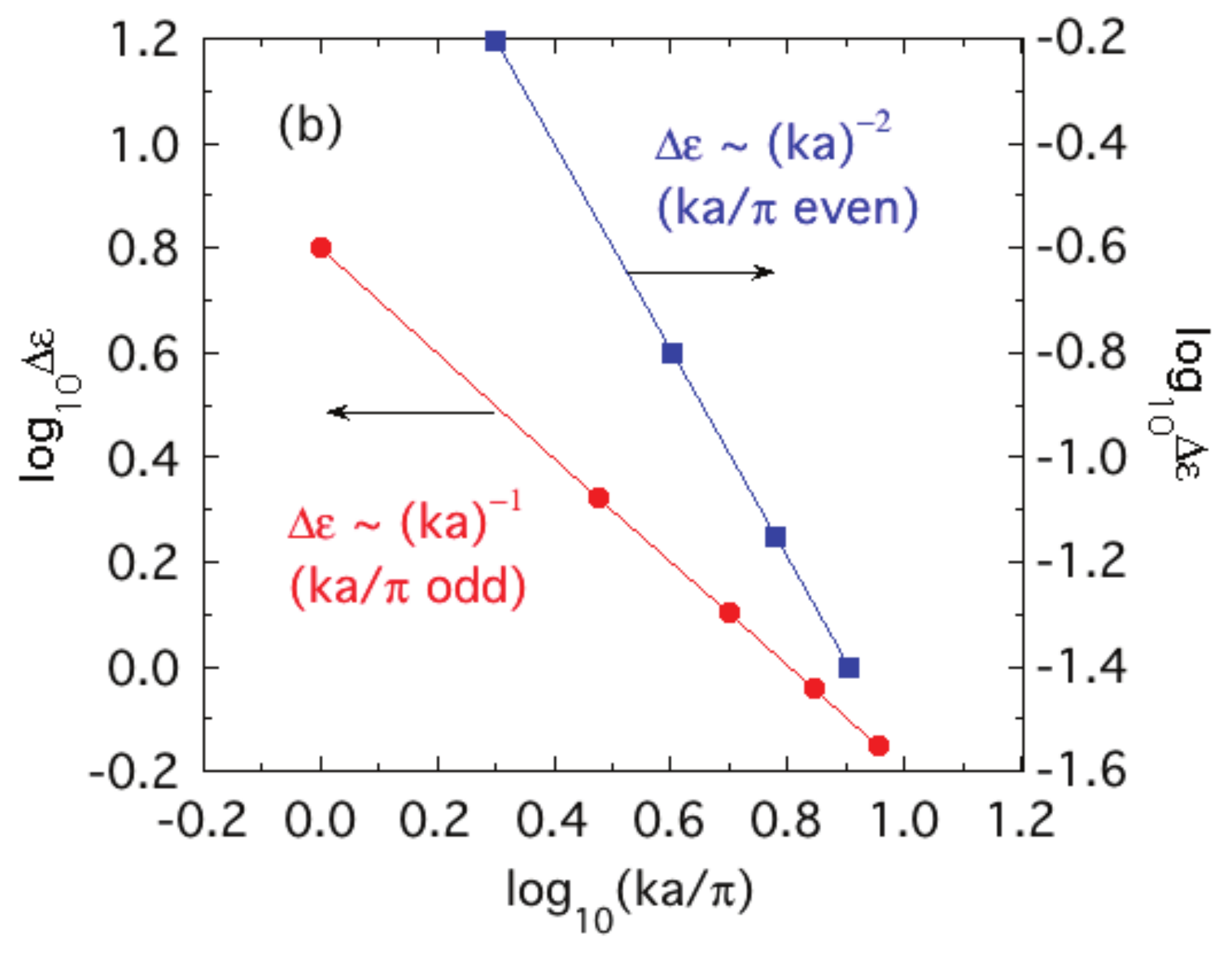}
\caption{(a) Band gap $\Delta\varepsilon$ versus $ka/\pi$ at the first nine Brillouin boundaries $ka=n\pi$ of the attractive KP model obtained using Eq.~(\ref{Eq:coska2}) with $b/a=1/2$ and strength~$Q=10$ of the potential well. (b)~Log-log plot of the data in~(a), where a clear difference between the power-law behavior of even-$n$ and odd-$n$ gaps is seen.}
\label{Fig:KPBandEdgesAvesGapsQ10}
\end{figure}

The band structure for specified values of the potential-well strength $Q$ and $b/a$ ratio is obtained by numerically solving Eq.~(\ref{Eq:coska2}). This is done in two steps.  First the energies of the bands at wave vectors at the centers of the respective Brillouin zones are determined by solving Eq.~(\ref{Eq:coska2}) for $\cos(ka) = 0$ with increasing energy.  Then using each of these energies as an initial parameter, we numerically solve for $\varepsilon$ as a function of $ka$ for band~$p$ which resides between $ka = (p-1)\pi$ and $p\pi$ in the extended-zone scheme.

The band structures for the first four Brillouin zones with $Q = 10$, 40, and 70 and $b/a=1/2$ are shown in Fig.~\ref{Fig:KP_Evska_Q10_40_70}.  One sees that  the attractive potential lowers the energies of all band states with respect to the free-electron dispersion and this influence is strongest for the lowest-energy bands.  Indeed, by $Q=70$ the lowest-two bands are submerged in the negative-energy sea.  Band gaps $\Delta\varepsilon$ occur at integer values of the reduced Brillouin-zone boundary wave vectors~$ka/\pi$  and the respective values increase with increasing~$Q$\@. For $Q=70$, the lowest-energy band has very little ddispersion. 

It is also apparent that the value of $\Delta\varepsilon$  decreases with increasing energy for $b/a=1/2$ and a given~$Q$\@. A plot of $\Delta\varepsilon$ versus $ka/\pi$ for $Q=10$ is shown in Fig.~\ref{Fig:KPBandEdgesAvesGapsQ10}(a) and a log-log plot is shown in Fig.~\ref{Fig:KPBandEdgesAvesGapsQ10}(b).  One sees that with increasing $ka/\pi$, the band gaps for even and odd values of $ka/\pi$ have different dependences on $ka/\pi$.  In particular, we obtain $\Delta\varepsilon \sim (ka)^{-1}$ for odd values of $ka/\pi$ whereas for even values we find $\Delta\varepsilon \sim (ka)^{-2}$.

\newpage

\subsection{\label{Sec:Wave Functions} Wave Functions $u(x)$ and $\psi(x)$}

Equations~(\ref{Eq:pmatrix}) allow the wave functions $u(x)$ and $\psi(x)$ in Bloch's theorem~(\ref{Eq:psi(x)u(x)}) to be obtained for given values of $b/a$ and $Q$.  The first three of Eqs.~(\ref{Eq:pmatrix}) give
\begin{widetext}
\bse
\label{Eqs:ApBBp}
\bea
\frac{A^\prime}{A} &=& \frac{k_1a - k_2a + 2 k_2a B^\prime/A}{k_1a+k_2a},\\
\frac{B}{A} &=& \frac{2 k_1a + (k_2a-k_1a) B^\prime/A}{k_1a+k_2a}\\
\frac{B^\prime}{A} &=& \frac{2k_1a e^{i(ka+k_1a)} + (k_2a - k_1a)e^{ib_a(k_1a+k_2a)} - (k_1a+k_2a)e^{i[2k_1a+b_a(k_2a-k_1a)]}}
{2k_2ae^{ib_a(k_1a+k_2a)}-e^{i(ka+k_1a)}\left[k_2a-k_1a + (k_1a+k_2a)e^{i2b_ak_2a}\right]}\label{Eq:BpA1}
\eea
\ese
The fourth of Eqs.~(\ref{Eq:pmatrix}) gives an additional equation for $B^\prime/A$ which when equated with Eq.~(\ref{Eq:BpA1})  and simplified gives the dispersion relation~(\ref{Eq:coska2}).  The three Eqs.~(\ref{Eqs:ApBBp}) leave the coefficient~$A$ unspecified.  That parameter is determined when the wave function is normalized (see below).  Changing variables according to Eqs.~(\ref{Eq:SymbolDefs}) gives
\bse
\label{Eqs:ApBBpRed}
\bea
\frac{A^\prime}{A} &=& \frac{\sqrt{\varepsilon} - \sqrt{\varepsilon+Q} + 2 \sqrt{\varepsilon+Q} B^\prime/A}{\sqrt{\varepsilon}+\sqrt{\varepsilon+Q}},\\
\frac{B}{A} &=& \frac{2 \sqrt{\varepsilon} + (\sqrt{\varepsilon+Q}-\sqrt{\varepsilon}) B^\prime/A}{\sqrt{\varepsilon}+\sqrt{\varepsilon+Q}},\\
\frac{B^\prime}{A} &=& \frac{2\sqrt{\varepsilon} e^{i(ka+\sqrt{\varepsilon})} + (\sqrt{\varepsilon+Q} - \sqrt{\varepsilon})e^{ib_a(\sqrt{\varepsilon}+\sqrt{\varepsilon+Q})} - (\sqrt{\varepsilon}+\sqrt{\varepsilon+Q})e^{i[2\sqrt{\varepsilon}+b_a(\sqrt{\varepsilon+Q}-\sqrt{\varepsilon})]}}
{2\sqrt{\varepsilon+Q}e^{ib_a(\sqrt{\varepsilon}+\sqrt{\varepsilon+Q})}-e^{i(ka + \sqrt{\varepsilon})}\left[\sqrt{\varepsilon+Q}-\sqrt{\varepsilon} + (\sqrt{\varepsilon}+\sqrt{\varepsilon+Q})e^{i2b_a\sqrt{\varepsilon+Q}}\right]},
\eea
\ese
\end{widetext}
where $ka$ and~$(\varepsilon)$ related by Eq.~(\ref{Eq:coska2}).

The coefficients $A^\prime$, $B$, and $B^\prime$ of the wave functions are determined for given values of $Q$, $b/a$ and $ka(\varepsilon)$, initially taking $A=1$. We first verify that the periodic function $u(x)$ and its slope $du/dx$ are continuous at $x=0$, $x=-b$, and $x=a-b$ and are indeed periodic in~$a$, where
\bea
u(x) &=& u_1(x) \qquad (0 \leq x/a \leq 1-b/a),\label{Eq:u(x)}\\
u(x) &=& u_2(x) \qquad (- b/a \leq x/a \leq 0)\nonumber.
\eea  
From Eqs.~(\ref{Eq:u1}) and~(\ref{Eq:u2}) one obtains
\bse
\label{Eqs:u1u2A}
\bea
u_1(x) &=& Ae^{-ika\, x_a}\left(e^{i k_1a\, x_a} + \frac{A^\prime}{A} e^{-ik_1a\,x_a}\right),\label{Eq:u1A}\\
u_2(x) &=& Ae^{-ika\, x_a}\left(\frac{B}{A} e^{i k_2a\, x_a} + \frac{B^\prime}{A} e^{-i k_2a\, x_a}\right),\hspace{0.4in} \label{Eq:u2A}
\eea
\ese
where the coefficients $A^\prime/A$, $B/A$, and $B^\prime/A$ are calculated using Eqs.~(\ref{Eqs:ApBBpRed}).  The coefficient~$A$ is determined from the normalization condition within one unit cell according to the prescrption
\bea
\frac{1}{A} &=& \sqrt{\int_{-b/a}^{1-b/a}u(x_a)u^*(x_a)dx_a},
\label{Eq:1OnA}
\eea
where $u^*(x_a)$ is the complex conjugate of $u(x_a)$.

\begin{figure*}
\includegraphics[width=7.in]{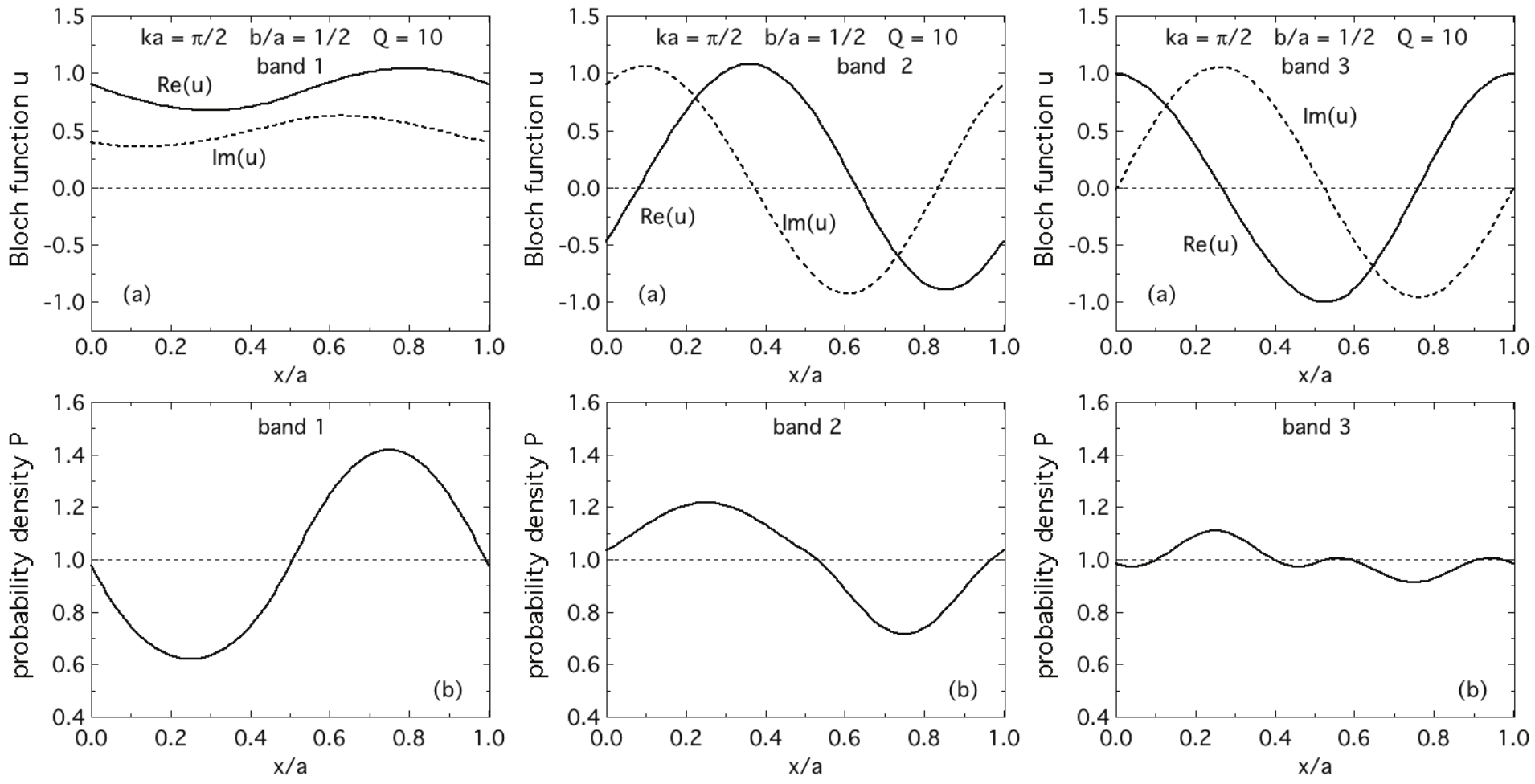}
\caption{(a) Wave functions~$u$ and (b) probability densities~$P$ for bands 1--3 in the attractive square-well KP model at the center \mbox{$ka=\pi/2$~rad} of the bands in the reduced-zone scheme versus position~$x/a$.  The plots were calculated using Eqs.~(\ref{Eqs:ApBBpRed})--(\ref{Eq:1OnA}) with the parameters $Q=10$ and $b/a = 2$.}
\label{Fig:Q10ba0.5ka0.5Bandu}
\end{figure*}

\begin{figure*}
\includegraphics[width=7.in]{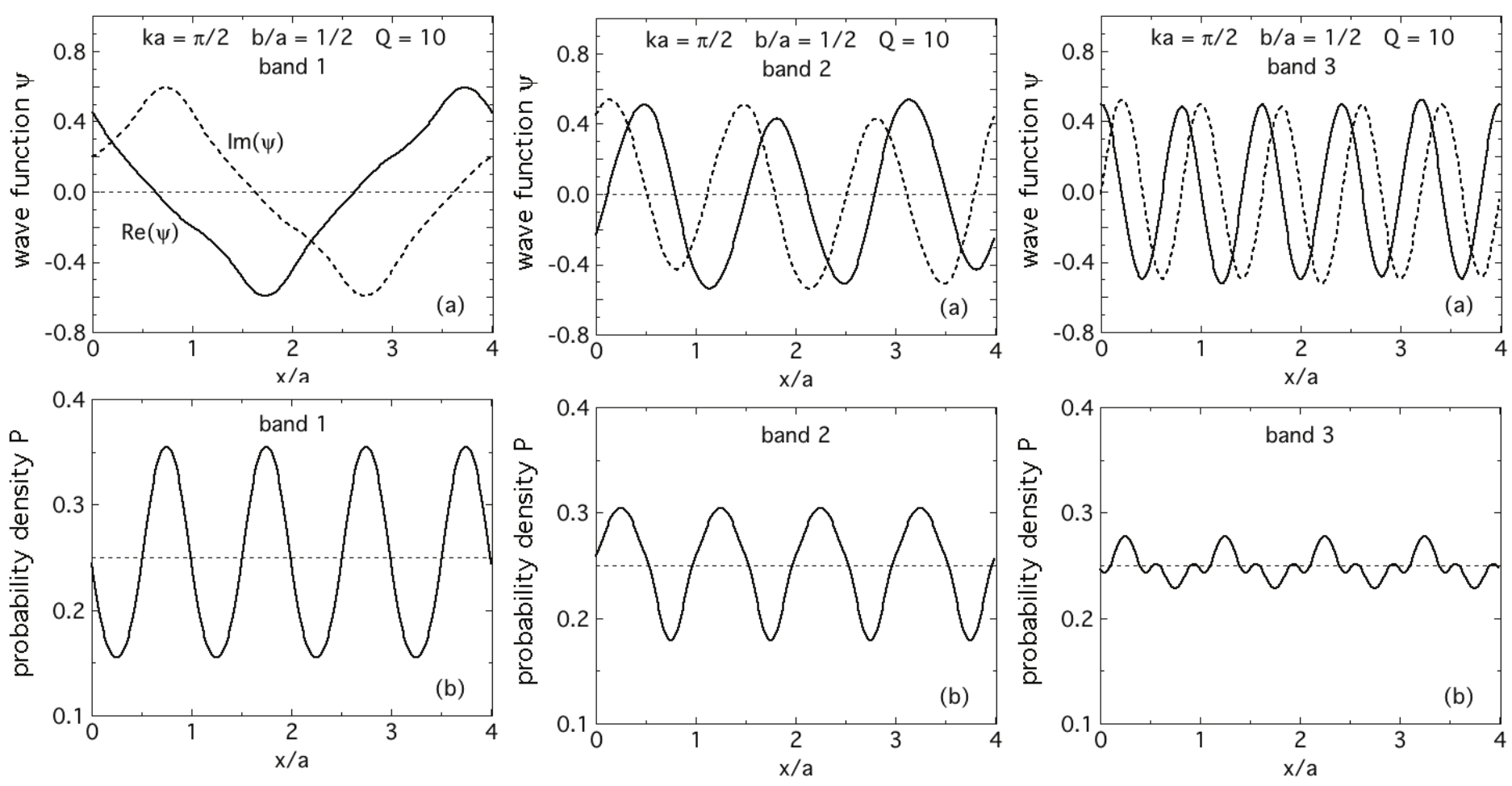}
\caption{(a) Wave functions~$\psi$ and (b)~probability densities~$P$ for bands 1--3 in the attractive square-well KP model at the center \mbox{$ka=\pi/2$~rad} of the bands in the reduced-zone scheme versus position~$x/a$.  The plots were calculated using Eqs.~(\ref{Eqs:ApBBpRed})--(\ref{Eq:psixaKP}) with the parameters $Q=10$ and $b/a = 2$.  The wavelength of $\psi(x_a)$ is $2\pi/ka = 4$ in units of $x/a$.}
\label{Fig:Q10ba0.5ka0.5Bands}
\end{figure*}

The wave functions $u(x/a)$ are plotted using Eqs.~(\ref{Eq:u(x)}) and~(\ref{Eqs:u1u2A}) in Fig.~\ref{Fig:Q10ba0.5ka0.5Bandu} at the middle of bands 1--3 with reduced-zone-scheme wave vector $ka/\pi = 1/2$ for $Q = 10$ and $b/a=1/2$.  The plots confirm that the period of $u(x/a)=1$ and that both $u(x)$ and $du(x)/dx$ are continuous at the unit-cell boundaries.

The wave function~$\psi$ versus normalized position~$x/a$ is given by the Bloch theorem~(\ref{Eq:psi(x)u(x)}) as
\bea
\psi(x/a) = e^{ika(x/a)}u(x/a).
\label{Eq:psixaKP}
\eea
Here we calculate $\psi$ versus~$x/a$ for the first three bands at wave vector $ka = \pi/2$~rad in the reduced-zone scheme.  The wavelength~$\lambda$ of~$\psi$ is given by
\bea
\lambda = \frac{2\pi}{ka}a.
\eea
Thus for $ka=\pi/2$~rad one obtains $\lambda = 4a$.  Figure~\ref{Fig:Q10ba0.5ka0.5Bands} shows plots of the real and imaginary parts of $\psi$ and the probability density versus position $x/a$ for the parameters $Q=10$ and $b/a=1/2$ for which the band structure in Fig.~\ref{Fig:KP_Evska_Q10_40_70}(a) was calculated.  The integrals of the probability densities between $x/a = 0$ and~4 were each normalized to unity.  One sees that the wave functions and probability densities are not necessarily sinusoidal.  In addition, with increasing band number (increasing energy) the dependence of the probability density on position becomes progressively weaker.  For a free-electron band, the probability density would be independent of position.

A particular case of the bound states of the attractive square-potential KP model is discussed in Appendix~\ref{Sec:KTBoundStates}.


\section{\label{Sec:DiracComb} Kronig-Penney Model with an Attractive Dirac-Comb Potential}

Kronig and Penney also considered what happens to the band structure if the width~$b$ of the potential well in Fig.~\ref{Fig:Kronig_Penney_Model} decreases to an infinitesimal value and at the same time $U_0$ increases to infinity, where
\bse
\label{Eqs:limits}
\begin{equation}
U_0\to\infty,\qquad b\to0,\quad {\rm but}\hspace{0.1in}\ \alpha = U_0b= {\rm constant}.
\label{Eq:alpha}
\end{equation}
For the attractive-potential case,  the strength of a Dirac $\delta$-function is $-\alpha$. The electric potential energy $U(x)$ between an electron and the lattice of positive ions is
\bea
U(x) = -\alpha\sum_{n=0}^{N-1}\delta(x -n a).
\label{Eq:DiracComb}
\eea
\ese
This so-called Dirac comb is similar to Fig.~\ref{Fig:Kronig_Penney_Model} but where the potential wells have infinitesimal width and infinite depth with finite area.

\subsection{\label{Sec:deltaFcnExtended} Band Structure}

One can derive the dispersion relation for the attractive Dirac comb potential energy by taking the following limits of Eq.~(\ref{Eq:coska}): 
\bea
k_2^2b&\to& {\rm constant},\quad k_2b\ll 1, \quad k_1/k_2\to0,\quad k_1^2b\to 0,\nonumber\\
&& \sin(k_2b)\to k_2b,\qquad \cos(k_2b)\to 1.\label{Eqs:KTlimits}
\eea
Using the limits in Eqs.~(\ref{Eq:alpha}) and~(\ref{Eqs:KTlimits}), Eq.~(\ref{Eq:coska}) becomes simply 
\begin{equation}
\cos(ka) = \cos(k_1a) - P\frac{\sin(k_1a)}{k_1a},
\label{Eq:coskak1a}
\end{equation}
where
\bea
P\equiv \frac{am\alpha}{\hbar^2}
\label{Eq:PDef}
\eea
is a dimensionless measure of the strength of a $\delta$~function.  Thus the value of $ka$ is closely tied to that of $k_1a$ and so is the energy~$E$ in Eq.~(\ref{Eq:k1Def}).

We now write $k_1a$ in Eq.~(\ref{Eq:coskak1a}) in terms of the energy to obtain the band structure.  First we define the dimensionless reduced energy $\varepsilon$ as
\begin{equation}
\varepsilon = \frac{2ma^2E}{\hbar^2}.
\label{epsDefdFcn}
\end{equation}
Using Eqs.~(\ref{Eq:k1Def}) and~(\ref{epsDefdFcn}) gives
\begin{equation}
k_1a=\sqrt{\varepsilon},
\label{Eq:k1aE}
\end{equation}
so Eq.~(\ref{Eq:coskak1a}) becomes
\begin{equation}
\cos(ka) = \cos(\sqrt{\varepsilon}) - P\frac{\sin(\sqrt{\varepsilon})}{\sqrt{\varepsilon}},
\label{Eq:coskak1a2}
\end{equation}
If the potential energy $U_0=0$, then one expects to find the free-electron dispersion relation. Setting $P\propto U_0 = 0$ in Eq.~(\ref{Eq:PDef}), Eq.~(\ref{Eq:coskak1a2}) becomes
\bse
\bea
\varepsilon = (ka)^2.
\label{Eq:FEepska2}
\eea
Inserting the definition of $\varepsilon$ in Eq.~(\ref{epsDefdFcn}) into this equation indeed gives the free-electron dispersion relation
\bea
E = \frac{\hbar^2 k^2}{2m}.
\label{Eq:FEDisp}
\eea
\ese

The band structure $\varepsilon$ versus $ka$ is calculated for a given value of~$P$ in two steps.  First we find energy of the bands at the center of the first Brillouin zone in the reduced-zone scheme by setting $\cos(ka) = 0$ and solving for $\varepsilon_0$.  Then we obtain $\varepsilon$ with $0 \leq ka \leq \pi$ using Eq.~(\ref{Eq:coskak1a2}) and a solution search routine with starting values $\varepsilon = \varepsilon_0$ by numerically solving for $\varepsilon(ka)$ from
\begin{equation}
ka = \arccos\left[\cos(\sqrt{\varepsilon}) - P\frac{\sin(\sqrt{\varepsilon})}{\sqrt{\varepsilon}}\right]
\label{Eq:caFromEps}
\end{equation}
that was obtained from Eq.~(\ref{Eq:coskak1a2}).  Carrying out the calculation this way, one can obtain numerically exact values for the energies of the band edges and then the band gaps versus~Brillouin-zone boundary number $n$.  In the band gaps, the value of $ka$ is complex with the real part alternating betweeen $\pi$ and~0 with increasing~$n$, starting with $\pi$ at $n=1$.  When plotting the band structure $\varepsilon$ versus~$ka$, one only includes $\varepsilon$ values for real $ka$. 

\begin{figure}
\includegraphics[width=3.in]{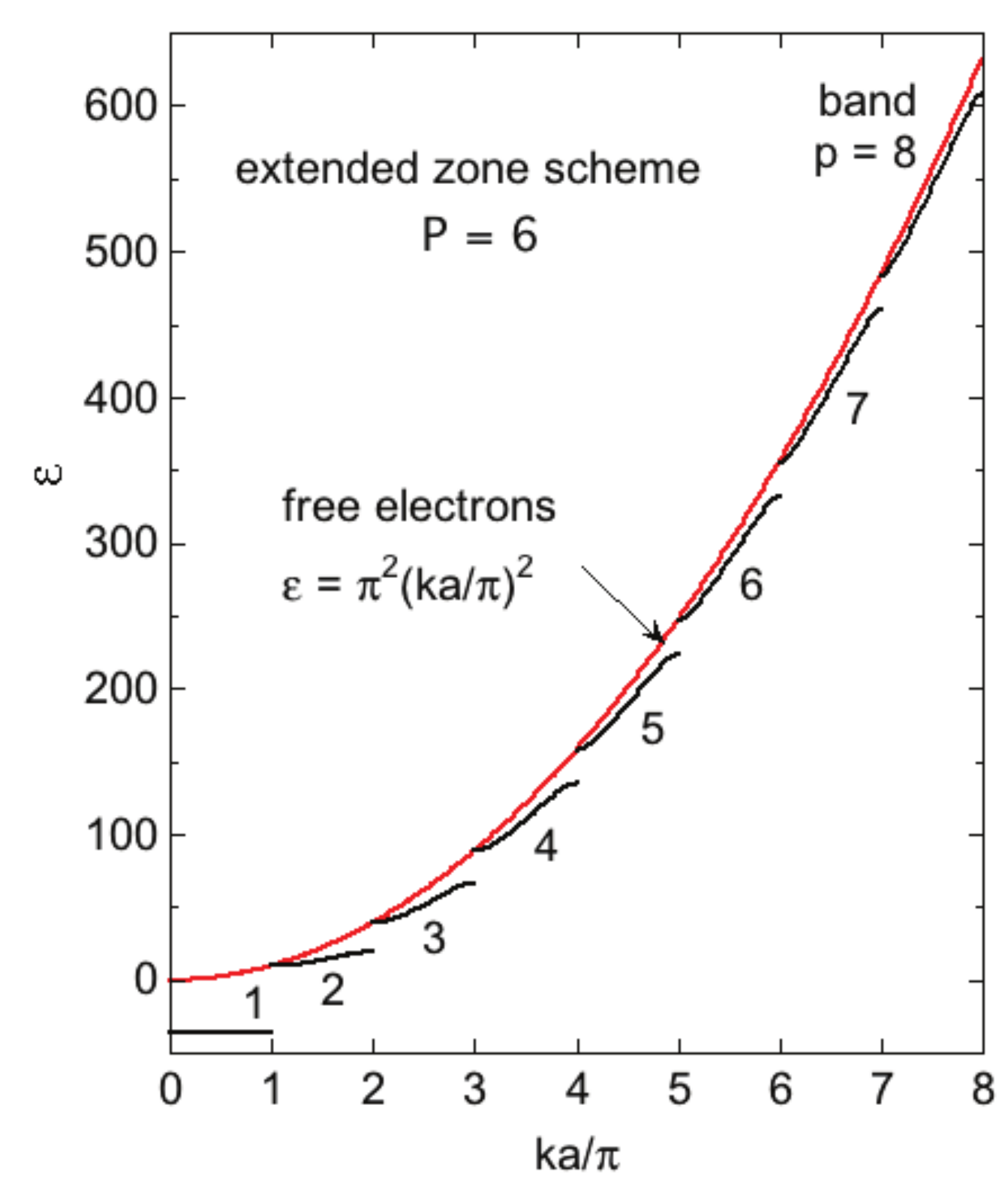}
\caption{Band structure for the attractive Dirac $\delta$-function potential energy of the one-dimensional KP model in the extended-zone scheme, expressed as the reduced energy $\varepsilon = 2ma^2E/\hbar^2$ versus reduced wave vector $ka/\pi$.  The data were calculated from Eq.~(\ref{Eq:coskak1a2}) using $P=6$.  The red curve is the free-electron expression~(\ref{Eq:FEepska2}), which intersects the tops of the bands $p\geq 2$ according to Eq.~(\ref{Eq:Eps-(p)}).  On the scale of the figure, band~1 appears dispersionless, but it is not; these band states do increase slightly in energy with increasing $ka$ as shown in Table~\ref{Tab:GapInfo}.}
\label{Fig:E_vs_ka_P6_extended_zone} 
\end{figure}

The band structure is plotted for $P=6$ in the extended-zone scheme in Fig.~\ref{Fig:E_vs_ka_P6_extended_zone} for bands $p=1$ to~8.  This corresponds to the so-called nearly-free-electron model when $P$ is ``small''.  One sees that band~1 has negative energy, as found for the square-well KP model in Fig.~\ref{Fig:KP_Evska_Q10_40_70}.  Band gaps occur at the Brillouin zone~(BZ)  boundaries in the figure at $ka = n\pi$ ($n=1$ to~8 here) with values listed in Table~\ref{Tab:GapInfo}, together with the reduced energies at the edges and midpoint of each band gap.  The negative-energy band~1 has a small but finite dispersion (see Table~\ref{Tab:GapInfo}).  This is reminiscent of flat bands in some three-dimensional solids ({\it e.g}, Ref.~\cite{Sangeetha2019}) and in twisted bilayer graphene~\cite{Bistritzer2011}.

\begin{figure}
\includegraphics[width=3.in]{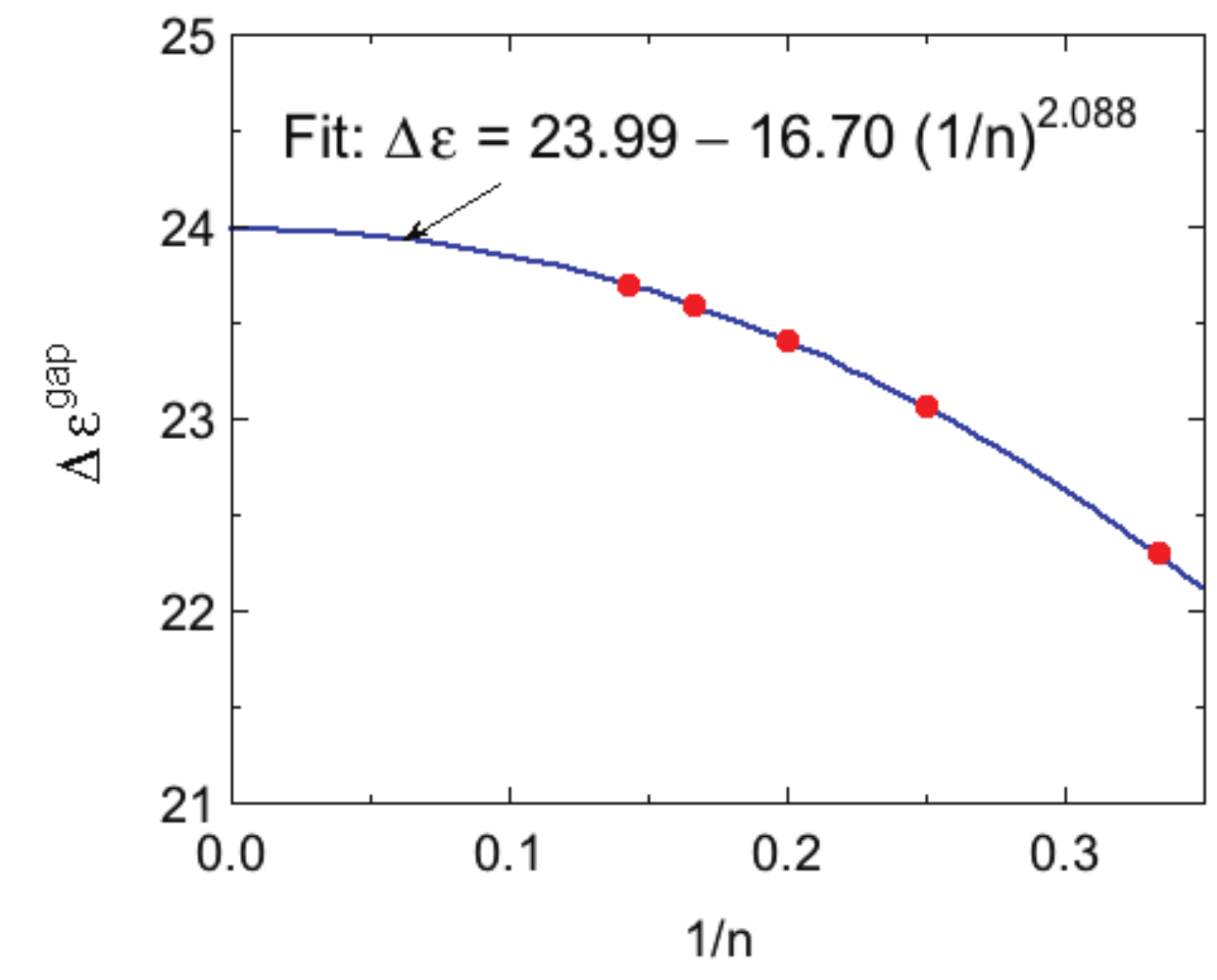}
\caption{Energy gap $\Delta\varepsilon^{\rm gap}$ versus the inverse of the Brillouin-zone number~$n$ (red circles) for $P=6$.  The fit is shown by the solid blue curve with the fit function and parameters given in the figure. }
\label{Eq:Gap_vs_n}
\end{figure}

The band gaps $\Delta\varepsilon^{\rm gap}$ in Table~\ref{Tab:GapInfo} appear to approach a constant value with increasing~$n$ instead of tending towards zero as in Fig.~\ref{Fig:KPBandEdgesAvesGapsQ10} for the finite square-well potential.  This is confirmed in Fig.~\ref{Eq:Gap_vs_n} in which $\Delta\varepsilon^{\rm gap} $ is plotted versus $1/n$ for $n=3$ to~7.  The fit shown yields $\Delta\varepsilon \sim 1/n^2$ and $\lim_{n\to\infty}\Delta\varepsilon \approx 24$ for $P=6$.  The exponent in $n^2$ is the same as for the square-well KP model at even values of $ka/\pi$ with $b/a=1/2$ and $Q=10$ in Fig.~\ref{Fig:KPBandEdgesAvesGapsQ10}(b).

\begin{table}
\caption{\label{Tab:GapInfo} Reduced band energies $\varepsilon_{\rm band}^-(p)$ and $\varepsilon_{\rm band}^+(p)$ at the lower and upper edges of band~$p$, respectively, band-center energy $\varepsilon^{\rm ave}_{\rm band}(p) \equiv [\varepsilon^+_{\rm band}(p)+\varepsilon^-_{\rm band}(p)]/2$, the band gap $\Delta \varepsilon^{\rm gap}(p) = \varepsilon_{\rm band}^-(p)-\varepsilon_{\rm band}^+(p-1)$ at Brillouin-zone boundary at $ka=n\pi$, and gap-center energy $\varepsilon^{\rm gap}_{\rm ave} =  [\varepsilon^+(p) + \varepsilon^-(p-1)]/2$ from the band structure obtained by numerically solving Eq.~(\ref{Eq:coskak1a2}) for $P=6$ and $ka=n\pi$.  The numerical values of $\varepsilon_{\rm band}^-(n)$ for $2\leq n\leq 8$ agree with the expression in Eq.~(\ref{Eq:Eps-(p)}). The large gap between bands $p=1$ and~2 at $n=1$ is seen to be anomalously large.}
\begin{ruledtabular}
\begin{tabular}{crrrccr}
band $p$ & 	$\varepsilon_{\rm band}^-$ 		&  $\varepsilon_{\rm band}^+$ & $\varepsilon_{\rm band}^{\rm ave}$ & $n$ & $\Delta \varepsilon^{\rm gap}$	& $\varepsilon^{\rm gap}_{\rm ave}$\\
\hline
1			& $-36.348$		& $-35.634$	& $-35.991$	&	---	&	---			&    ---				\\
2			&   9.870			& 19.440		& 14.655		&	1	& 45.504			&	$-12.882$		\\
3			&  39.478			& 66.525		& 53.002		&	2	& 20.039			&  	29.459	\\
4			&  88.826 		& 134.852		& 111.839		&	3	& 22.301			&	77.676	\\
5			& 157.914			& 223.335		& 190.624		&	4	& 23.062			&	146.383	\\
6			& 246.740 		& 331.716		& 289.228		&	5	& 23.406			&	235.037	\\
7			& 355.306			& 459.911		& 407.609		&	6	& 23.589			&	343.511	\\
8			& 483.611			& 607.884		& 545.747		&	7	& 23.699			&	471.761	\\
\end{tabular}
\end{ruledtabular}
\end{table}

To obtain the dispersions of the reduced energy $\varepsilon$ near the BZ boundaries~$ka = n\pi$, we Taylor-expand $\cos(ka)$ in Eq.~(\ref{Eq:coskak1a2}) about $ka = n\pi$ in Fig.~\ref{Fig:E_vs_ka_P6_extended_zone} to second order in $ka-n\pi$ and $\epsilon$ to first-order in~$\varepsilon-\varepsilon_n$ and equate the two expressions.  Equating the constant terms in each expansion gives Eq.~(\ref{Eq:coskak1a2}) with $\cos(ka)=(-1)^n$ from which the values~$\varepsilon_n$ were derived as listed in Table~\ref{Tab:GapInfo}.  Equating the remaining two terms gives the dispersion relation near the BZ boundaries as
\begin{equation}
\varepsilon-\varepsilon_n = \frac{\varepsilon_n^{3/2}}{P\sqrt{\varepsilon_n}\cos(\sqrt{\varepsilon_n})+(\varepsilon_n-P)\sin({\sqrt{\varepsilon_n}})}(ka-n\pi)^2.
\label{Eq:DispRelnNearBZboundaries}
\end{equation}
Thus this quadratic dispersion is free-electron-like at the BZ boundaries and \mbox{$d\varepsilon/d(ka)=  0$} at the band edges.  The latter result is confirmed by taking the derivative of both sides of Eq.~(\ref{Eq:coskak1a2}) with respect to $ka$ and then setting $ka=n\pi$ as appropriate for a band edge.  However, as seen in Eq.~(\ref{Eq:DispRelnNearBZboundaries}) and Fig.~\ref{Fig:E_vs_ka_P6_extended_zone}, the coefficients of the quadratic dispersion at the bands at the BZ boundaries do not match the free-electron behavior in Eq.~(\ref{Eq:FEepska2}).  This arises because the band effective mass of the band electrons near the BZ boundaries is not the same as the free-electron mass~$m$, as discussed later in Sec.~\ref{Sec:Effective mass}.

For a free-electron band the dispersion relation for the reduced energy is given in Eq.~(\ref{Eq:FEepska2}) and is plotted as the red curve in Fig.~\ref{Fig:E_vs_ka_P6_extended_zone}.  From Fig.~\ref{Fig:E_vs_ka_P6_extended_zone}, one sees that the band energy at the lower edge of the $p^{\rm th}$ band at $ka=n\pi$ for $p=2$--8 is given by the free-electron prediction
\begin{equation}
\varepsilon_{\rm band}^-(n) = (n\pi)^2.
\label{Eq:Eps-(p)}
\end{equation}
The numerical data in Table~\ref{Tab:GapInfo} for $\varepsilon_{\rm band}^-(n)$ for bands 2 to~8 at $n=1$ to 7 are indeed in precise agreement with Eq.~(\ref{Eq:Eps-(p)}).

\begin{figure}
\includegraphics[width=3.in]{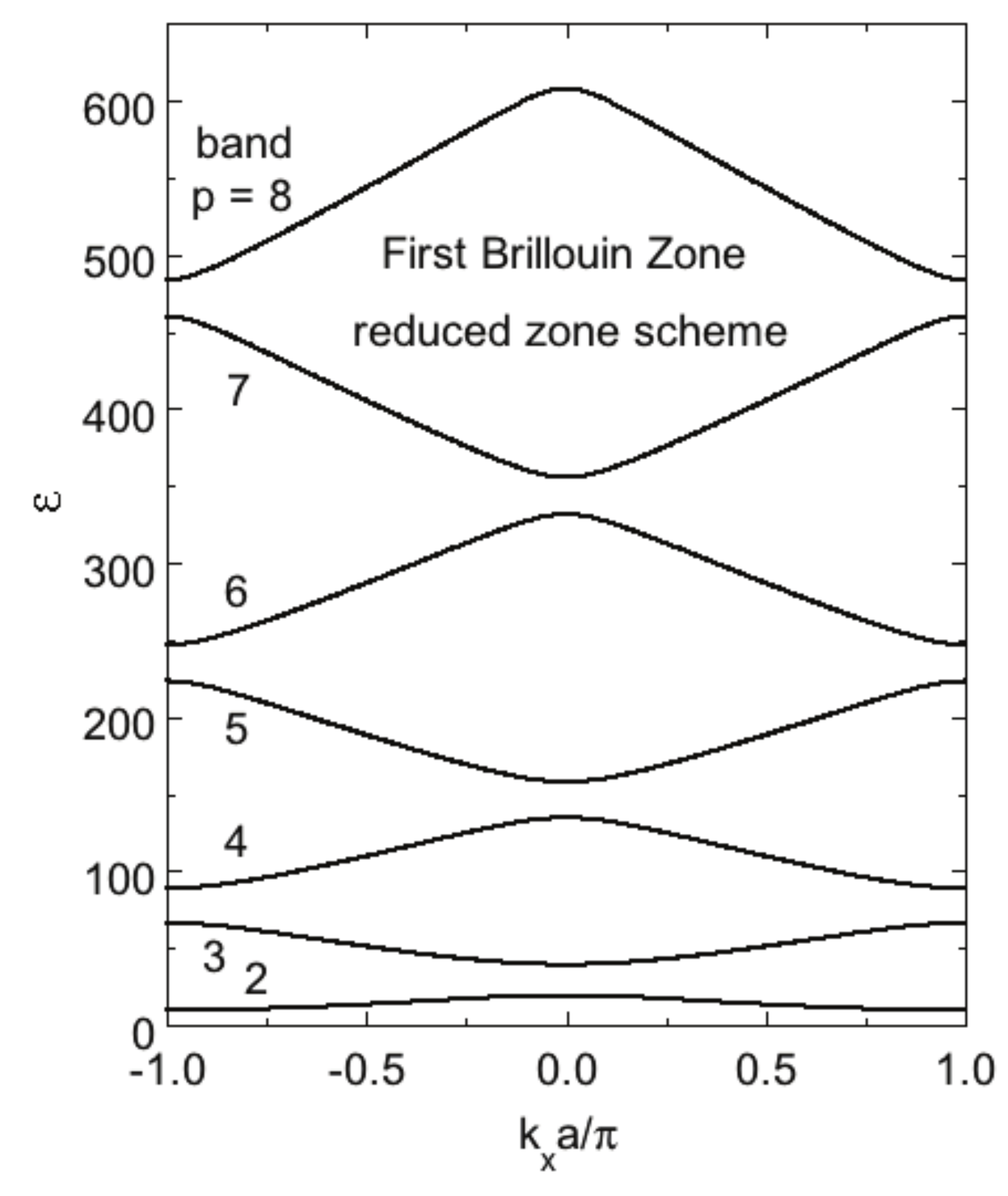}
\caption{Band structure $\varepsilon$ versus reduced electron wave vector $k_xa/\pi$ in the reduced-zone scheme for bands $p=2$ to~8 with $P=6$.}
\label{Fig:Kronig_Penney_Reduced_Band}
\end{figure}

The band structure for $P=6$ in the reduced-zone scheme is shown in Fig.~\ref{Fig:Kronig_Penney_Reduced_Band}.

\subsection{\label{Sec:WveFcnsDiracPlus} Wave Functions for Positive Electron-Band Energies}

To determine the wave functions for an electron with a positive energy we follow the outline given in Ref.~\cite{Wolfe1978}.  The attractive potential energy is given in Eq.~(\ref{Eq:DiracComb}).  Periodic boundary conditions on the wave function for a ring containing a large number $N$ of potential wells will be applied.

We first solve for the boundary conditions on the wave function $\psi(x)$ for one unit cell.  
We need to find $\psi(x)$ only for the region $0\leq x\leq a$ since $\psi(x)$ in other unit cells can then be obtained using Bloch's theorem~(\ref{Eq:Psixplusa2}).  Equation~(\ref{Eq:Psixplusa2}) gives the first boundary condition on $\psi(x)$ as
\bea
\psi(0^+) = e^{- ika}\psi(a^+) = e^{- ika}\psi(a^-),
\label{Eq:BC1}
\eea
where the second equality follows from continuity of $\psi(x)$ across the $\delta$~function at $x = a$.  The notations $0^+$, $a^+$, and $a^-$ mean taking the limit as $x$ approaches 0 from $x>0$ and $x$ approaches $a$ from $x>a$ and $x<a$, respectively.

Setting $x=0^-$ in Eq.~(\ref{Eq:dpsixpma2}) gives
\bea
\frac{d\psi(x)}{dx}(0^-) = e^{-ika}\frac{d\psi(x)}{dx}(a^-).
\label{Eq:dps0m}
\eea
For the $\delta$-function potential well in Eq.~(\ref{Eq:DiracComb}) at $x=0$, one has
\bea
\int_{0^-}^{0^+}\frac{d^2\psi(x)}{dx^2}dx = \frac{d\psi(x)}{dx}(0^+) - \frac{d\psi(x)}{dx}(0^-).
\label{Eq:Intdpsi200}
\eea
The time-independent Schr\"odinger equation is
\bea
\frac{d^2\psi(x)}{dx^2} = -\frac{2m}{\hbar^2}E\psi(x) +  \frac{2m}{\hbar^2}U(x)\psi(x).
\label{Eq:d2psix}
\eea
When the right side of Eq.~(\ref{Eq:d2psix}) is inserted into the left side of Eq.~(\ref{Eq:Intdpsi200}), the integral over the first term is zero due to continuity of $\psi(x)$.  However, using the potential energy function in Eq.~(\ref{Eq:DiracComb}), the integral over the second term gives the nonzero value
\bea
\int_{0^-}^{0^+}\frac{d^2\psi(x)}{dx^2} = -\frac{2m\alpha}{\hbar^2}\psi(0) = -\frac{2P}{a}\psi(0^+),
\eea
where the dimensionless variable $P$ was previously defined in Eq.~(\ref{Eq:PDef}).  Equating the far right side of this equation to the right side of Eq.~(\ref{Eq:Intdpsi200}) gives
\bea
\frac{d\psi(x)}{dx}(0^+) - \frac{d\psi(x)}{dx}(0^-) = -\frac{2P}{a}\psi(0^+).
\label{Eq:dpsidx+-}
\eea
Therefore the derivative $d\psi(x)/dx$ is not the same on both sides of a $\delta$-function potential well.  Finally, substituting $\frac{d\psi(x)}{dx}(0^-)$ in Eq.~(\ref{Eq:dps0m}) into this equation gives the second boundary condition on $\psi(x)$ as
\bea
\frac{d\psi(x)}{dx}(0^+) +\frac{2P}{a}\psi(0^+) = e^{-ika} \frac{d\psi(x)}{dx}(a^-),
\label{Eq:BC2}
\eea
where only the positions inside the unit cell \mbox{$0<x<a$} are considered here as in the first boundary condition~(\ref{Eq:BC1}).

In region~1 of Fig.~\ref{Fig:Kronig_Penney_Model} in which $0< x < a$ and $U(x)=0$, the Schr\"odinger equation has the free-electron form
\bea
\frac{d^2\psi(x)}{dx^2} = -\frac{2mE}{\hbar^2}\psi(x)
\eea
with eigenvalue $E$ and general solution
\bea
\psi(x) = C e^{ik_1x} + D e^{-ik_1x},
\label{Eq:psiCD}
\eea
where $k_1=\sqrt{2mE/\hbar^2}$ as in Eq.~(\ref{Eq:k1Def}).  

Equation~(\ref{Eq:psiCD}) together with the boundary conditions~(\ref{Eq:BC1}) and~(\ref{Eq:BC2}) respectively yield the homogeneous equations
\bse
\bea
\left[1-e^{i(k_1a-ka)}\right]C + \left[1-e^{-i(k_1a+ka)}\right]D = 0,\label{Eq:CD1}\\*
&& \hspace{-2.85in} \left\{ik_1a[1-e^{i(k_1a-ka)}]+2P/a\right\}C \\*
&& \hspace{-2.65in} -\  \left\{ik_1a[1-e^{-i(k_1a+ka)}] - 2P/a\right\}D = 0.\nonumber
\eea
\ese
In order for nonzero solutions for $C$ and $D$ to exist, the determinant of the coefficients must vanish, yielding the same Kramers-Kronig dispersion relation already found in Eq.~(\ref{Eq:coskak1a}).  

The wave function $\psi(x)$ in Eq.~(\ref{Eq:psiCD}) is found using Eq.~(\ref{Eq:CD1}), which yields
\bea
\frac{C}{D} = -\frac{1-e^{-i(k_1a + ka)}}{1 - e^{i(k_1a-ka)}}.
\label{Eq:FracDC}
\eea
We set
\bea
C &=& \frac{1}{2i}\left[1-e^{-i(k_1a + ka)}\right],\\
D &=& -\frac{1}{2i}\left[1 - e^{i(k_1a-ka)}\right],
\eea
which satisfy Eq.~(\ref{Eq:FracDC}).  Substituting these expressions into Eq.~(\ref{Eq:psiCD}) and using the definition
\bea
x_a\equiv x/a,
\eea
Eq.~(\ref{Eq:psiCD}) gives
\bea
\psi(x_a) &=&  A\frac{1}{2i} \bigg\{\left[1-e^{-i(k_1a+ka)}\right]e^{ik_1a x_a} \\
&& -\ \left[1 - e^{i(k_1a - ka)}\right]e^{-ik_1a x_a}\bigg\},\nonumber
\eea
where $A$ is the energy- and wave vector-dependent normalization factor.  This simplifies to
\bse
\label{Eqs:psixa99}
\bea
\psi(x_a) = A\left\{\sin(k_1a x_a) + e^{-ika}\sin[k_1a(1-x_a)]\right\}.\nonumber\\
\label{Eq:psixa}
\eea
According to Eq.~(\ref{Eq:k1aE}) one has $k_1a= \sqrt{\varepsilon}$, so the wave function~(\ref{Eq:psixa}) can be written
\bea
\psi(x_a) = A\left\{\sin(\sqrt{\varepsilon} x_a) + e^{-ika}\sin[\sqrt{\varepsilon}(1-x_a)]\right\},\nonumber\\
\label{Eq:psixa2}
\eea
\ese
where $ka$ and $\varepsilon$ are related to each other by Eq.~(\ref{Eq:coskak1a2}).

As shown in Eq.~(\ref{Eq:sqrtNA}) below, the normalization factor $A$ is a real positive function of the real and/or imaginary parts of $ka$ in addition to $\varepsilon$.  For electron energies within the energy bands, $ka$ is real.  However, in the energy gaps, Eq.~(\ref{Eq:coskak1a2}) gives complex values of $ka$. We will see that electron states exist within the energy gaps.   In order to calculate $\psi(x_a)$ for energies both within the energy bands and within the energy gaps using the same expression we write
\bea
ka = \Re(ka) + i\, \Im(ka),
\label{Eq:kaReIm}
\eea
where $\Re(ka)$ and $\Im(ka)$ are the real and imaginary parts of $ka$, respectively.  Therefore using Eq.~(\ref{Eq:kaReIm}) the wave function~(\ref{Eq:psixa2}) becomes
\bea
\psi(x_a) &=& A\Big\{\sin(\sqrt{\varepsilon} x_a) \label{Eq:psixa3}\\
&& \hspace{0.3in} +\ e^{-i\Re[ka(\varepsilon)]}e^{\Im[ka(\varepsilon)]}\sin[\sqrt{\varepsilon}(1-x_a)]\Big\}.\nonumber
\eea
For complex in-gap wave vectors, we find that $\Re(ka)$ alternates with increasing Brillouin-zone number~$n=ka/\pi$ between $\pi$ and 0 starting with $\pi$ at $n=1$.

\subsubsection{Probability Density and Wave Function Normalization Factor}

The energy- and position-dependent electron probability density ${\cal P}(x_a)$ is
\bea
{\cal P}(x_a) = \psi(x_a)\psi^*(x_a),
\label{Eq:CalP0}
\eea
so using Eq.~(\ref{Eq:psixa3}) one obtains
\bea
{\cal P}(x_a) &=& A^2\Big\{\sin^2(\sqrt{\varepsilon}x_a) + e^{2\Im[ka(\varepsilon)] }\sin^2[\sqrt{\varepsilon}(1-x_a)] \nonumber\\*
&& \hspace{-0.55in}+\ 2e^{\Im(ka)}\cos\big\{[\Re[ka(\varepsilon)]\big\}\sin(\sqrt{\varepsilon}x_a)\sin[\sqrt{\varepsilon}(1-x_a)]\Big\}.\nonumber\\
\label{Eq:CalPxa}
\eea
This expression is applicable to energies both within the energy bands and within the energy gaps.

One requires that the electron be somewhere in the ring of attractive potentials.  Thus the normalization factor~$A$ in $\psi(x_a)$ is determined for a ring of $N$ positive ions according to
\bea
\lim_{N\to\infty}\int_0^{N-1} {\cal P}(x_a)dx_a = 1. 
\label{Eq:Norm}
\eea
Inserting Eq.~(\ref{Eq:CalPxa}) into~(\ref{Eq:Norm}) and solving the resultant equation gives
\bea
NA^2 &=& 2\Big\{1+e^{2\Im[ka(\varepsilon)]} \label{Eq:NAeps}\\
&& \hspace{0.25in} -\ 2e^{\Im[ka(\varepsilon)]}\cos(\sqrt{\varepsilon}) \cos\{\Re[ka(\varepsilon)]\} \Big\}^{-1},\nonumber
\eea
and hence the wave-function normalization factor is 
\bea
\sqrt{N}A &=& \sqrt{2}\Big\{1+e^{2\Im[ka(\varepsilon)]} \label{Eq:sqrtNA}\\
&& \hspace{0.25in} -\ 2e^{\Im[ka(\varepsilon)]}\cos(\sqrt{\varepsilon}) \cos\{\Re[ka(\varepsilon)]\} \Big\}^{-1/2}.\nonumber
\eea
Then Eq.~(\ref{Eq:CalPxa}) becomes
\bea
N{\cal P}(x_a) &=& \nonumber\\
&&\hspace{-0.5in} NA^2\Big\{\sin^2(\sqrt{\varepsilon}x_a) + e^{2\Im[ka(\varepsilon)] }\sin^2[\sqrt{\varepsilon}(1-x_a)] \nonumber\\
&& \hspace{-0.6in}+\ 2e^{\Im(ka)}\cos\big\{[\Re[ka(\varepsilon)]\big\}\sin(\sqrt{\varepsilon}x_a)\sin[\sqrt{\varepsilon}(1-x_a)]\Big\}.\nonumber\\
\label{Eq:NAeps2}
\eea

\begin{figure}
\includegraphics[width=2.75in]{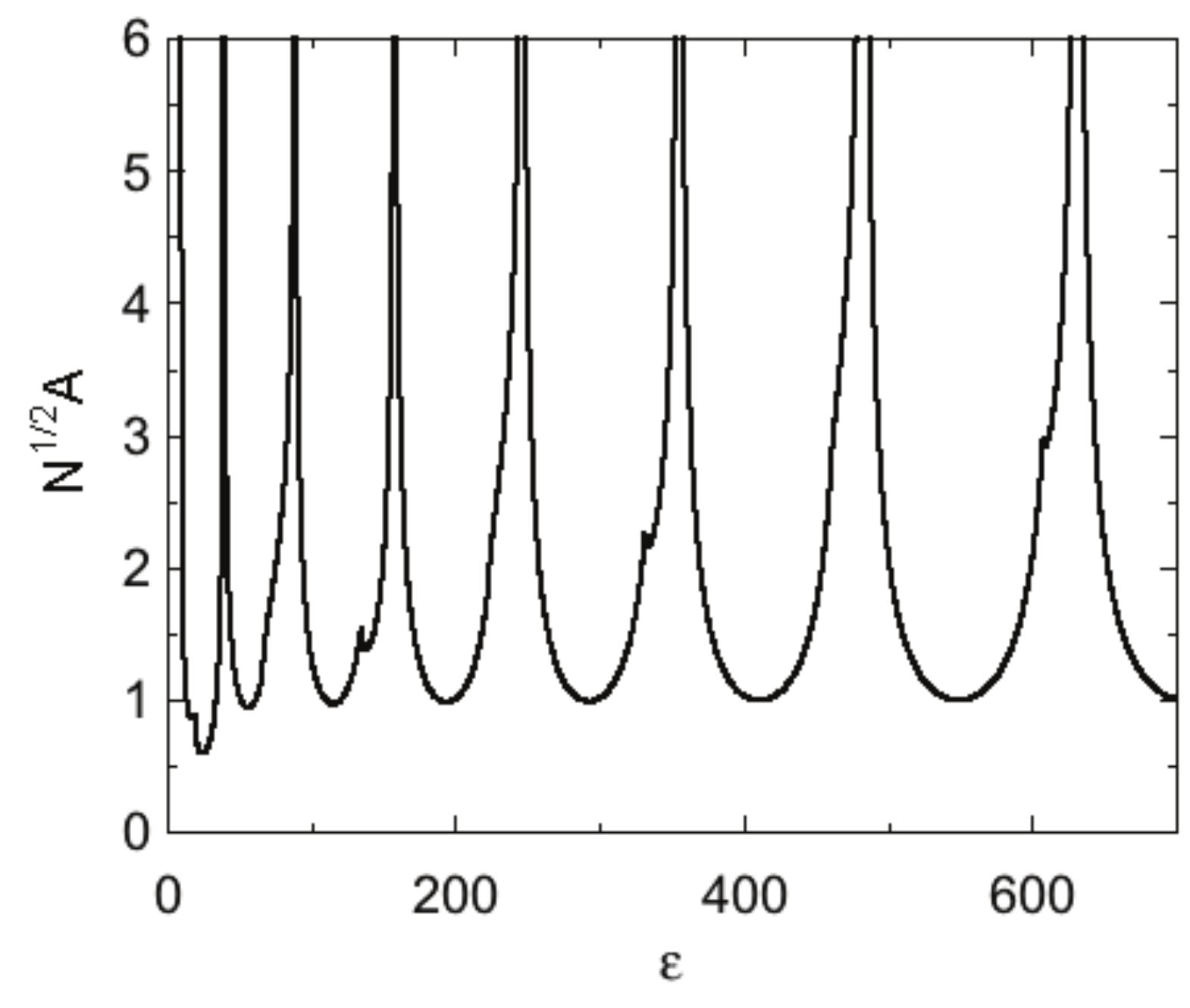}
\caption{Wave function normalization factor~$\sqrt{N}A$ versus reduced energy $\varepsilon$ calculated using  Eq.~(\ref{Eq:sqrtNA}).  The sharp peaks are at the bottom of bands $p = 2$ through 8, whereas the cusps or shoulders just below the energies of the peaks for bands 2 through 8  are at the tops of bands $p =3$, 5, and~7, and bands $p=2$, 4, and~6, respectively  (see Table~\ref{Tab:GapInfo}).  The minima of $\sqrt{N}A_{\varepsilon}$ versus~$\varepsilon$ occur close to the midband energies of bands $p=2$ through~7 as given in Table~\ref{Tab:GapInfo}.}
\label{Fig:Sqrt_NA2_vs_eps} 
\end{figure}

A plot of $\sqrt{N}A$ versus $\varepsilon$ for $\varepsilon$ obtained for $P=6$ using Eqs.~(\ref{Eq:coskak1a2}) and (\ref{Eq:sqrtNA}) in the range $1\leq\varepsilon\leq700$ that was utilized in constructing Fig.~\ref{Fig:Kronig_Penney_Reduced_Band} is shown in Fig.~\ref{Fig:Sqrt_NA2_vs_eps}.  Here $ka(\varepsilon)$ was computed using Eq.~(\ref{Eq:caFromEps}).  The sharp maxima in Fig.~\ref{Fig:Sqrt_NA2_vs_eps} occur at the bottoms of bands $p=2$ to~8 where free-electron-like dispersions $\varepsilon\sim (ka)^2$ occur as discussed above, whereas the cusps or shoulders just below the sharp maxima occur at the tops of the respective bands \mbox{$p-1$} (\emph{i.e.}, at the bottoms of the energy gaps below band~$p$).  The minima with \mbox{$\sqrt{N}A_{\varepsilon}\approx 1$} occur at reduced energies corresponding approximately to the midband energies in Table~\ref{Tab:GapInfo}.  Figure~\ref{Fig:Sqrt_NA2_vs_eps} also shows that the probability density is continuous on passing through the energy gaps.

\subsubsection{Wave Function and Probability Density Plots versus Position for Positive Energies}

\begin{figure}
\includegraphics[width=3.4in]{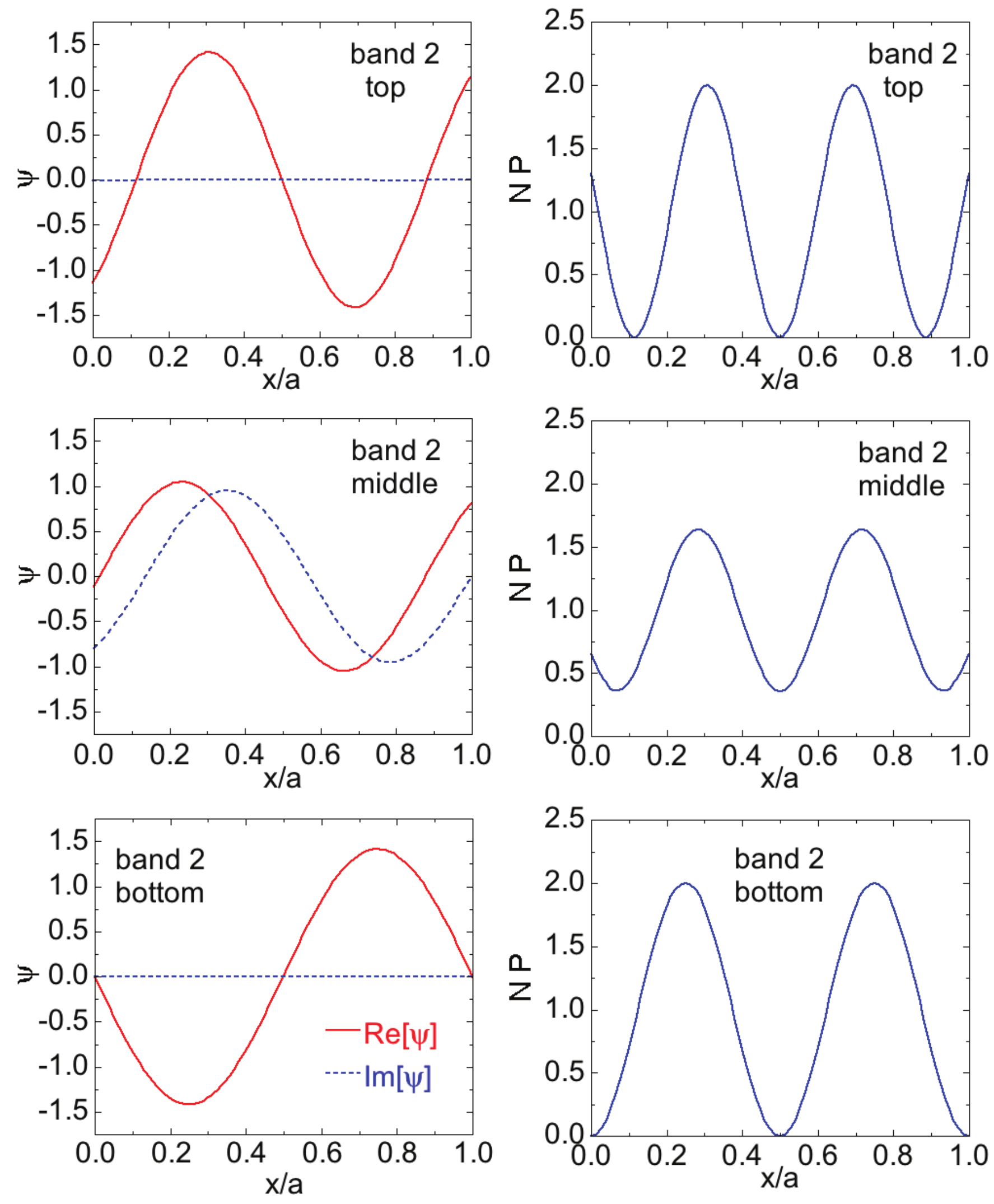}
\includegraphics[width=3.4in]{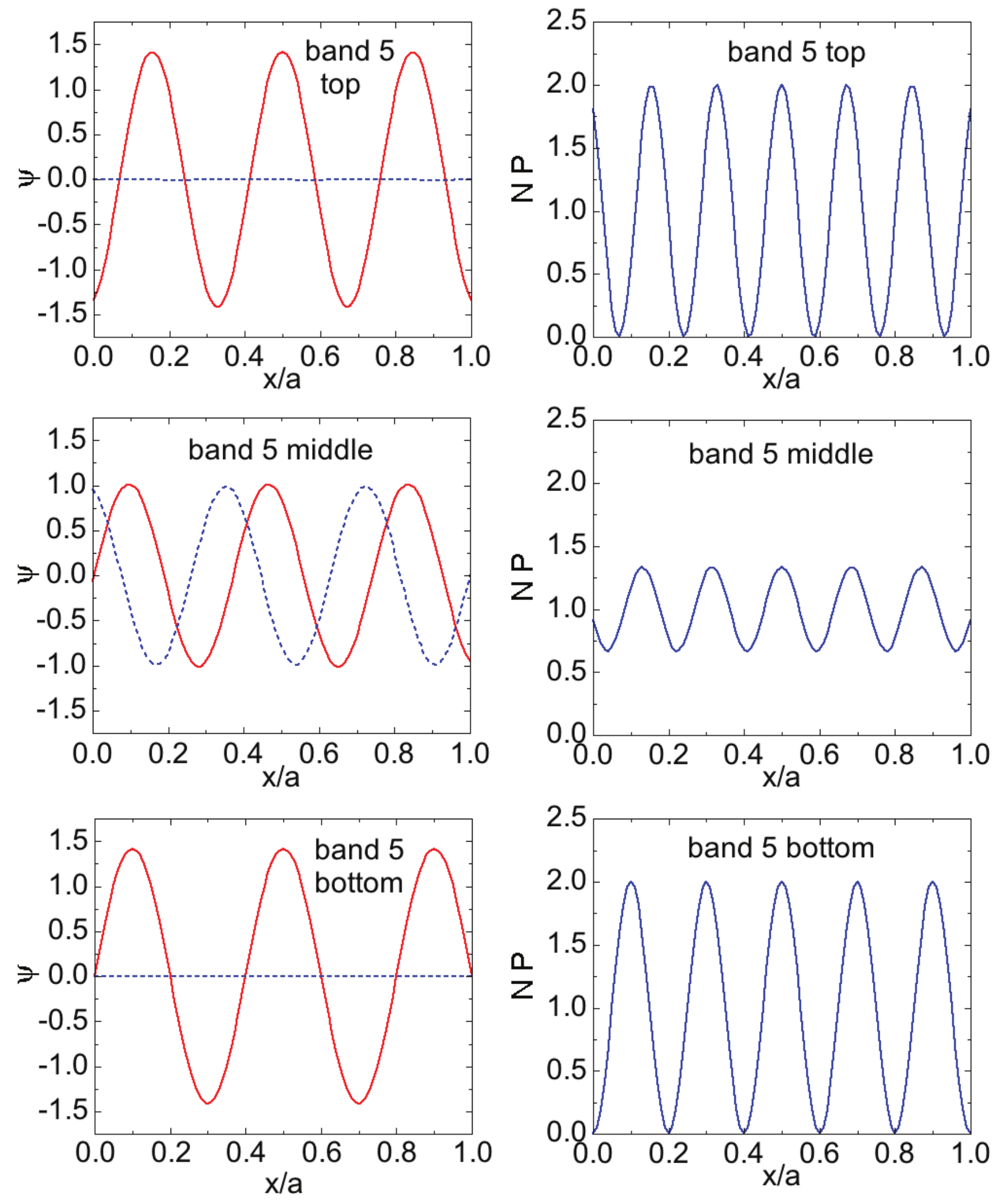}
\caption{Panels on left: plots of the real [Re($\psi$), solid red curves] and imaginary [Im($\psi$), dashed blue curves] parts of the wave function $\psi$ versus position $x/a$ in a unit cell at the top, middle, and bottom of bands 2 and~5 in Fig.~\ref{Fig:E_vs_ka_P6_extended_zone}, respectively.  Panels on right: probability density $NP \equiv NA^2$ for bands~2 and~5.   The data were calculated using Eqs.~(\ref{Eq:coskak1a2}), (\ref{Eq:psixa3}), (\ref{Eq:CalPxa}), and~(\ref{Eq:sqrtNA}).}
\label{Fig:Band2PsiProb} 
\end{figure}

The wave functions~$\psi$ are calculated using Eqs.~(\ref{Eq:psixa3}) and~(\ref{Eq:sqrtNA}).  Shown in the left panels of Fig.~\ref{Fig:Band2PsiProb} are plots of the real and imaginary parts of $\psi$ versus position $x/a$ for representative bands~$p=2$ and~5 in Fig.~\ref{Fig:E_vs_ka_P6_extended_zone}.  One sees that $\psi$ is real at the very tops and bottoms of the two bands, thus representing standing waves as was already well known for one-dimensional band structures in the nearly-free-electron approximation. $\psi$ at the bottoms of the two bands exhibit nodes at the edges $x/a = 0$ and~1 of the unit cell with $p/2$ oscillation periods in between.  We find that for bands 4, 6, and~7, the wavefunction at the lower band edges is purely imaginary (not shown), again representing standing waves.

The result that $\psi$ at the bottoms of the bands is zero at $x/a = 0$ and~1 in Fig.~\ref{Fig:Band2PsiProb} is explicable in terms of Eq.~(\ref{Eq:psixa2}).  Substituting $\sqrt{\varepsilon} = p\pi$ at the bottoms of the bands from Eq.~(\ref{Eq:Eps-(p)}) gives $\psi(x/a = 0) = \psi(x/a=1) = 0$ identically, explaining the nodes in the wave functions at these positions.  On the other hand, for $ka$ values in the interior of the bands, both real and imaginary parts of the wave functions occur, indicating propagating solutions.   The associated probability densities $N {\cal P}$ versus $x/a$ are plotted in the respective right-hand panels of Fig.~\ref{Fig:Band2PsiProb}.

\begin{figure}
\includegraphics[width=3.4in]{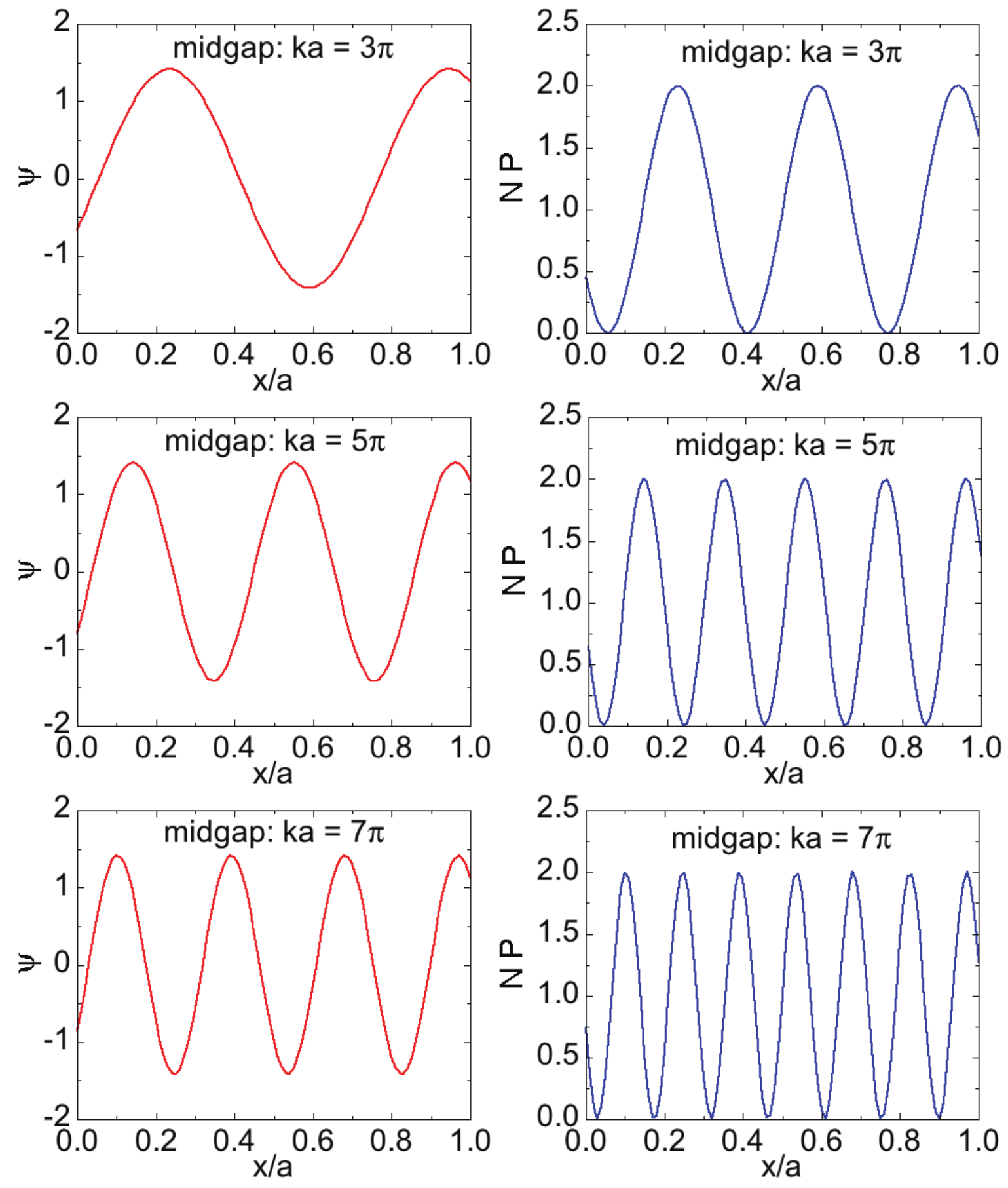}
\caption{Same as Fig.~\ref{Fig:Band2PsiProb} except that the energies considered are the midgap energies at the Brillouin-zone boundaries  at $ka=3\pi,\ 5\pi$, and $7\pi$ in the extended-zone scheme as listed (see Fig.~\ref{Fig:E_vs_ka_P6_extended_zone} and Table~\ref{Tab:GapInfo}).  The mid-gap wave functions are real, indicating the presence of standing waves.}
\label{Fig:Midgap_PsiProb} 
\end{figure}

Despite the fact that the wavevector $k$ is complex in the energy-gap regions of the extended-zone band structure in  Fig.~\ref{Fig:E_vs_ka_P6_extended_zone}, the wavefunctions for these in-gap energies are found to be real.  For example, shown in Fig.~\ref{Fig:Midgap_PsiProb} are plots of both $\psi$ and $N{\cal P}$ versus $x/a$ for the midgap states at the Brillouin-zone boundaries $ka = 3\pi,\ 5\pi$, and~$7\pi$.  The fact that the in-gap wave functions are real indicates that these wave functions are standing waves, corresponding to the occurrence of Bragg reflections of the propagating electron waves from the lattice of ions at the reciprocal-lattice wave vector $ka = \pi$ in reduced-zone notation.

\begin{figure}
\includegraphics[width=2.5in]{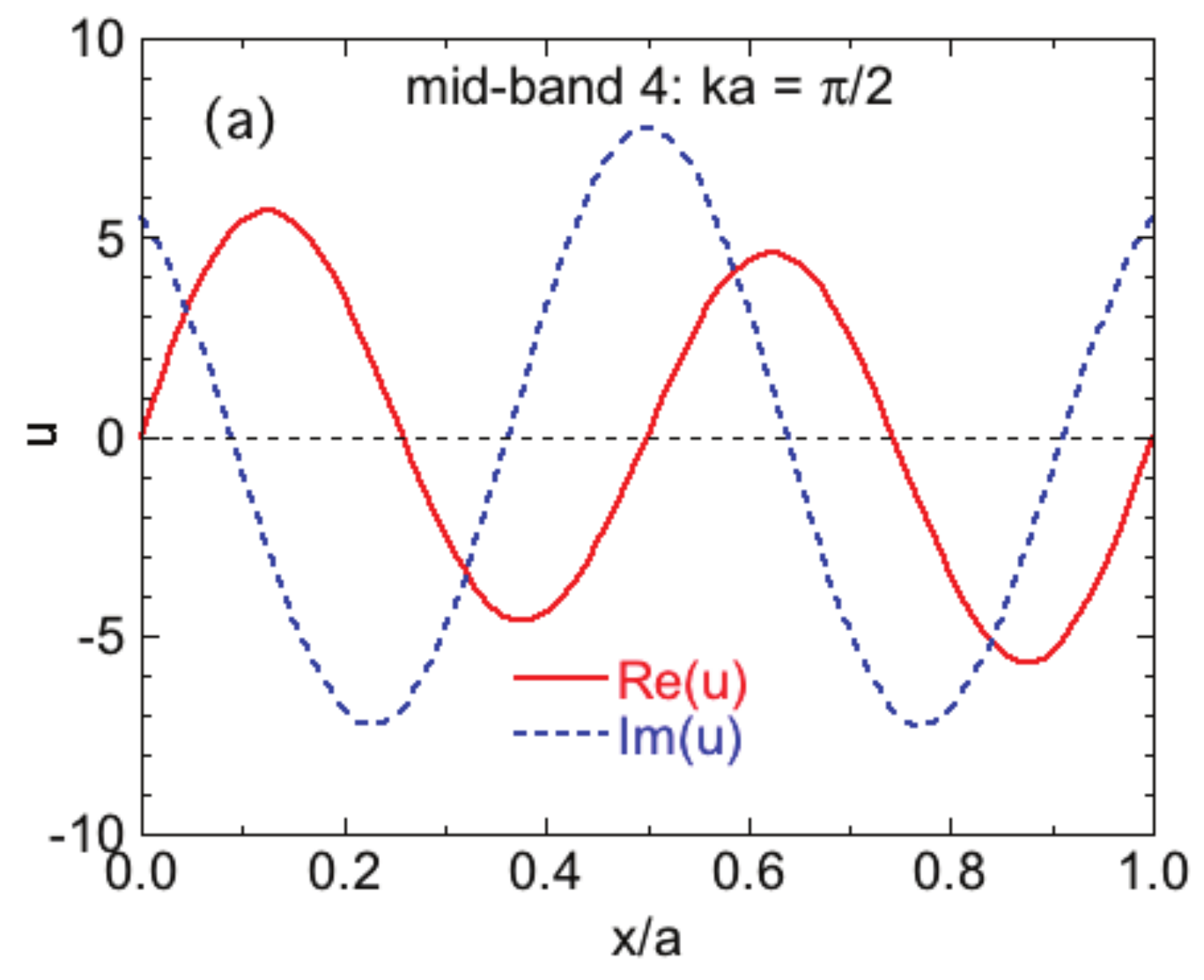}
\includegraphics[width=2.5in]{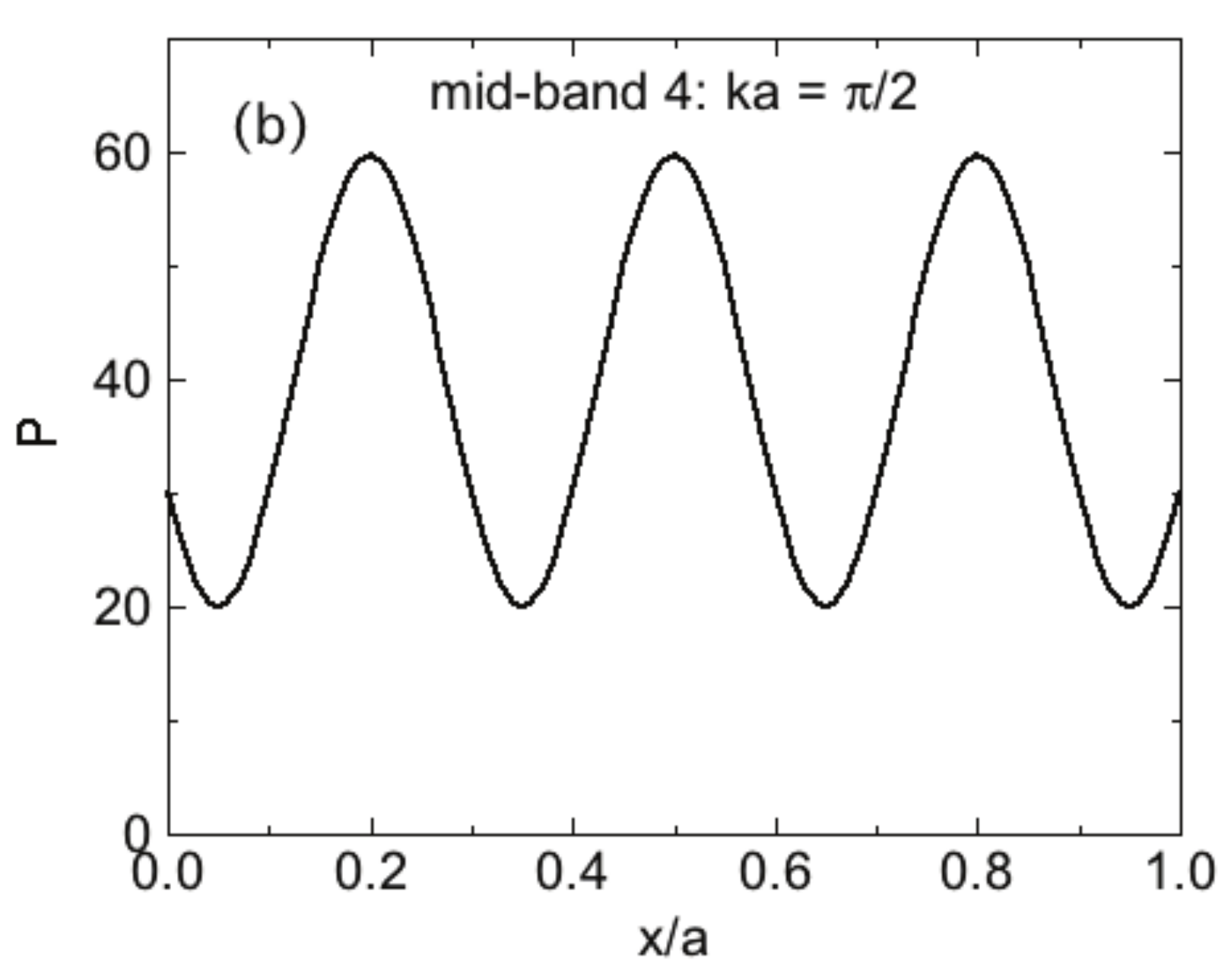}
\caption{Real and imaginary parts of the unnormalized periodic function $u(x)$ in Eq.~(\ref{Eq:psi(x)u(x)}) versus $x/a$ for (a)~the state at mid-band~4 with $ka = \pi/2$ in the reduced-zone scheme and~(b)~the associated probability density ${\cal P}(x)$.  The parameter $P$ is $P=6$ and the reduced energy of the state is $\varepsilon = 109.734$.  All three quantities are periodic in the lattice parameter~$a$.}  
\label{Fig:u(xa)midband_p4_P6} 
\end{figure}

The periodic wave functions $u(x)$ in Eq.~(\ref{Eq:psi(x)u(x)}) are obtained from the normalized wave functions $\psi(x)$ according to Bloch's theorem~(\ref{Eq:psi(x)u(x)}).  The real and imaginary parts of $u(x)$ for the parameter $P=6$ are plotted versus $x/a$ for the mid-band state of band~$p=4$ with wave vector $ka=\pi/2$ in the reduced-zone scheme in Fig.~\ref{Fig:u(xa)midband_p4_P6}(a).  The corresponding probability density ${\cal P} = u^*(x)u(x)$ is plotted in Fig.~\ref{Fig:u(xa)midband_p4_P6}(b).  One sees that the real and imaginary parts of $u(x)$ as well as ${\cal P}(x)$ are each periodic in the lattice parameter~$a$ as required and that $u(x)$ is not normalized even though $\psi(x)$ is.  $u(x/a)$ and ${\cal P}(x/a)$ are respectively found to be the same at the mid-energy of each of the energy gaps in Fig.~\ref{Fig:E_vs_ka_P6_extended_zone} with $2 \leq n\leq 7$ (not shown).

\subsection{\label{Sec:DOS} Electron Density of States versus Energy for Positive Energies}

The density of electron states versus energy $D(E)$ is defined as
\begin{subequations}
\label{Eqs:D(E)2}
\begin{equation}
D(E) = \frac{dN_{\rm e}}{dE} = \frac{dN_{\rm e}/dk}{dE/dk}.
\label{Eq:D(E)Def}
\end{equation}
which is the rate of change of the number $N_{\rm e}$ of electron states with increasing energy~$E$\@.  The term $dN_{\rm e}/dk$ is the number of electron states per unit $k$ range.  This has a fixed value for a one-dimensional lattice.  Including the electron Zeeman degeneracy of~2, one obtains
\begin{equation}
\frac{dN_{\rm e}}{dk} = \frac{\Delta N_{\rm e}}{\Delta k} = \frac{2}{2\pi/L} = \frac{L}{\pi} = \frac{Na}{\pi},
\label{Eq:dNedk}
\end{equation}
where $L = Na$ is the circumference of the ring in the KP model and $N\gg1$ is the number of ions in the ring.
\end{subequations}

\begin{figure}
\includegraphics[width=3.3in]{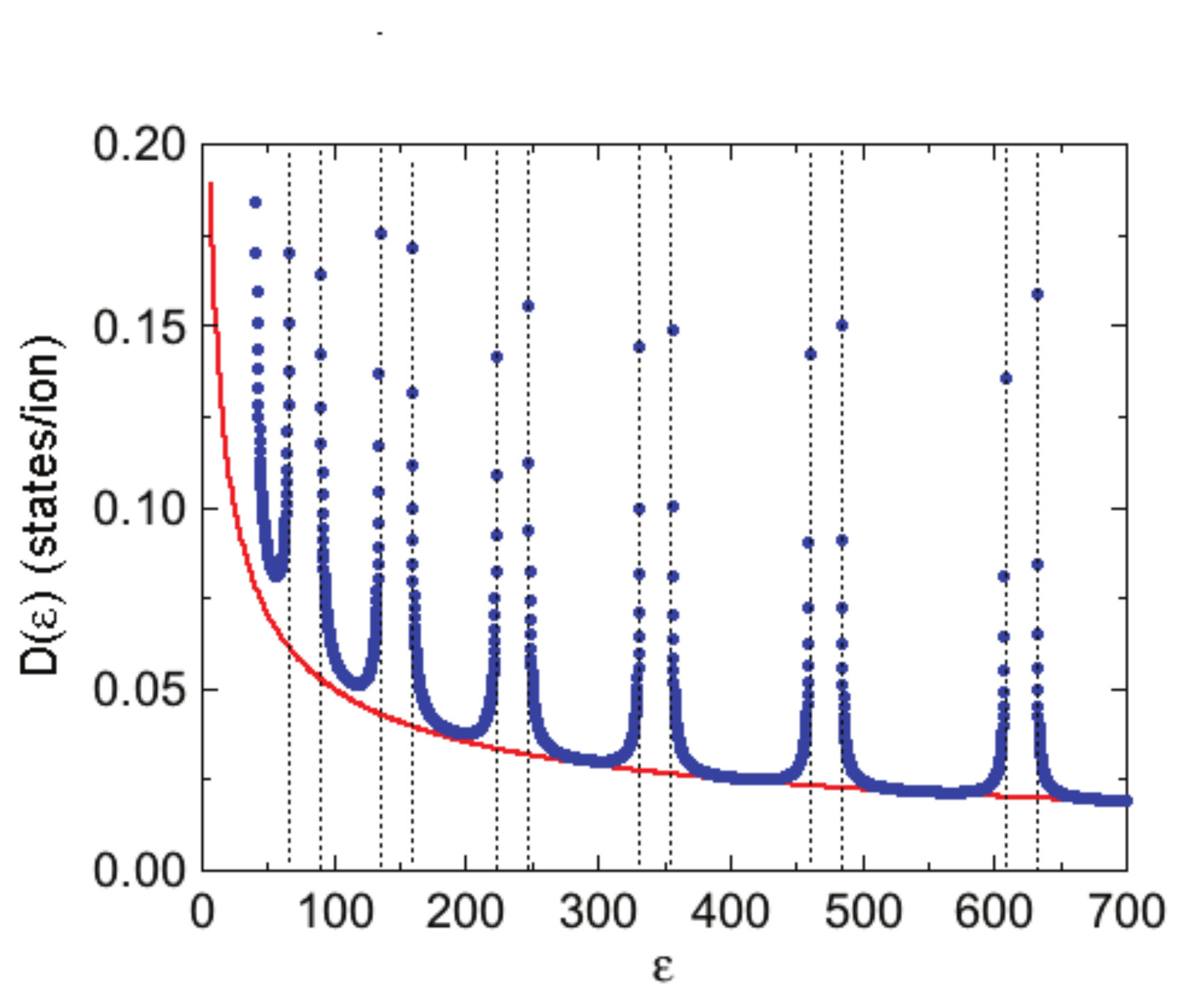}
\caption{Reduced density of states {\tt D} versus reduced energy~$\varepsilon$ for the Korig-Penney Dirac-comb model with $P=6$ (blue circles) calculated using Eqs.~(\ref{Eq:caFromEps}), (\ref{Eq:depsdka}), and~(\ref{Eq:D(varepsilon2)}).  The edges of the Brillouin zones are indicated by black vertical dashed lines.  The free-electron density of states in Eq.~(\ref{Eq:FEDOS}) is plotted as the solid red curve.}
\label{Fig:DOS_P6_Dirac_KP} 
\end{figure}

We now introduce reduced parameters.  Using Eq.~(\ref{epsDefdFcn}) the electron energy~$E$ is expressed in terms of the reduced energy~$\varepsilon$ as
\begin{subequations}
\label{Eqs:EdEdk}
\begin{equation}
E = \frac{\hbar^2\varepsilon}{2ma^2}.
\end{equation}
Thus
\begin{equation}
\frac{dE}{dk} = \frac{\hbar^2}{2ma}\frac{d\varepsilon}{d(ka)}.
\label{Eq:dEdk}
\end{equation}
\end{subequations}
Using Eqs.~(\ref{Eqs:D(E)2}) and~(\ref{Eq:dEdk}) one then obtains
\begin{equation}
D(E) = \frac{N(2ma^2)/(\pi\hbar^2)}{d\varepsilon/d(ka)}.
\label{Eq:reducedD(E)}
\end{equation}
Defining the dimensionless density of states per ion ${\tt D}(\varepsilon)$ as
\begin{equation}
{\tt D}(\varepsilon) = \frac{\pi\hbar^2}{2ma^2N}D(E),
\end{equation}
Eq.~(\ref{Eq:reducedD(E)}) gives
\begin{subequations}
\label{Eqs:DofEpsilon}
\begin{equation}
{\tt D}(\varepsilon) = \frac{d[ka(\varepsilon)]}{d\varepsilon},
\label{Eq:D(varepsilon)}
\end{equation}
where we use this expression in order that $\varepsilon$ is the independent variable and $ka(\varepsilon)$ is obtained numerically  from  Eq.~(\ref{Eq:caFromEps}).

Taking the first derivative $d/d\varepsilon$ of both sides of Eq.~(\ref{Eq:caFromEps}) and solving for $d[ka(\varepsilon)]/d\varepsilon$ gives
\begin{equation}
\frac{d[ka(\varepsilon)]}{d\varepsilon} = \frac{P\sqrt{\varepsilon}\cos(\sqrt{\varepsilon}) + (\varepsilon - P)\sin(\sqrt{\varepsilon})}{2\varepsilon^{3/2}\sin[ka(\varepsilon)]},
\label{Eq:epsprime(eps)}
\end{equation}
yielding
\begin{equation}
\frac{d\varepsilon}{d(ka)} = \frac{2\varepsilon^{3/2}\sin[ka(\varepsilon)]}{P\sqrt{\varepsilon}\cos(\sqrt{\varepsilon}) + (\varepsilon - P)\sin(\sqrt{\varepsilon})}
\label{Eq:depsdka}
\end{equation}
and
\begin{equation}
{\tt D}(\varepsilon) = \frac{1}{d\varepsilon/d(ka)}.
\label{Eq:D(varepsilon2)}
\end{equation}
\end{subequations}
The ${\tt D}(\varepsilon)$ for $P=6$ obtained using Eqs.~(\ref{Eq:caFromEps}), (\ref{Eq:depsdka}), and~(\ref{Eq:D(varepsilon2)}) is plotted in Fig.~\ref{Fig:DOS_P6_Dirac_KP} in the range \mbox{$1\leq\varepsilon\leq700$} as in Figs.~\ref{Fig:Kronig_Penney_Reduced_Band} and~\ref{Fig:Sqrt_NA2_vs_eps}.  The {\tt D}($\varepsilon)$ of the bands are seen to diverge at the Brillouin-zone boundaries (van Hove singularities).

The free-electron expression for $\varepsilon(k)$ in Eq.~(\ref{Eq:FEepska2}) gives
\begin{equation}
\varepsilon_{\rm free~el}^\prime(ka) = 2(ka) = 2\sqrt{\varepsilon},
\end{equation}
yielding the free-electron density of states
\begin{equation}
{\tt D}_{\rm free\ el}(\epsilon) = \frac{1}{\varepsilon_{\rm free~el}^\prime(ka)} = \frac{1}{2\sqrt{\varepsilon}}.
\label{Eq:FEDOS}
\end{equation}
A plot of ${\tt D}_{\rm free\ el}(\epsilon)$ is shown in Fig.~\ref{Fig:DOS_P6_Dirac_KP} as a solid red curve.  One sees that the Dirac-comb KP model data are in approximate agreement with the free-electron density of states near the middle of the bands at the higher energies.

\subsection{\label{Sec:Effective mass} Electron Effective Mass for Positive Energies}

In a single-band free-electron model one has the dispersion relation given in Eq.~(\ref{Eq:FEDisp}), from which one obtains
\begin{equation}
\frac{d^2E_{\rm free\ el}}{dk^2} = \frac{\hbar^2}{m}.
\end{equation}
Hence the free-electron mass~$m$ is obtained as
\begin{equation}
\frac{1}{m} = \frac{1}{\hbar^2}\frac{d^2E_{\rm free\ el}}{dk^2}.
\end{equation}
When considering electron bands such as shown in Fig.~\ref{Fig:E_vs_ka_P6_extended_zone} that are not parabolic in~$k$ due to the presence of the ionic potential energy, the band effective mass $m^*$ at each~$k$ is defined analogously as
\begin{equation}
\frac{1}{m^*} = \frac{1}{\hbar^2}\frac{d^2E}{dk^2},
\label{Eq:meff1}
\end{equation}
where $d^2E/dk^2$ at a given~$k$ is the curvature of the band at that $k$ value.

From the definition of the reduced energy for the \mbox{$\delta$-function} KP model in Eq.~(\ref{epsDefdFcn}), one obtains the curvature in reduced parameters as 
\begin{equation}
\frac{d^2E}{dk^2} = \frac{\hbar^2}{2m}\frac{d^2\varepsilon}{d(ka)^2}.
\label{Eq:d2Edk2}
\end{equation}
Substituting the right side of Eq.~(\ref{Eq:d2Edk2}) into Eq.~(\ref{Eq:meff1}) gives the ratio $m/m^*$ as
\begin{subequations}
\begin{equation}
\frac{m}{m^*} = \frac{1}{2}\frac{d^2\varepsilon}{d(ka)^2},
\label{Eq:d2epsdka2}
\end{equation}
or equivalently,
\begin{equation}
\frac{m^*}{m} = \frac{2}{d^2\varepsilon/d(ka)^2}.
\label{Eq:m*/m}
\end{equation}
\end{subequations}
In the attractive Dirac-comb KP band structure in Fig.~\ref{Fig:E_vs_ka_P6_extended_zone}, $m^*/m$ is not equal to unity near Brillouin-zone boundaries as briefly discussed below Eq.~(\ref{Eq:DispRelnNearBZboundaries}), and diverges at $ka$ values at which inflection points in $\varepsilon(ka)$ occur.  Furthermore, from Fig.~\ref{Fig:E_vs_ka_P6_extended_zone} one sees that the curvature is positive near the bottom of a band, corresponding to a positive value of $m^*$, whereas it is negative near the top of the same band.  These are electron-hole states as opposed to electron states because they arise from holes (missing electrons) in a filled band.  

Taking the derivative $d/d(ka)$ of Eq.~(\ref{Eq:depsdka}) gives
\begin{subequations}
\label{Eqs:d2epsdka2}
\begin{equation}
\frac{d^2\varepsilon}{d(ka)^2} = \frac{\rm num}{\rm denom},
\end{equation}
where
\begin{eqnarray}
{\rm num} &=& 2\epsilon\Big\{ 2P^3 - 3P(P+1)\varepsilon + \varepsilon^2\\
&& -\ (P-4)P^2\varepsilon \csc^2(\sqrt{\varepsilon}) - P\sqrt{\varepsilon} \cot(\sqrt{\varepsilon}) \nonumber\\
&& \times\ \left[P(P + 5) - 3\varepsilon - P\varepsilon \csc^2(\sqrt{\varepsilon})\right] \Big\}, \nonumber\\
{\rm denom} &=& \Big\{\varepsilon + P\left[\sqrt{\varepsilon}\cot(\sqrt{\varepsilon}) -1\right]\Big\}^3.
\end{eqnarray}
\end{subequations}

\begin{figure}
\includegraphics[width=3.3in]{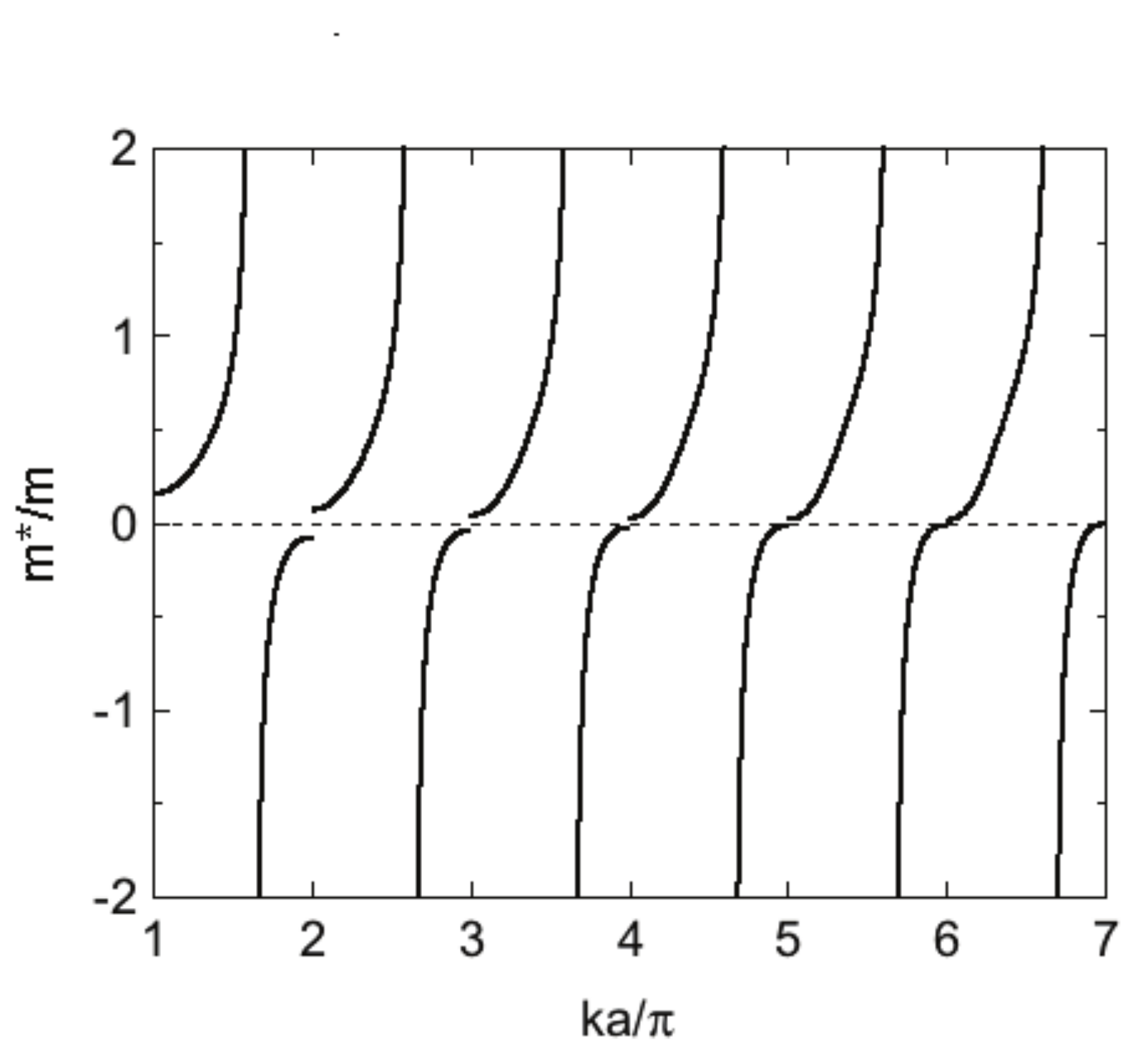}
\caption{Ratio $m^*/m$ of the effective electron band mass~$m^*$ to the free-electron mass~$m$ versus reduced wave vector $ka/\pi$ calculated using Eqs.~(\ref{Eq:m*/m}) and~(\ref{Eqs:d2epsdka2}).  The band mass is seen to be positive in the first part of each Brillouin zone and negative in the second part.  The ratio $m^*/m=1$ occurs near the midpoint of each Brillouin zone.}
\label{Fig:mStarONm_Dirac_KP} 
\end{figure}

Shown in Fig.~\ref{Fig:mStarONm_Dirac_KP} is a plot of $m^*/m$ versus $ka/\pi$ obtained in the extended-zone scheme using Eqs.~(\ref{Eq:m*/m}) and~(\ref{Eqs:d2epsdka2}).  One sees that $m^*/m\ll1$ at the Brillouin-zone boundaries.  In contrast, $m^*/m$ diverges at the inflection points of the band structure in Fig.~\ref{Fig:E_vs_ka_P6_extended_zone} where the curvature is zero.  This ratio becomes equal to unity around the midpoint of each band.  One also sees negative values of $m^*/m$ in the second part of each Brillouin zone for electron-hole states.

\begin{table}
\caption{\label{Tab:m*/m data} The first number lists the band number $p = 2$ to~8.  Also listed are the reduced energies~$\varepsilon_{\rm band}^- = [(p-1)pi)]^2$ and $\varepsilon_{\rm band}^+$ from Table~\ref{Tab:GapInfo}, and the ratios of the effective mass~$m^\ast$ to the free-electron mass~$m$ obtained using Eq.~(\ref{Eq:m*/m}) at the bottoms of bands~$p$ calculated using the parameter~$P=6$.}
\begin{ruledtabular}
\begin{tabular}{ccccc}
band $p$ 	& 	$\varepsilon_{\rm band}^-$ 	& $m_{\rm elec}^\ast/m$ &  $\varepsilon_{\rm band}^+$ & $m_{\rm hole}^\ast/m$ \\
\hline
 2			&   9.870				& 0.151982				& 19.440				& $-0.079640$ \\
 3			&  39.478				& 0.0675475				& 66.525				& $-0.041367$ \\
 4			&  88.826 			& 0.0379954				& 134.852				& $-0.025616$\\
 5			& 157.914				& 0.0243171				& 223.335				& $-0.017498$\\
 6			& 246.740 			& 0.0168869				& 331.716				& $-0.012731$\\
 7			& 355.306				& 0.0124067				& 459.911				& $-0.009687$\\
 8			& 483.611				& 0.0094989				& 607.884				&  \\
\end{tabular}
\end{ruledtabular}
\end{table}

Table~\ref{Tab:m*/m data} lists data for the band numbers~$p=2$ to~8 with $P = 6$ of the reduced energy $\varepsilon^-$ at the bottom of the band from Table~\ref{Tab:GapInfo}, ratio of the electron effective mass at the bottom of the band, $m_{\rm elec}^\ast$, to the free-electron mass~$m$, the reduced energy $\varepsilon^+$ at the top of the band from Table~\ref{Tab:GapInfo}, and the ratio of the electron-hole effective mass at the top of the band to the free-electron mass, $m_{\rm hole}^\ast/m$.  The magnitudes of $m_{\rm elec}^\ast/m$ and $m_{\rm hole}^\ast/m$ at each Brillouin zone boundary are seen to converge with increasing energy.  The magnitudes of the $m^*/m$ ratios in Table~\ref{Tab:m*/m data} are much smaller than unity, consistent with the data at the Brillouin-zone boundaries in Fig.~\ref{Fig:mStarONm_Dirac_KP}, because the curvatures of the bands at the bottoms of the bands are much larger than for the free-electron band as illustrated above in Fig.~\ref{Fig:E_vs_ka_P6_extended_zone}.

\subsection{\label{Sec:group velocity} Influence of the Band Structure on Electron Motion for Positive Energies}

We consider an electric field applied to an electron in the Dirac-comb KP band structure in a direction along the circumference of the large ring of periodic potentials, defined as the $x$ axis.  The electron wave function is a wave that permeates the entire ring.  However, in a particle picture, one can localize an electron to a finite width $\Delta x$ by forming a wave packet (pulse) consisting of the superposition of plane waves with variable wave vectors~$k$.  When the electron pulse moves towards the positive $x$~direction, it can be shown that it moves at the group velocity $v_{\rm g}=d\omega/dk$, whereas the constituent sinusoidal waves making up the pulse each move at the phase velocity $v_{\rm p} = \omega/k$.  These are not the same if the dispersion relation does not satisfy $\omega\propto k$.

For an electron pulse moving in a band structure such as the KP attractive Dirac-comb band structure, the pulse moves at the group velocity in the $x$ direction with $x$~component
\begin{equation}
v_{\rm g} = \frac{d\omega}{dk} = \frac{a}{\hbar}\frac{dE}{d(ka)},
\end{equation}
where $\omega = E/\hbar$ and $E$ is the energy of the electron.  Using Eq.~(\ref{epsDefdFcn}), this can be written in terms of the reduced energy $\varepsilon$ as
\begin{equation}
v_{\rm g} = \frac{\hbar}{2ma}\frac{d\varepsilon}{d(ka)},
\label{Eq:vg}
\end{equation}
where $d\varepsilon/d(ka)$ is the dimensionless slope of the band structure in reduced variables.

From Eq.~(\ref{Eq:vg}), one can define a dimensionless group velocity~$v_{\rm g}^*$ as
\begin{equation}
v_{\rm g}^*(ka) = \frac{2ma}{\hbar}v_{\rm g}(ka) = \frac{d\varepsilon(ka)}{d(ka)},
\label{Eq:vgStar}
\end{equation}
where the band structure~$\varepsilon(ka)$ is given in Fig.~\ref{Fig:E_vs_ka_P6_extended_zone}.  Using Eq.~(\ref{Eq:depsdka}) one has
\begin{eqnarray}
v_{\rm g}^*(ka) &=& \label{Eq:depsdka2}\\
&& \hspace{-0.5in}\frac{2\varepsilon^{3/2}(ka)\sin(ka)}{P\sqrt{\varepsilon(ka)}\cos[\sqrt{\varepsilon(ka)}] + [\varepsilon(ka) - P]\sin[\sqrt{\varepsilon(ka)}]},\nonumber
\end{eqnarray}
where $\varepsilon(ka)$ is obtained by numerically solving Eq.~(\ref{Eq:caFromEps}), and here the range $1\leq ka/\pi\leq 7$ is used as input values for $ka$.  

\begin{figure}
\includegraphics[width=3.in]{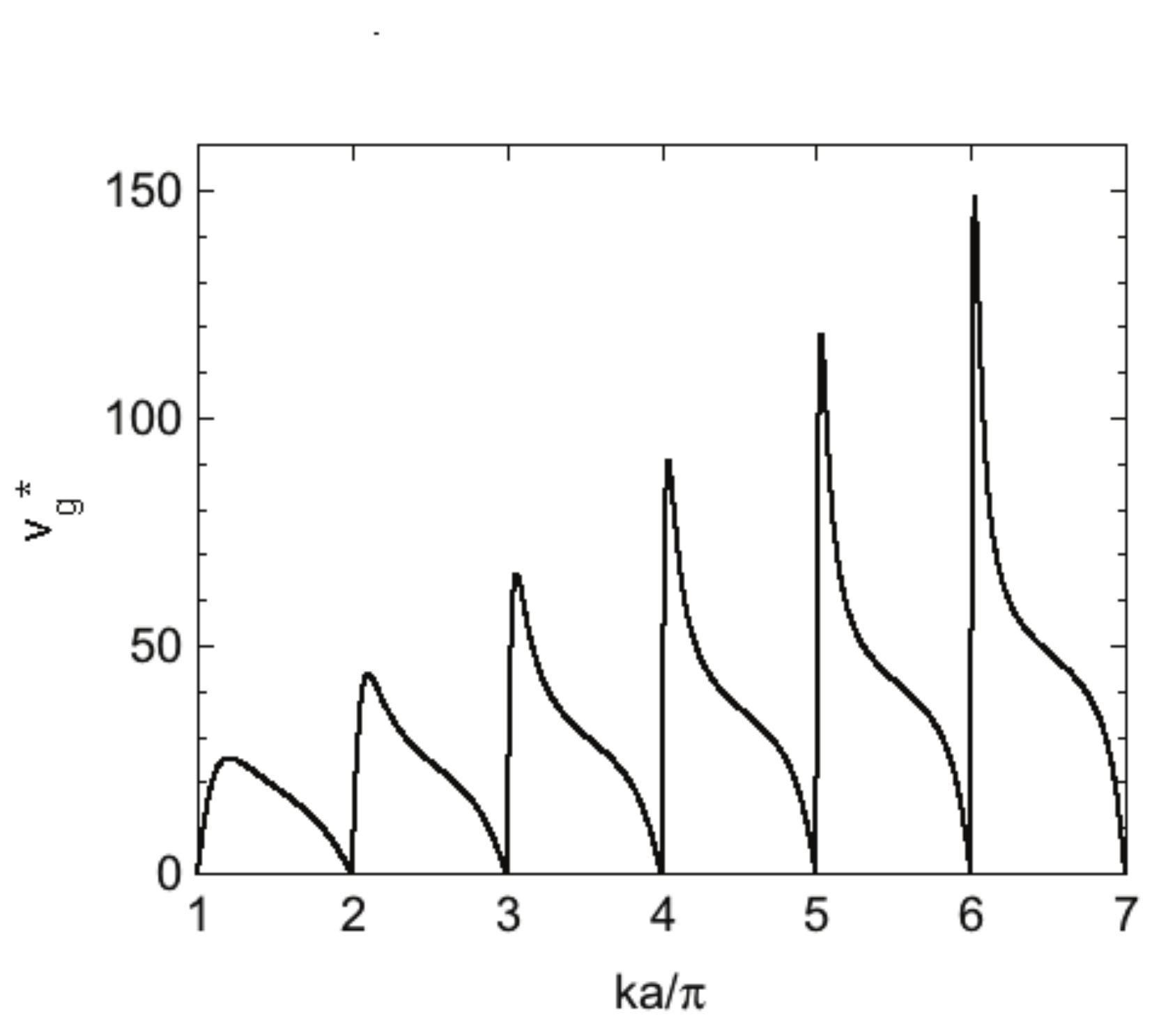}
\caption{Reduced electron group velocity $v_{\rm g}^*$ versus wave vector $ka/\pi$ obtained by solving Eqs.~(\ref{Eq:depsdka2}) with $P=6$ for bands~2 to~7 in Fig.~\ref{Fig:E_vs_ka_P6_extended_zone}.}
\label{Fig:vg_vs_ka_Dirac_KP}
\end{figure}

Figure~\ref{Fig:vg_vs_ka_Dirac_KP} shows a plot of $v_{\rm g}^*$ versus $ka/\pi$ in the extended-zone scheme as previously used for the band structure in Fig.~\ref{Fig:E_vs_ka_P6_extended_zone}.  One sees that $v_{\rm g}^*=0$ at the Brillouin-zone boundaries, as already discussed below Eq.~(\ref{Eq:DispRelnNearBZboundaries}).  Furthermore, with increasing $ka$, $v_{\rm g}^*$ increases rapidly from zero with increasing $ka$, levels out, and then decreases. The initial increase occurs increasingly rapidly with increasing band number~$p$.  The initial increases in Fig.~\ref{Fig:vg_vs_ka_Dirac_KP} are thus all continuous, although the ones with the larger band numbers~$p$ may look discontinuous in the figure.

Using Eq.~(\ref{Eq:vg}), the $x$~component $a_{\rm g}$ of the acceleration of the electron pulse is
\begin{equation}
a_{\rm g} = \frac{dv_{\rm g}}{dt} = \frac{dv_{\rm g}}{d(ka)}\frac{d(ka)}{dt} = \frac{\hbar}{2ma}\frac{d^2\varepsilon}{d(ka)^2}\frac{d(ka)}{dt}.
\label{Eq:ag} 
\end{equation}
Then using Eq.~(\ref{Eq:m*/m}) for $d^2\varepsilon/d(ka)^2$ one obtains 
\begin{equation}
a_{\rm g} = \frac{\hbar}{m^*(t)a}\frac{d(ka)}{dt},
\label{Eq:ag2}
\end{equation}
where $m^*$ is the band effective mass at the particular value of $ka$ under consideration in the extended-zone band structure which varies with time as $ka$ does.

If an electric field with $x$~component~$E$ is applied to an electron pulse, the $x$~component of the force on it is \mbox{$F=qE = -eE$}, where \mbox{$q=-e$} is the charge on the electron.  The $x$~component of the acceleration of the electron pulse is
\begin{equation}
a_{\rm g} = \frac{qE}{m^*} = -\frac{eE}{m^*}.
\label{Eq:afeE}
\end{equation}
Thus when treating an electron as a particle, the mass associated with its acceleration is the band mass~$m^*$ instead of the free-electron mass~$m$.
Equating Eqs.~(\ref{Eq:ag2}) and~(\ref{Eq:afeE}) gives
\begin{equation}
\frac{\hbar}{a}\frac{d(ka)}{dt} = -eE,
\label{Eq:dkadteE}
\end{equation}
where $m^\ast$ has dropped out.  This is still a quantum-mechanical expression due to the presence of~$\hbar$.  However, using the quantum expression $p = \hbar k$ for the momentum~$p$ of an electron, the classical expression
\begin{equation}
dp/dt=F_{\rm net}
\label{Eq:dpdtFnet}
\end{equation}
is obtained, which is Newton's 2$^{\rm nd}$ law.  This suggests that for some applications Eq.~(\ref{Eq:dpdtFnet}) can be useful such as in the derivation of Ohm's law in introductory textbooks.

\subsection{\label{Sec:NegativeEnergyStates} Wave Functions of the Negative-Energy Band States}

The reduced energies $\varepsilon$ in band~1 in Fig.~\ref{Fig:E_vs_ka_P6_extended_zone} for $P=6$ are negative.  Defining a corresponding reduced energy $\epsilon$ by
\bea
\epsilon = -\varepsilon,
\label{Eq:NewEpsDef}
\eea
Eq.~(\ref{Eq:coskak1a2}) becomes
\bea
\cos(ka) = \cosh(\sqrt{\epsilon}) - P \frac{\sinh(\sqrt{\epsilon})}{\sqrt{\epsilon}},
\label{coskcoshsinh}
\eea
where $ka$ is real for electron-band states.  This equation was implicitly employed by {\tt Mathematica} to calculate the dispersion relation of band~1 in Fig.~\ref{Fig:E_vs_ka_P6_extended_zone} using Eq.~(\ref{Eq:coskak1a2}).

The energy- and wave vector-dependent wave function in the region between $x/a=0$ and $x/a=1$ is obtained from Eq.~(\ref{Eq:psixa2}) for negative-energy band states as
\bse
\bea
\psi(x_a) &=& A\left\{ \sinh(\sqrt{\epsilon}x_a) + e^{-ika}\sinh[\sqrt{\epsilon}(1 -x_a)] \right\},\nonumber\\
\eea
and the corresponding probability density is
\bea
{\cal P}(x_a) &=& \psi^*(x_a)\psi(x_a)=u^*(x_a)u(x_a) \nonumber \\
&=& A^2\big\{\sinh^2(\sqrt{\epsilon}x_a) + \sinh^2[\sqrt{\epsilon}(1-x_a)] \nonumber\\
&&\hspace{-0.in} +\ 2\cos(ka) \sinh(\sqrt{\epsilon}x_a)\sinh[\sqrt{\epsilon}(1-x_a)]  \big\}.\nonumber\\
\eea
\ese
Normalizing $\psi(x_a)$ in the unit cell with $x/a = 0$ to~1 gives the amplitude $A$ as
\bea
A &=& \left\{(\epsilon +P)\Big[\frac{\sinh(\sqrt{\epsilon})}{\sqrt{\epsilon}}\Big]^2-P\cosh(\sqrt{\epsilon})\frac{\sinh(\sqrt{\epsilon})}{\sqrt{\epsilon}} \right\}^{-1/2},\nonumber\\
\label{Eq:AmplitudeA} 
\eea
where the relationship between $ka$ and~$\epsilon$ in Eq.~(\ref{coskcoshsinh}) was used.

The periodic function $u(x_a)$ is obtained from $\psi(x_a)$ using Bloch's theorem~(\ref{Eq:psi(x)u(x)}) as
\bea
u(x_a) = e^{-ika\,x_a} \psi(x_a)
\label{Eq:psixaKPdelta}
\eea
where here $u(x_a)$ is understood to mean $u[x_a,{\rm modulo}(1)]$ which expresses that the periodicity of $u(x_a)$ is the same as that of the lattice.
\begin{figure}
\includegraphics[width=2.75in]{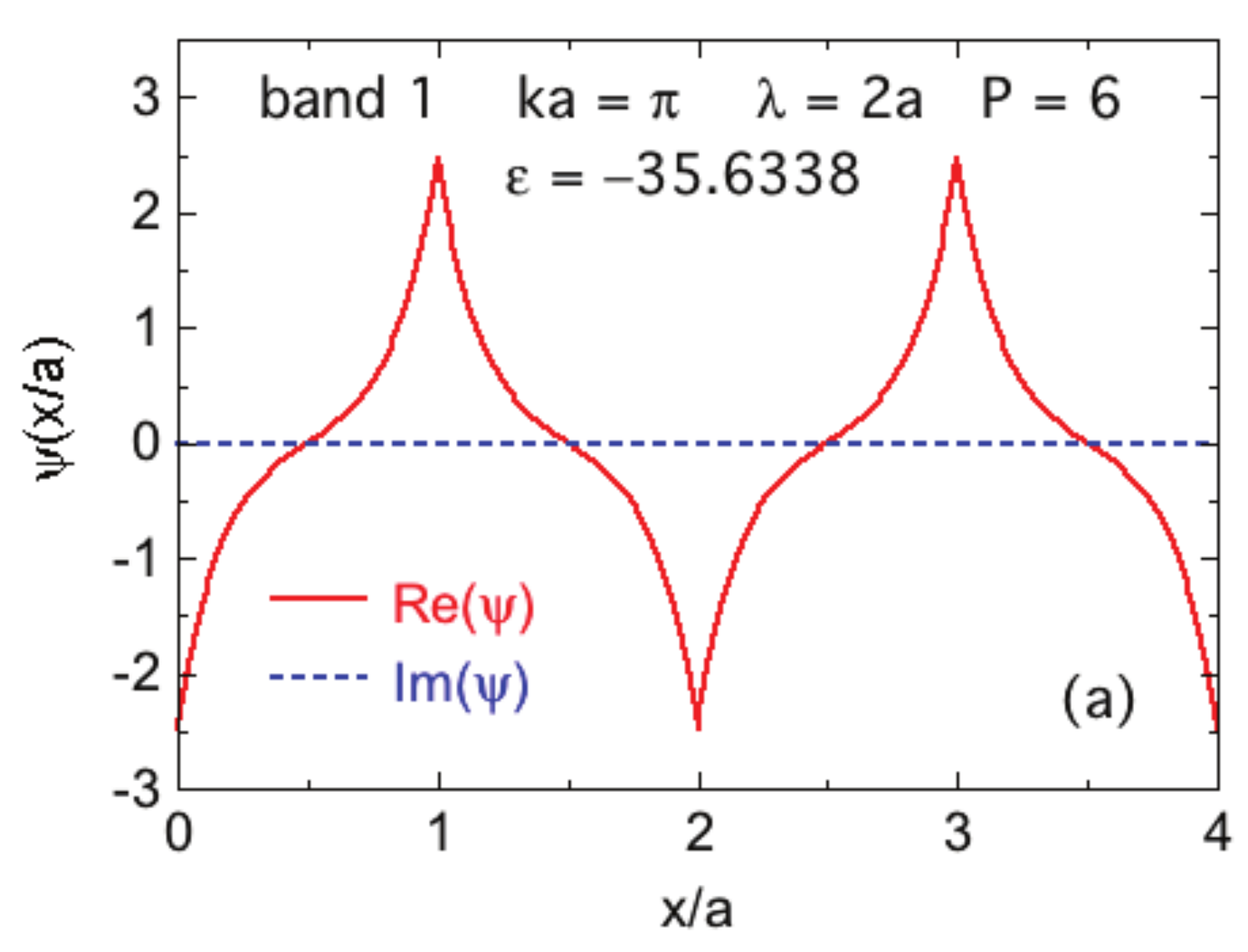}
\includegraphics[width=2.75in]{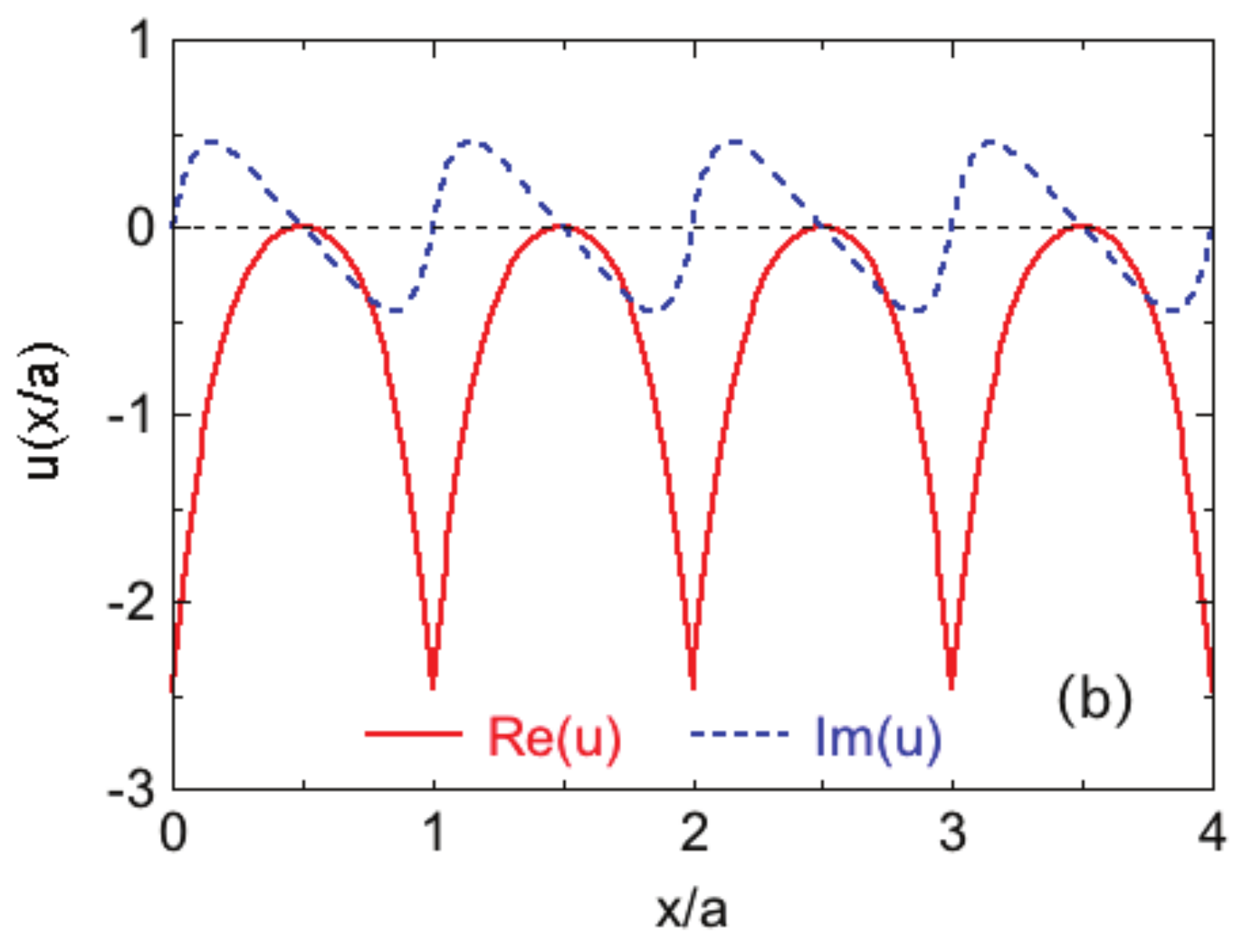}
\includegraphics[width=2.75in]{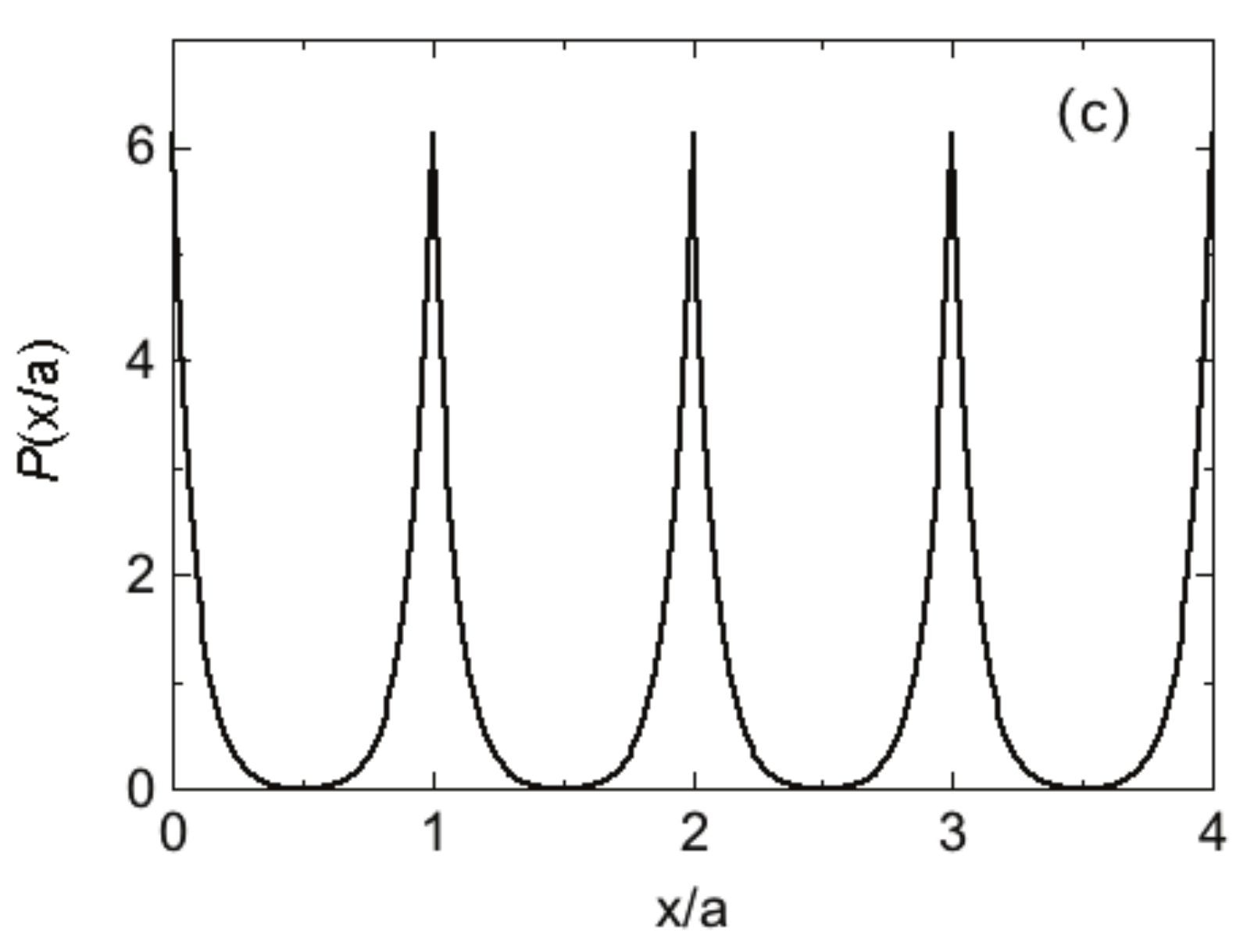}
\caption{(a) Real and imaginary parts of (a) the wave function $\psi(x/a)$ and (b)~$u(x/a)$ for the negative-energy band~1 in Fig.~\ref{Fig:E_vs_ka_P6_extended_zone} with \mbox{$ka = \pi$}.  (c)~Probability density ${\cal P}(x/a)$ associated with the wave function in~(a).}
\label{Fig:KPDirac_PsikaPi} 
\end{figure}

\begin{figure}
\includegraphics[width=2.75in]{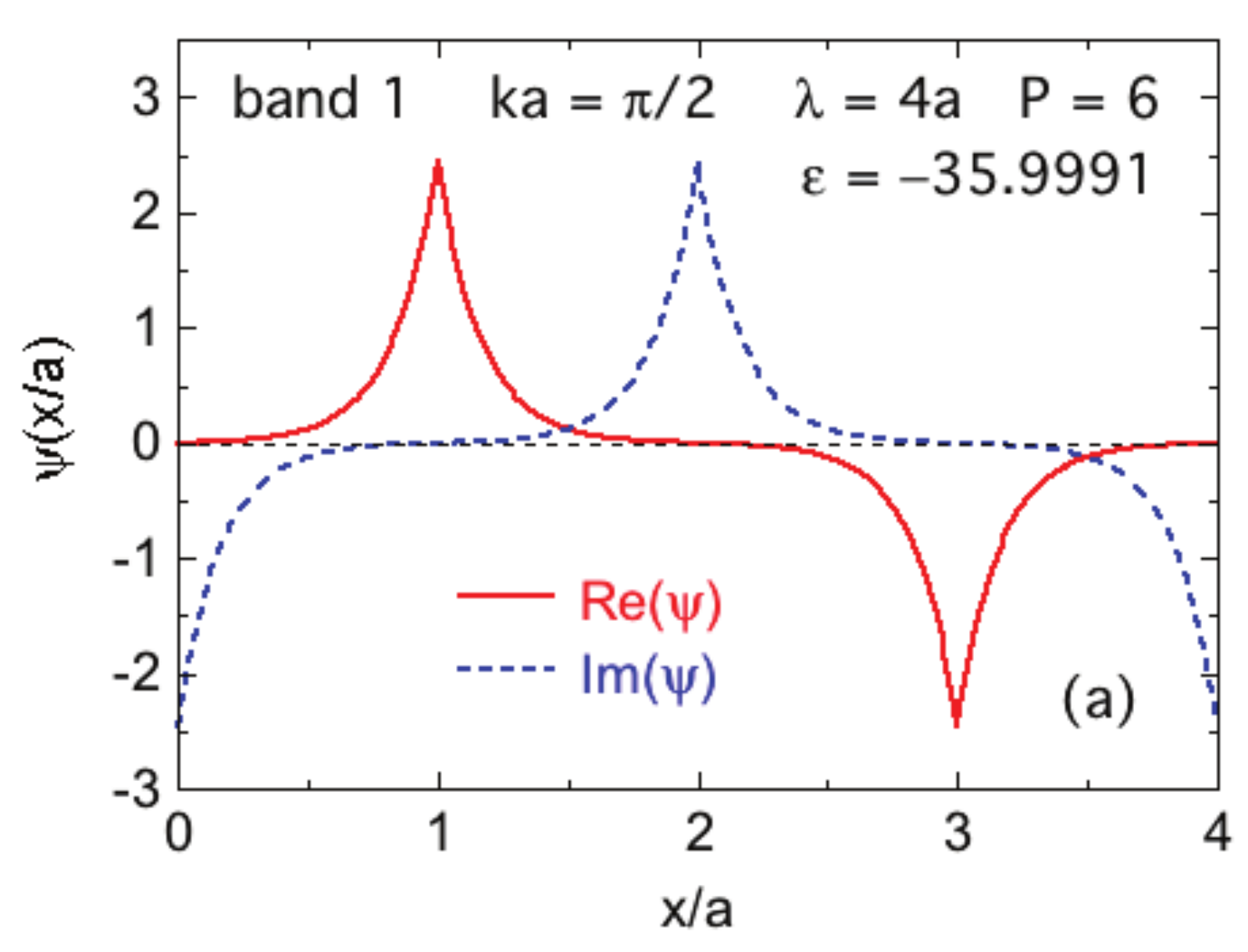}
\includegraphics[width=2.75in]{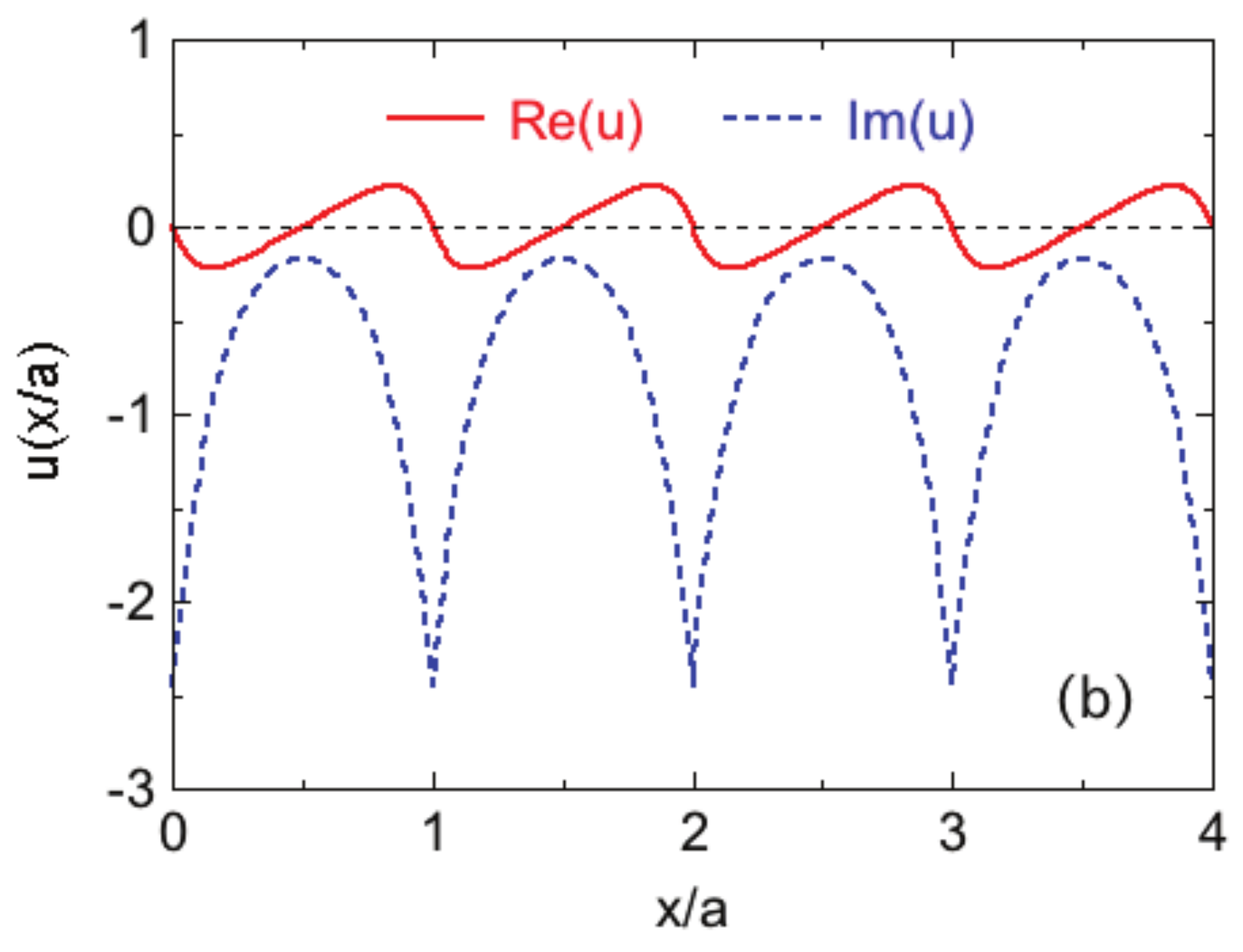}
\caption{Same as Fig.~\ref{Fig:KPDirac_PsikaPi} but with $ka = \pi/2$.  Here the wavelength of $\psi(x/a)$ is four instead of two as in Fig.~\ref{Fig:KPDirac_PsikaPi}(a).  The probability density versus position (not shown)  is the same as in Fig.~\ref{Fig:KPDirac_PsikaPi}(c).}
\label{Fig:KPDirac_PsikaPiOn2} 
\end{figure}

Within the first Brillouin zone, the wavelength~$\lambda$ of $\psi(x_a)$ is given by
\bea
\frac{\lambda}{a} = \frac{2\pi}{ka}.
\label{Eq:lambdaFromk}
\eea
Figures~\ref{Fig:KPDirac_PsikaPi}(a,b)  and~\ref{Fig:KPDirac_PsikaPiOn2}(a,b) show the real and imaginary parts of $\psi(x)$ and of the periodic function $u(x)$ versus $x/a$ for $ka=\pi$ and $ka = \pi/2$, respectively, with wavelengths $\lambda/a = 2$ and 4, respectively.  These wave functions have very different character compared to the sinusoidal behavior for positive-energy band and energy-gap wave functions such as shown above in Figs.~\ref{Fig:Band2PsiProb} and~\ref{Fig:Midgap_PsiProb}, respectively.

Shown in Fig.~\ref{Fig:KPDirac_PsikaPi}(c) is the probability density ${\cal P}$ versus $x/a$.  Because $\psi(x_a)$ and $u(x/a)$ are normalized over one unit cell, the integral of the probability density ${\cal P}(x_a)$ of either $u(x_a)$ or $\psi(x_a)$ from $x/a = 0$ to~$\lambda/a$ is
\bea
\int_0^{\lambda/a}{\cal P}(x_a)dx_a = \lambda/a. 
\eea
To normalize $\psi(x_a)$ or $u(x_a)$ to unity over the a length $\lambda/a = 2\pi/(ka)$, one would therefore divide the amplitude~$A$ in Eq.~(\ref{Eq:AmplitudeA}) by~$\sqrt{\lambda/a}$.

\subsection{\label{Sec:KTBoundDiracCombStates} Bound States}

\subsubsection{Introduction: A Single Attractive Dirac $\delta$ Function Potential Energy}

\begin{figure}
\includegraphics[width=3.in]{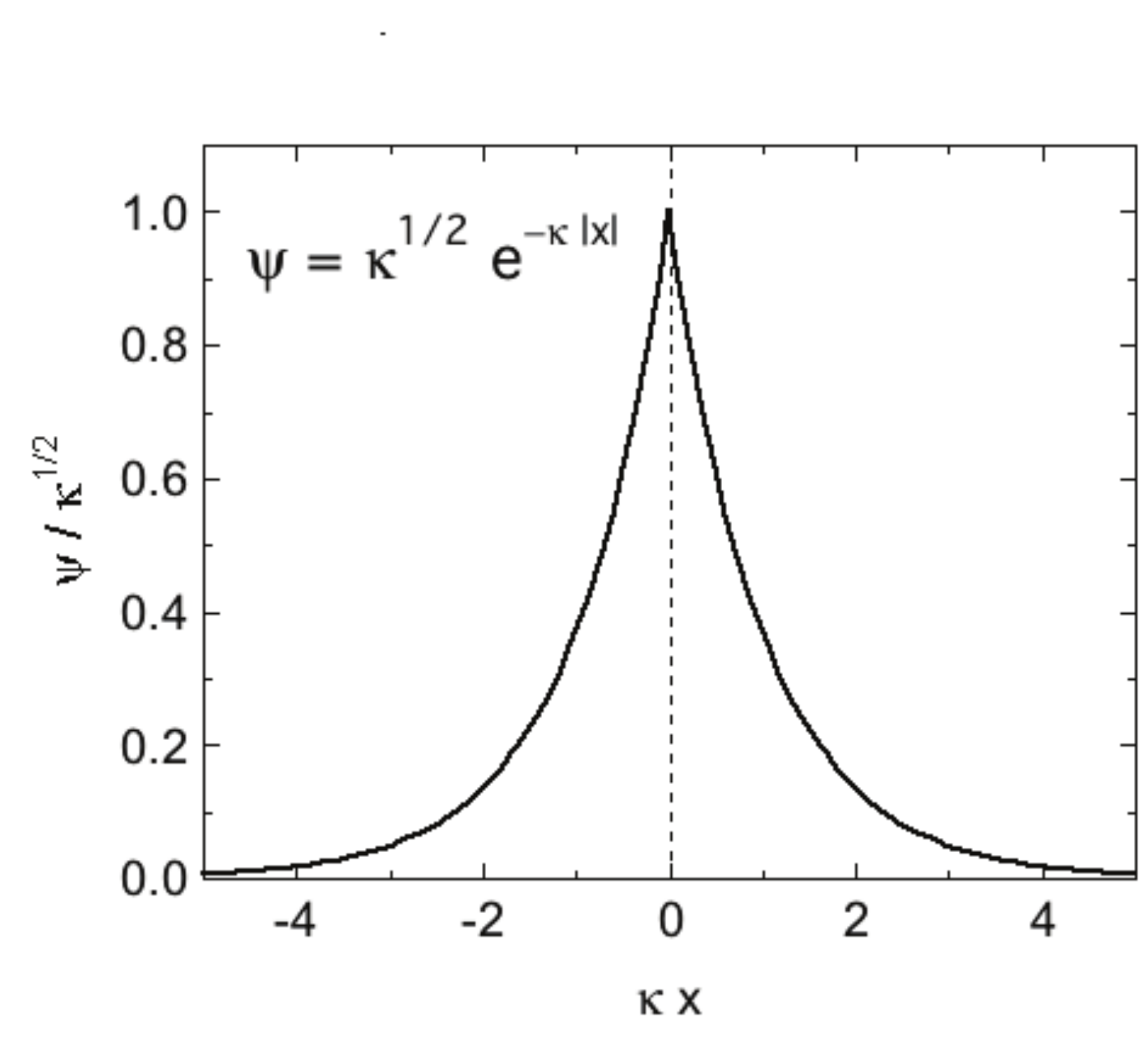}
\caption{Wave function $\psi$ normalized by the square root of the decay constant~$\sqrt{\kappa}$ versus $\kappa x$ described by Eq.~(\ref{Eq:deltaone}).}
\label{Fig:Psi_delta_fcn} 
\end{figure}

Consider first a single attractive $\delta$ function at $x=0$ with strength~$-\alpha$ where the potential energy $U(x< 0) = 0$ and $U(x> 0) = 0$ .  The Schr\"odinger equation gives a single bound state with negative energy~$E$ described by~\cite{Griffiths2015}
\bse
\label{Eqs:1deltaFcn}
\bea
E &=& -\frac{\hbar^2 \kappa^2}{2m} = -\frac{m\alpha^2}{2\hbar^2},\label{Eq:EBoundDef}\\
\kappa &=& \sqrt{\frac{-2mE}{\hbar^2}} = \frac{m\alpha}{\hbar^2},\label{Eq:kappaBound}\\
\psi(x) &=& \sqrt{\kappa}\,e^{-\kappa|x|},\label{Eq:deltaone}
\eea
\ese
where the prefactor of the wave function is determined from the normalization condition.  Thus $E$ is proportional to the square of the strength~$\alpha$ of the $\delta$~function.  The wave function must have even parity so that it does not diverge on either side of the $\delta$~function with increasing $|x|$.  A plot of Eq.~(\ref{Eq:deltaone}) using dimensionless normalized coordinates is shown in Fig.~\ref{Fig:Psi_delta_fcn}.  Thus if $\kappa |x| \gtrsim 5$ and there are multiple $\delta$~functions along a line in one dimension separated by distances~$\Delta x$ satisfying $\kappa\Delta x\gtrsim 10$, then the solution for $\psi$ associated with each $\delta$~function is given by Eq.~(\ref{Eq:deltaone}).  According to the definition of $P$ in Eq.~(\ref{Eq:PDef}) and of the reduced energy~$\varepsilon$ in Eq.~(\ref{epsDefdFcn}) for an attractive Dirac comb, the reduced energy for the bound state in Eq.~(\ref{Eq:EBoundDef}) becomes simply
\bea
\varepsilon = -P.
\label{Eq:epsInPunits}
\eea

On crossing a $\delta$ function at position $x$ with increasing~$x$, $\psi(x)$ is continuous,
\bea
\psi(x)^+ = \psi(x)^- = \psi(x),
\label{Eq:PsiDirac+-}
\eea
but the slope $\psi^\prime(x)\equiv d\psi(x)/dx$ is not.  Equation~(\ref{Eq:dpsidx+-}) gives
\bse
\label{Eqs:PsiPrime}
\bea
\psi^\prime(x)^+ -\psi^\prime(x)^- &=& -z \psi(x),\label{PsiPrime(a)}\\
z = \frac{2m\alpha}{\hbar^2} &=& 2\kappa,\label{Eq:zDef}
\eea
\ese
where the superscripts + and~$-$ indicate that the quantity in question is taken on approaching the $\delta$~function position from the right and left, respectively.  One can easily show that the wave function for the $\delta$~function at $x=0$ in Fig.~\ref{Fig:Psi_delta_fcn} satisfies Eqs.~(\ref{Eqs:PsiPrime}).

\subsubsection{Attractive Dirac Comb Potential Energy}

\begin{figure}
\includegraphics[width=2.in]{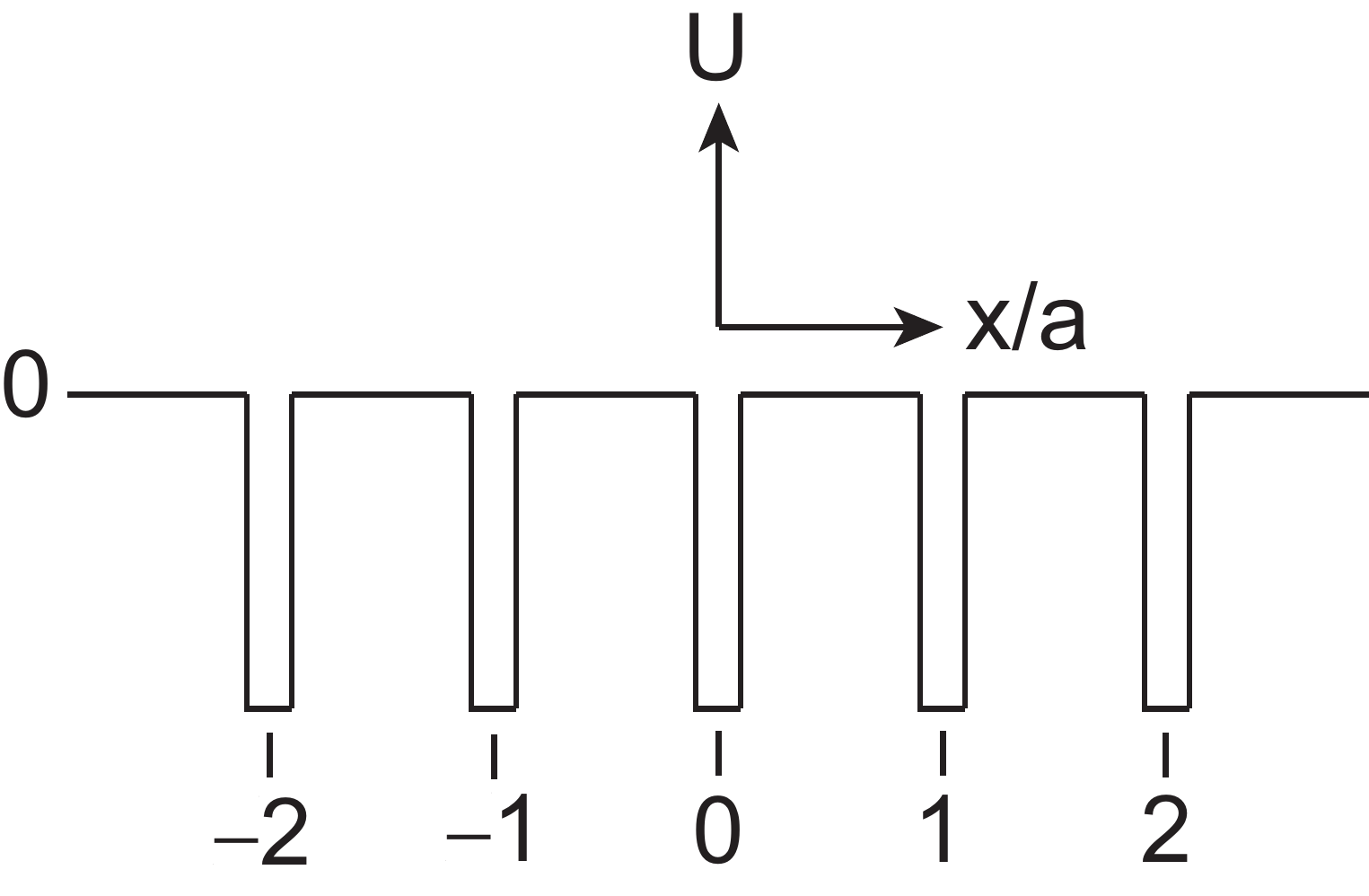}
\caption{Dirac comb with an attractive $\delta$~function potential energy. The labeling of the positions of the $\delta$ functions is shown.  The potential wells are infinitely narrow but also infinitely deep, resulting in a nonzero area (strength) of~$-\alpha$. }
\label{Fig:Kronig_Penney_Model_Dirac} 
\end{figure}

Now we extend the discussion to the attractive Dirac comb described in Eq.~(\ref{Eq:DiracComb}) with lattice parameter~$a$.  A part of the Dirac comb near $x=0$ is shown in Fig.~\ref{Fig:Kronig_Penney_Model_Dirac}, where the reduced position coordinates of the $\delta$~functions are 
\bea
x/a = 0,\ \pm 1,\ \pm 2, \ldots,
\eea
as shown. The wave function $\psi(x)$ is defined to be centered at $x=0$.  Here we only derive the wave function for $x/a \geq 0$ because as noted above, the parity of the wave function is even.  The solution for $\psi(x)$ for $x \leq 0$ can thus be obtained as $\psi(x\leq 0) = \psi(x\geq0)|_{x\to -x}$.

According to Fig.~\ref{Fig:Psi_delta_fcn}, if $\kappa a \gtrsim 10$, $\psi(x)$ is, for all practical purposes, confined to the region $-a/2<x<a/2$ and hence is given simply by Eq.~(\ref{Eq:deltaone}).  However, if $\kappa a \lesssim 10$, the wave functions in adjacent unit cells overlap one must compute the bound-state  wave function for this case using Eqs.~(\ref{Eq:PsiDirac+-}) and~(\ref{Eqs:PsiPrime}).  A  regression algorithm can be used until $\psi$ decreases to a negligible value with increasing $x/a$.  In either case, the wave function is \mbox{$N$-fold} degenerate because the position of the maximum in $\psi$ can be placed at any of the positions of the $N$ positive ions in the ring.  

To obtain the regression series, we redefine the reduced parameter~$x_a$ for each unit cell with $x \geq 0$ and $n = 0,\ 1,\, 2,\ \ldots$ as 
\bea
x/a = n + x_a,\label{xaDef}\\
0 \leq x_a \leq 1.\nonumber 
\eea
The wave functions and their derivatives with respect to~$x$ are then written as
\bea
\psi(n,x_a) &=& A(n)e^{-B(n)(n + x_a)},\\
\psi^\prime(n,x_a) &\equiv& \frac{d\psi(n,x_a)}{dx_a} = - A(n) B(n) e^{-B(n)(n+x_a)},\nonumber
\eea
where the inital parameters $A(0)$ and $B(0)$ are
\bea
A(0) = 1,\quad B(0) = \kappa_0 a.
\label{Eqs:InitialPars}
\eea
The condition~(\ref{Eq:PsiDirac+-}) gives
\bse
\bea
\psi(n+1,0) = \psi(n,1),
\label{Eq:Psi0+}
\eea
and condition~(\ref{PsiPrime(a)}) requires
\bea
\psi^\prime(n+1,0) = \psi^\prime(n,1) - 2\kappa_0 a,
\label{Eq:Psip0+}
\label{Eq:psiSlopeChange}
\eea
\ese
where $za=2\kappa_0 a$ from Eq.~(\ref{Eq:zDef}) and $\kappa_0a$ is the decay constant for $0<x/a<1$ as noted above.  These two conditions and the initial conditions~(\ref{Eqs:InitialPars}) allow $A(n)$ and $B(n)$ and hence $\psi(n,x_a)$ to be calculated by regression for a given initial value of $\kappa_0 a$.  The analyic regression series quickly becomes too cumbersome to use with increasing~$n$.  Therefore, we carried out numerical calculations of the $A(n)$ and~$B(n)$ values for an initial value $\kappa_0 a = 0.1$.  This initial value was chosen so that a bound-state wave function has a total width of about ten unit cells.  The normalization factor is also determined numerically.

\begin{figure}
\includegraphics[width=3.in]{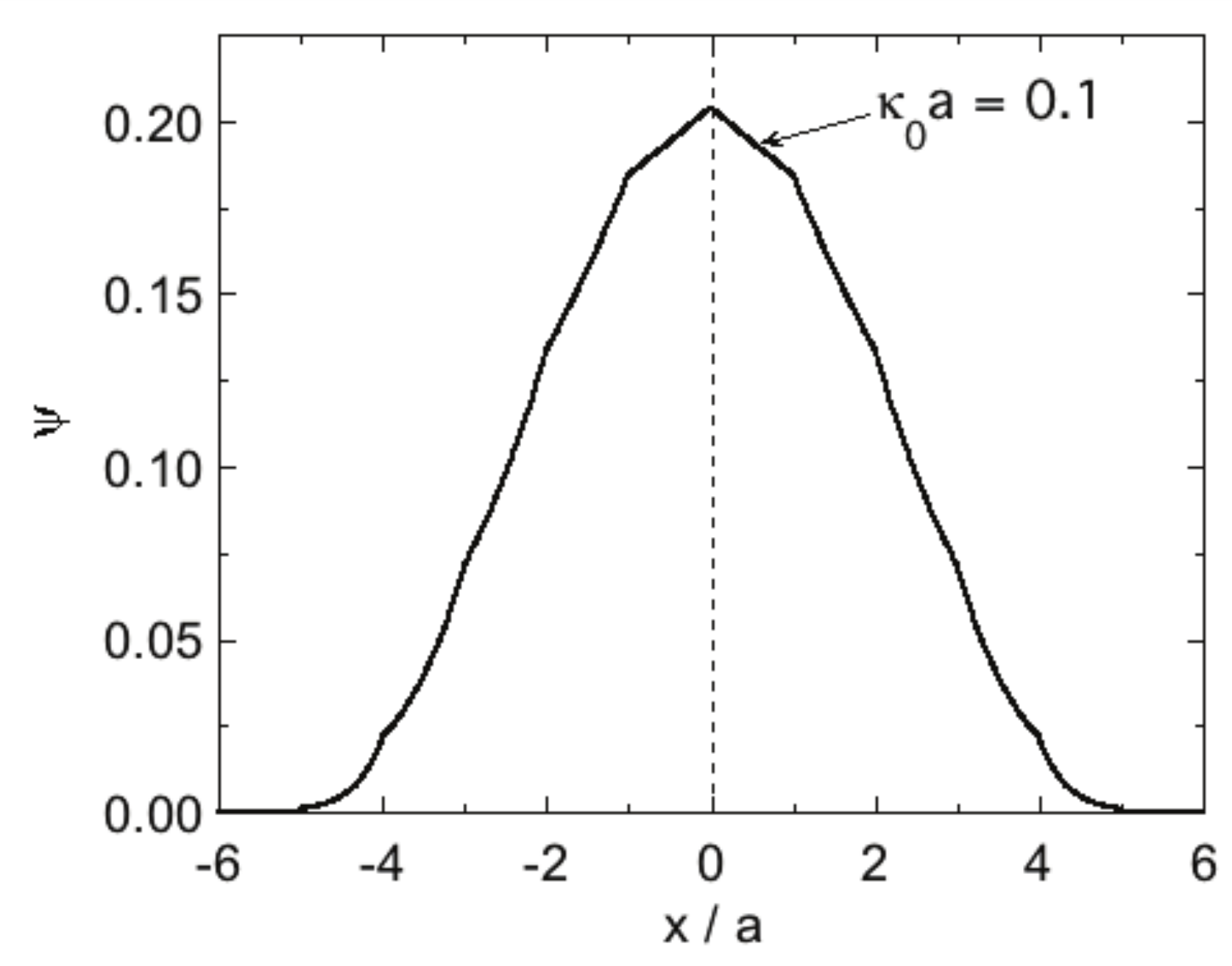}
\caption{Normalized bound-state wave function $\psi$ for an attractive Dirac-comb potential energy versus normalized position $x/a$ for initial $\kappa_0 a = 0.1$.}
\label{Fig:KPdeltaFcn_kappaa0.1} 
\end{figure}

A plot of the normalized bound-state wave function~$\psi$  versus reduced position $x/a$ is shown in Fig.~\ref{Fig:KPdeltaFcn_kappaa0.1} for $\kappa_0 a = 0.1$.  The discontinuities in slope of $\psi(x)$ predicted by Eqs.~(\ref{Eqs:PsiPrime}) are clearly visible in the figure.  The decay rate discontinuously increases as $|x/a|$ increases past each $\delta$~function.  For this reason, the inital decay rate is much smaller than the final one before the wave function becomes exponentially small in magnitude.

\subsubsection{Energy of the Bound State}

To obtain the expectation value of the Hamiltonian $\langle{\cal H}\rangle$ associated with a (real) wave function $\psi(x)$ for an attractive Dirac comb potential energy such as in Fig.~\ref{Fig:KPdeltaFcn_kappaa0.1}, we use the conventional expression
\bse
\label{Eqs:DiracEnergy}
\bea
\langle{\cal H}\rangle \equiv E = \int_{{\rm all}\ x}\psi(x)\hat{\cal H}\psi(x)dx,
\eea
where the Hamiltonian operator $\hat{\cal H}$ for the regions between the $\delta$ functions is
\bea
\hat{\cal H}=-\frac{\hbar^2}{2m}\frac{d^2}{dx^2}.
\label{Eq:hatcalH}
\eea
\ese
The presence of the Dirac-comb potential energy is taken into accoount via the slope changes of $\psi(x_a)$ at the positions $x_a = n$ of the $\delta$~functions described in Eq.~(\ref{Eq:psiSlopeChange}) and shown for initial $\kappa_0 a = 0.1$ in Fig.~\ref{Fig:KPdeltaFcn_kappaa0.1}.

In between the $\delta$ functions at $x/a = n$ and~$n+1$ and using the definition $x_a = x/a$ in Eq.~(\ref{xaDef}), the wave function $\psi(n,x_a)$ for $x>0$ is
\bea
\psi(n,x_a) = \psi(n,0) e^{-\kappa_n a x_a},
\label{Eq:psinxa}
\eea
where according to Eqs.~(\ref{Eqs:PsiPrime}) one has
\bea
\kappa_n a = \kappa_0a + 2n\kappa_0a =\kappa_0a(1+2n).
\label{Eq:kappan0}
\eea
The contribution $E_n$ to $E$ from the wave function between $x = na$ and $(n+1)a$ obtained from Eqs.~(\ref{Eqs:DiracEnergy}), (\ref{Eq:psinxa}), and~(\ref{Eq:kappan0}) is
\bea
E_n &=& -\frac{\hbar^2}{2ma^2}(\kappa_na)^2 a\int_0^1 \psi^2(n,x_a)dx_a \nonumber\\
&=& -\frac{\hbar^2}{2ma^2}(\kappa_na)^2\psi^2(n,0)a \int_0^1 e^{-\kappa_nax_a}dx_a \nonumber\\
&=& -\frac{\hbar^2}{2ma^2}(\kappa_0a)^2(1+2n)^2\psi^2(n,0) a \label{EnDiracComb}\\
&&\hspace{0.5in}\times\ \left[1-e^{-\kappa_0a(1+2n)}\right],\nonumber
\eea
where $\kappa_0a$ and $\psi^2(n,0)a$ are dimensionless.  The energy of the bound state is then
\bea
E = 2\sum_{n=0}^{n_{\rm max}}E_n,
\label{Eq:EDiracComb}
\eea
where the factor of two takes into account the wave function at $x<0$ and $n^{\rm max}$ is the value of $n$ at which $\psi(x_a)$ becomes exponentially small.  In Fig.~\ref{Fig:KPdeltaFcn_kappaa0.1}, $n_{\rm max} = 5$.

In order to compare the energy of a state arising from a Dirac comb with the energy~$E_1$ of an electron exposed to a single attractive Dirac $\delta$~function with the same initial value of $\kappa_0a$ we divide $E_n$ in Eq.~(\ref{EnDiracComb}) by the energy $E_1$ in Eq.~(\ref{Eq:EBoundDef}) given in the present notation by
\bea
E_1 = -\frac{\hbar^2 (\kappa_0a)^2}{2ma^2}.
\eea
Using Eq.~(\ref{EnDiracComb}) one obtains
\bse
\bea
\frac{E_n}{E_1} = (1+2n)^2\psi^2(n,0) a \left[1-e^{-\kappa_0a(1+2n)}\right].
\eea
The only input needed here are the values of the normalized $\psi^2(n,0)a$ which were already calculated in order to construct a plot such as given in  Fig.~\ref{Fig:KPdeltaFcn_kappaa0.1} for $\kappa_0a=0.1$. Then using Eq.~(\ref{Eq:EDiracComb}) one obtains
\bea
\frac{E}{E_1} = 2\sum_{n=0}^{n_{\rm max}}(1+2n)^2\psi^2(n,0) a \left[1-e^{-\kappa_0a(1+2n)}\right].\nonumber\\
\eea
\ese
Using $\kappa_0a=0.1$, we numerically obtain
\bea
\frac{E}{E_1} = 0.818.
\eea
Thus this energy of an electron localized in an attractive Dirac-comb potential is on the order of that for a single Dirac $\delta$~function with $\kappa a = \kappa_0 a$.  Using Eq.~(\ref{Eq:epsInPunits}) gives the reduced energy of the bound state as
\bea
\varepsilon = -0.818 P.
\eea
This gives the negative value $\varepsilon = -4.91$ for $P=6$, which is above band~1 in Fig.~\ref{Fig:E_vs_ka_P6_extended_zone} and below the minimum of the free-electron dispersion. To reiterate, the state is $N$-fold degenerate because the maximum in the wave function can be placed at the position of any of the $N$ positive ions in the ring of ions.  Thus the degeneracy of the localized state is the same as the number of states in each band.  This discussion suggests that localized states may generally be present in solids as previously intimated in Ref.~\cite{Borowitz1967}.

\section{\label{Summary} Concluding Remarks}

Due to the importance of the KP model which is usually the first topic studied with respect to the band structure of solids, we have presented results for we believe is a more realistic version of the KP model corresponding to attractive rather than the repulsive ionic potentials usually considered.  The primary new results are that the lowest-energy band of electron states is at negative energy and that these states have wave functions qualitatively different from the sinusoidal states of the positive-energy band and band-gap states.  In addition, a negative-energy bound-state solution was found and the wave function given.  This state has $N$-fold degeneracy where $N$ is the number of ions in the ring, which is the same number of states as occurs in each propagating-electron band.

Despite its pedagogical importance, the attractive Kronig-Penney model has attracted little attention in the past.  Even for the repulsive KP model, usually only the band structure is discussed together with related quantities such as the density of states and band effective mass versus energy.  Here quantitative calculations of the wave functions of both band and energy-gap states of the KP model are reported.  A perhaps unexpected result is that in-gap states with complex crystal momenta have standing wave functions, just as the states at the tops and bottoms of the positive-energy bands do as is well known.

\appendix

\section{\label{Sec:KTBoundStates} Bound States of the Attractive Square-Well Kronig-Penney Potential}

\subsection{Bound-State Energies}
        
When the energy is negative, bound states may occur as suggested from Fig.~\ref{Fig:Kronig_Penney_Model}.  If the potential wells are suffciently far apart, then there is no overlap of the bound-state wave functions associated with different wells and the problem reduces to finding the bound-state energies and wave functions of a single square well.  This is the case considered here.  Textbooks usually give the energies of the even- and odd-parity wave functions in terms of graphical solutions.  However, solutions for the bound-state wave functions and energies versus the quantum number~$n$ of the state have been known for a long time~\cite{Cantrell1971, Sprung1992, Aronstein2000} but have not been widely discussed.  Here this solution is reviewed, the bound-state energies and wave functions described in terms of the quantum number~$n$, and then the results are considered within the context of the attractive square-potential Kronig-Penney model.  

\begin{figure}
\includegraphics[width=1.75in]{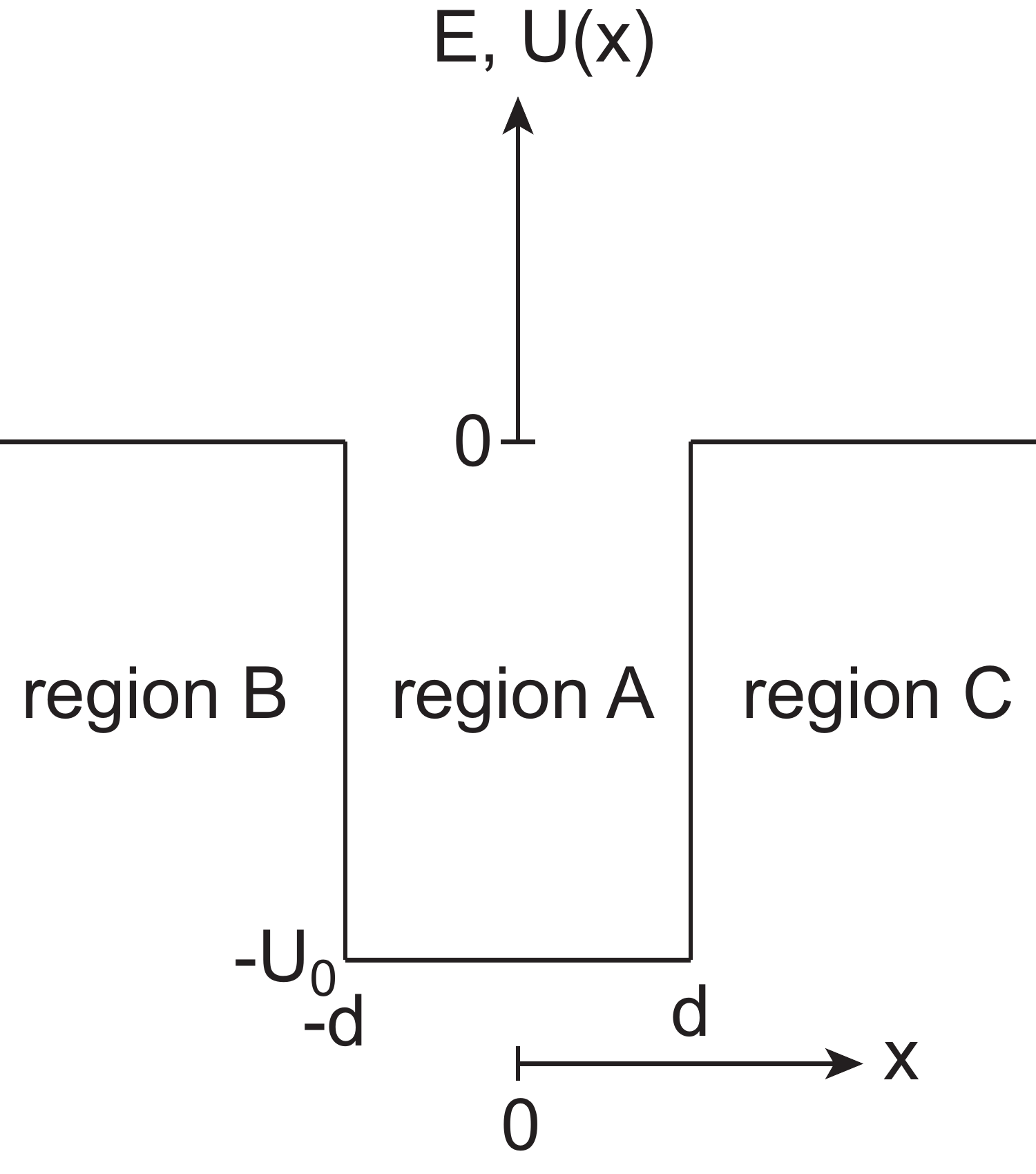}
\caption{Picture of a potential well of depth $U_0$ and width $2d$.}
\label{Fig:PotentialWell}
\end{figure}

A picture of a potential well with the notation used here is shown in Fig.~\ref{Fig:PotentialWell}.  We define the dimensionless reduced $x$-axis position parameter 
\bea
x_d = x/d.
\eea
The width of a well is taken as $2d$, with the potential energy seen by an electron with mass~$m$ given by
\bea
U(x) = 
\begin{cases}
0, & x_d < -1\ {\rm (region~B)}\\
-U_0, & -1 < x_d < 1 \ {\rm (region~A)}\\
0, & x_d > 1\ {\rm (region~C)}
\end{cases}
\eea

The time-independent Schr\"odinger equation yields the even-parity wave functions
\bse
\label{Eqs:WaveFunctions}
\bea
\psi^{\rm even}_{\rm B} &=& B e^{\kappa d x_d},\nonumber\\
\psi^{\rm even}_{\rm A} &=& \cos(kd), \label{Eqs:PsiEven}\\
\psi^{\rm even}_{\rm C} &=& B e^{-\kappa d x_d},\nonumber
\eea
and odd-parity wave functions
\bea
\psi^{\rm odd}_{\rm B} &=& B e^{\kappa d x_d},\nonumber\\
\psi^{\rm odd}_{\rm A} &=& \sin(kd), \label{Eqs:PsiOdd}\\
\psi^{\rm odd}_{\rm C} &=& -B e^{-\kappa d x_d},\nonumber
\eea
where $B$, $k$, and $\kappa$ are all real for bound-state wave functions and the normalization factor for the wave functions will be determined later.  The amplitude $B$ is found to be
\bea
B = 
\begin{cases}
\frac{\cos(kd)}{e^{-\kappa d}} & {\rm (even~parity)},\\
\\
\frac{\sin(kd)}{e^{-\kappa d}} & {\rm (odd~parity)}.
\end{cases}
\eea
\ese
Also obtained are the relationships
\bse
\label{Eqs:kdkdP}
\bea
(kd)^2 &=& \frac{2m(E+U_0)d^2}{\hbar^2},\label{Eq:kd2}\\
(\kappa d)^2 &=& \frac{2m(-E)d^2}{\hbar^2},\\
(kd)^2+(\kappa d)^2 &=& \frac{2m U_0d^2}{\hbar^2} = R^2,\label{Eq:kd2kd2}\\
R &\equiv& \sqrt{ \frac{2m U_0d^2}{\hbar^2}}.\label{Eq:PDef22}
\eea
\ese
For calculation and plotting purposes we define the reduced energy $\varepsilon$ obtained using Eqs.~(\ref{Eqs:kdkdP}) as
\bse
\label{Eqs:epskap}
\bea
\varepsilon \equiv -\frac{2m(-E)d^2}{\hbar^2} =  -(\kappa d)^2 = -\left[R^2 - (kd)^2\right] \label{Eq:epsFromkd},\hspace{0.2in}
\label{Eq:EpsFromKd}
\eea
where
\bea
\kappa d = \sqrt{R^2-(kd)^2}.
\label{Eq:kappad}
\eea
\ese

Utilizing the boundary conditions that the wave functions and their derivatives must match at the boundaries of $U(x)$ at $x = \pm d$ gives the relationships between $kd$ and $\kappa d$ as
\bse
\bea
\kappa d &=& kd \tan(kd)   \quad{\rm (even~parity)},\\*
\kappa d &=& -kd\, {\rm cotan}(kd)\quad{\rm (odd~parity)}.
\eea
\ese
These can be respectively written as
\bse
\bea
kd\sin(kd)-\kappa d \cos(kd) &=& 0,\\
kd\cos(kd)+\kappa d\sin(kd) &=& 0.
\eea
\ese

In order to obtain a single equation for both even and odd parity wave functions, one can multipy these two equations~\cite{Cantrell1971, Sprung1992, Aronstein2000}, yielding after trigonometric manipulations
\bea
(kd^2-\kappa d^2)\sin(2kd)+2kd\,\kappa d \cos(2kd) = 0.
\label{Eq:kd2-kd2}
\eea
Using Eq.~(\ref{Eq:kappad}) gives
\bea
(R^2-2kd^2)\sin(2kd) + 2kd\sqrt{P^2-kd^2}\cos(2kd)= 0,\nonumber\\
\eea
which only contains $kd$ and the specified value of~$R$\@.  Now dividing both sides by $R$ gives
\bea
\left[1-2\left(\frac{kd}{R}\right)^2\right]\sin(2kd)\hspace{1.5in}\label{Eq:InterimEq}\\
 +\ \left[2\frac{kd}{R}\sqrt{1-\left(\frac{kd}{R}\right)^2}\right]\cos(2kd) = 0.\nonumber
\eea

One can show that the left side of Eq.~(\ref{Eq:InterimEq}) is the sine of the sum of two angles
\bse
\bea
\sin(2kd+2\Phi)=0,
\label{Eq:sin2kd2Phi}
\eea
where 
\bea
\Phi = \arcsin\left(\frac{kd}{R}\right).
\label{Eq:Phiarcsin}
\eea
\ese
The solution of Eq.~(\ref{Eq:sin2kd2Phi}) is
\bea
2(kd+\Phi) = n\pi,
\eea
where $n\geq 1$ is the integer quantum number of the bound state with the ground state having~$n=1$.  Using Eq.~(\ref{Eq:Phiarcsin}), one obtains an  equation from which $kd$ can be determined for $n = 1,\ 2, \ldots,\ n^{\rm max}$~\cite{Cantrell1971, Sprung1992, Aronstein2000},
\bse
\label{Eqs:Solve_knvsn}
\bea
kd + \arcsin\left(\frac{kd}{R}\right) = \frac{n\pi}{2},
\label{Eq:kdCalc}
\eea
where the maximum $n$ for a bound state is
\bea
n^{\rm max} = {\rm Integer}\left(\frac{R}{\pi/2}\right) + 1
\label{Eq:nmax}
\eea
\ese
and ${\rm Integer}(x)$ is the largest integer less than~$x$.  When solving for $kd(n)$ using Eq.~(\ref{Eq:kdCalc}), we used a numerical search routine with an initial value $kd_0=0.1$. 

\begin{figure}
\includegraphics[width=2.5in]{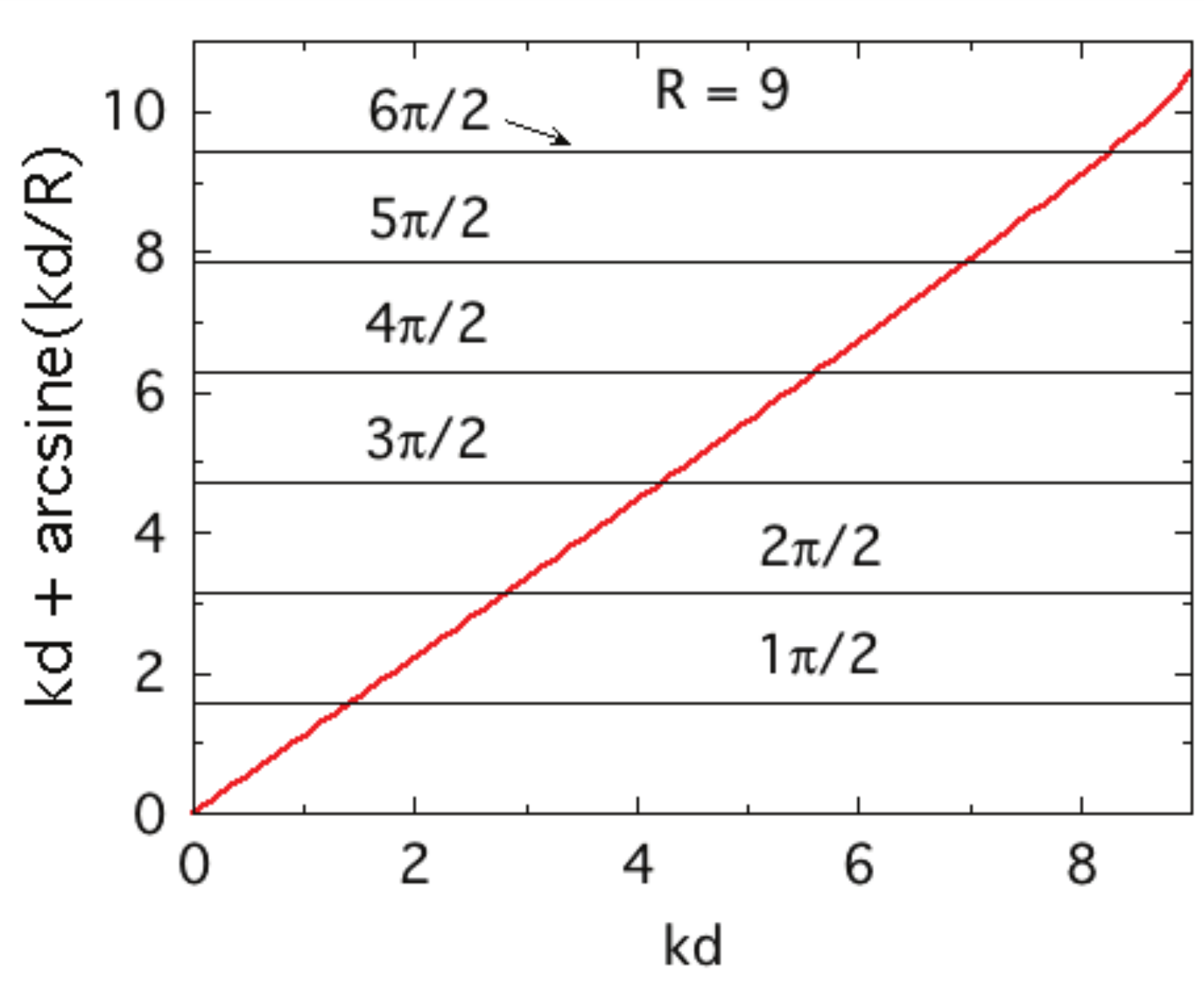}
\caption{Plot of $kd + {\rm arcsine}(kd/R)$ in Eq.~(\ref{Eq:kdCalc}) versus~$kd$ for $R=9$ (red curve).  Although the plot looks proportional over much of its length, the plot has a small positive curvature.  The values of $kd$ at the intersections of this curve with the horizontal values of $n\pi/2$ are the allowed values of~$kd$.  From these, the values of the reduced bound-state energies~$\varepsilon$ can be found using Eq.~(\ref{Eq:EpsFromKd}).}
\label{Fig:KPepsVSkd_R9}
\end{figure}

One can also obtain graphical solutions of the discrete allowed values of $kd$ for a given value of~$R$\@.  Figure~\ref{Fig:KPepsVSkd_R9} shows a plot of \mbox{$kd + \arcsin(kd/R)$} on the left side of Eq.~(\ref{Eq:kdCalc}) versus~$kd$ for $R=9$.  According to right side of Eq.~(\ref{Eq:kdCalc}), the solution is given by the intersection of this plot with horizontal lines with $y$-axis values  $n\pi/2$, as shown in the figure.  Once the allowed $kd$ values are obtained, the bound-state reduced energies $\varepsilon$ are calculated from Eq.~(\ref{Eq:EpsFromKd}).  One sees from Fig.~\ref{Fig:KPepsVSkd_R9} that there are $n^{\rm max} = 6$ discrete bound-state energies for $R=9$, consistent with Eq.~(\ref{Eq:nmax}).

\begin{figure}
\includegraphics[width=3.4in]{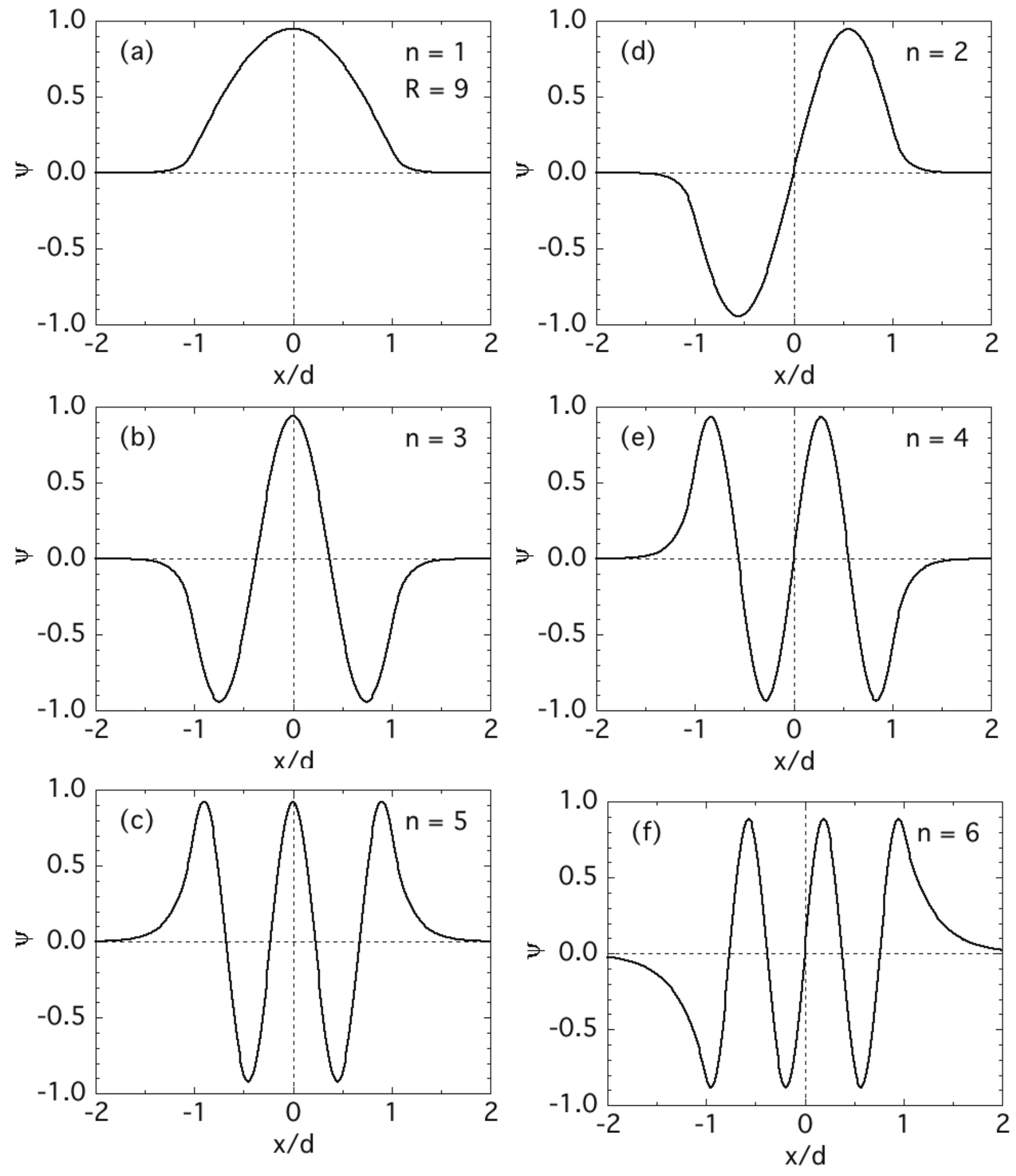}
\caption{Normalized bound-state wave functions $\psi$ versus position~$x_d\equiv x/d$, where $2d$ is the width of the potential well centered at $x/d = 0$.  Here the parameter $R = 9$ in Eq.~(\ref{Eq:PDef22}).  The data were calculated using Eqs.~(\ref{Eqs:WaveFunctions}) and the normalization factor in Eq.~(\ref{Eq:AnP}).  Panels~(a)--(c) with quantum numbers $n=1$, 3, and~5 show the three even-parity wave functions for $R=9$, whereas panels~(d)--(f) with $n=2$, 4, and~6 show the three odd-parity wave functions.  }
\label{Fig:SingleWellWaveFcnsP9} 
\end{figure}

\subsection{Wave Functions}

Once the values of $kd$ are calculated from Eq.~(\ref{Eq:kdCalc}) for a given value of~$R$ and~$n$, the value of $\kappa d$ is found using Eq.~(\ref{Eq:kappad}).  The odd values of $n$ correspond to even-parity wave functions, and the even values to odd-parity wave functions.  Since $kd(R,n)$ and $\kappa d(R,n)$ are now known, the even- and odd-parity wave functions $\psi(x_d,n,R)$ in Eqs.~(\ref{Eqs:PsiEven}) and~(\ref{Eqs:PsiOdd}) can be calculated for the respective $n$ and $R$ values.  The ground bound-state wave function $(n=1)$ always has even parity.

Before calculating and plotting the (real) bound-state wave functions versus position using Eqs.~(\ref{Eqs:WaveFunctions}), they must be multiplied by the normalization factor $A(n,R)$ which is obtained from the normalization condition as
\bea
A(n,R) =\left[\int_{-\infty}^\infty\psi^2(x_d,n,R)dx_d\right]^{-1/2},
\label{Eq:AnP}
\eea
where the $\psi(x_d,n,R)$ are given in Eqs.~(\ref{Eqs:WaveFunctions}).  Numerical calculations of the wave functions were carried out for $R=9$ for which there are six bound states according to Eq.~(\ref{Eq:nmax}).  Figures~\ref{Fig:SingleWellWaveFcnsP9}(a)--\ref{Fig:SingleWellWaveFcnsP9}(c) show plots of the even-parity wave functions versus~$x_d$ with $n=1$, 3, and~5, and Figs.~\ref{Fig:SingleWellWaveFcnsP9}(d)--\ref{Fig:SingleWellWaveFcnsP9}(f) show plots of the odd-parity wave functions versus~$x_d$ with $n=2$, 4, and~6.  The respective probability densities are ${\cal P}(x_d) = \psi^2(x_d,n,R)$ (not shown).

\begin{table}
\caption{\label{Tab:kavarP9} Real wave vector $k$ times the well half-width~$d$ and the reduced energy $\varepsilon \equiv -2m(-E)d^2/\hbar^2$ for the six bound states with quantum numbers $n$ in a square well with strength $R=9$. The odd $n$ values correspond to even-parity wave functions, whereas the even values correspond to odd-parity wave functions. }
\begin{ruledtabular}
\begin{tabular}{lcc}
$n$ & 	$kd$ 		&  $\varepsilon$ \\
\hline
1		& 1.41313		& $-0.975346$	\\
2		& 2.82259		& $-0.901642$		\\
3		& 4.22387 	& $-0.779740$		\\
4		& 5.61016		& $-0.611433$		\\
5		& 6.96842		& $-0.400507$		\\
6		& 8.26182 	& $-0.157313$		\\
\end{tabular}
\end{ruledtabular}
\end{table}

The eigenvectors $kd$ and reduced energies~$\varepsilon$ for the well with $R=9$ and six bound states are listed versus their quantum number in Table~\ref{Tab:kavarP9}.  Both $kd$ and $\varepsilon$ increase with increasing~$n$.  The slope of a plot of $\varepsilon$ versus~$(kd)^2$ is found to satisfy the dispersion relation for a free electron.  This is not surprising, because in the region \mbox{$-d < x < d$} the wave function of the electron in Eqs.~(\ref{Eqs:WaveFunctions}) is that of an electron standing wave that consists of a superposition of right- and left-moving traveling waves.

\subsection{\label{Sec:SWvsKP} Relationship of the Bound States at Negative Energies in the Attractive Square Well and in the Attractive Square-Well Kronig-Penney Model}

To connect the single-square-well model with the attractive Kronig-Penney model, one must first establish the relationship of the $R$ parameter in the former with the $Q$ parameter in the latter.  From the definitions of $R$ and $Q$ in Eqs.~(\ref{Eq:PDef22}) and~(\ref{Eq:SymbolDefs}), respectively, one has
\bea
\frac{R^2}{Q} = \left(\frac{d}{a}\right)^2.
\label{Eq:fracR2Q}
\eea
From Figs.~\ref{Fig:Kronig_Penney_Model}  and~\ref{Fig:PotentialWell}, one has 
\bea
d = \frac{b}{2} = a \frac{b/a}{2},
\eea
so Eq.~(\ref{Eq:fracR2Q}) gives
\bea
R = {\frac{b/a}{2}}\sqrt{Q}.
\eea
Similarly, Eqs.~(\ref{Eqs:Solve_knvsn}) become
\bse
\label{Eqs:Solve_kd23}
\bea
\frac{b/a}{2}ka + \arcsin\left(\frac{ka}{\sqrt{Q}}\right) = \frac{n\pi}{2},
\label{Eq:kdCalc2}
\eea
where the maximum quantum number for a bound state is
\bea
n^{\rm max} = {\rm Integer}\left[\frac{(b/a)\sqrt{Q}}{\pi}\right] + 1.
\label{Eq:nmax2}
\eea
\ese
The reduced energy~$\varepsilon$ and decay constant $\kappa a$ in Eqs.~(\ref{Eqs:epskap}) become
\bse
\bea
\varepsilon &=& -\left(\frac{b/a}{2}\right)^2\left[Q-(ka)^2\right],\\
\kappa a &=& \sqrt{Q - (ka)^2},\\
Q &=& (ka)^2 + (\kappa a)^2.
\eea
\ese

\begin{table}
\caption{\label{Tab:SqWellPars} Bound-state parameters of an attractive square well with strength~$Q$ appropriate to the Kronig-Penney model. The values of $Q$, quantum number~$n$, parity, $ka$, $\kappa a$, and the reduced energy~$\varepsilon$ are listed.}
\begin{ruledtabular}
\begin{tabular}{lccccc}
$Q$ 		& $n$ 	& parity 	& $ka$ 		& $\kappa a$	&  $\varepsilon$ \\
\hline
5  		&	1	& even	& 1.9703		& 1.0574		& $-0.0699$	\\
10 		&	1	& even	& 2.5440		& 1.8783		& $-0.2205$  	\\
20 		&	1	& even	& 3.1531	 	& 3.1714		& $-0.6286$  	\\
30 		&	1	& even	& 3.5053		& 4.2087		& $-1.1071$  	\\
40 		&	1	& even	& 3.7469		& 5.0952		& $-1.6226$  	\\
 		&	2	& odd	& 6.3242		& 0.0649		& $-0.0003$  	\\
50 		&	1	& even	& 3.9276	 	& 5.8800		& $-2.1609$  	\\
 		&	2	& odd	& 6.9678	 	& 1.2043		& $-0.0906$  	\\
60 		&	1	& even	& 4.0702 		& 6.5904		& $-2.7146$  	\\
 		&	2	& odd	& 7.4299	 	& 2.1902		& $-0.2998$  	\\
70 		&	1	& even	& 4.1868	 	& 7.2437		& $-3.2794$  	\\
 		&	2	& odd	& 7.7843	 	& 3.0667		& $-0.5878$  	\\
\end{tabular}
\end{ruledtabular}
\end{table}

For the Kronig-Penney band structures in Fig.~\ref{Fig:KP_Evska_Q10_40_70}, one has $b/a = 1/2$.  Using this ratio, Table~\ref{Tab:SqWellPars} lists the values of the bound-state quantum number~$n$, parity, $ka$, $\kappa a$, and of the reduced energy~$\varepsilon$ for eight $Q$ values including those in Fig.~\ref{Fig:KP_Evska_Q10_40_70}. 

\newpage
\acknowledgments

The author is grateful to \mbox{D. D. Johnson} and L.-L. Wang for helpful comments on the manuscript.  This research was supported by the U.S. Department of Energy, Office of Basic Energy Sciences, Division of Materials Sciences and Engineering.  Ames Laboratory is operated for the U.S. Department of Energy by Iowa State University under Contract No.~DE-AC02-07CH11358.


\end{document}